\documentclass[a4paper,11pt]{article}
\pdfoutput=1

\usepackage{jheppub2}
\bibstyle{JHEP}

\usepackage[english]{babel}
\usepackage[utf8]{inputenc}
\usepackage[T1]{fontenc}
\usepackage{amsfonts}
\usepackage{amsmath}         
\usepackage{amsfonts}          
\usepackage{amssymb}
\usepackage{mathtools}
\usepackage{comment}
\usepackage{tikz}
\usetikzlibrary{decorations.markings}
\usepackage{physics}
\usepackage{dsfont}

\newcommand{\overbar}[1]{\mkern 1.5mu\overline{\mkern-1.5mu#1\mkern-1.5mu}\mkern 1.5mu}

\renewcommand{\exp}[1]{\text{exp}\,{#1}}
\usepackage[shortlabels]{enumitem}
\usepackage{subcaption}
\title{Towards complexity of primary-deformed Virasoro circuits}

\author[1]{Johanna Erdmenger,}
\author[1]{Jani Kastikainen,}
\author[1,2]{and Tim Schuhmann}
\affiliation[1]{Institute for Theoretical Physics and Astrophysics and Würzburg-Dresden Cluster of Excellence
ct.qmat, Julius-Maximilians-Universität Würzburg, Am Hubland, 97074 Würzburg, Germany}
\affiliation[2]{Department of Physics and Astronomy, Ghent University, 9000 Ghent, Belgium}

\emailAdd{erdmenger@physik.uni-wuerzburg.de}
\emailAdd{jani.kastikainen@uni-wuerzburg.de}
\emailAdd{tim.schuhmann@ugent.be}


\abstract{The Fubini--Study metric is a central element of information geometry. We explore the role played by information geometry for determining the circuit complexity of Virasoro circuits and their deformations. 
To this effect, we study unitary quantum circuits generated by the Virasoro algebra and Fourier modes of a primary operator. Such primary-deformed Virasoro circuits can be realized in two-dimensional conformal field theories, where they provide models of inhomogeneous global quenches. We consider a cost function induced by the Fubini--Study metric and provide a universal expression for its time-evolution to quadratic order in the primary deformation for general source profiles. For circuits generated by the Virasoro zero mode and a primary, we obtain a non-zero cost only if spatial inhomogeneities are sufficiently large. In this case, we find that the cost saturates when the source becomes time-independent. The exact saturation value is determined by the history of the source profile. As a byproduct, returning to undeformed circuits, we relate the Fubini--Study metric to the Kähler metric on a coadjoint orbit of the Virasoro group.}

\preprint{$\begin{array}{rr}
	\text{ }\end{array}$
}

\begin{document}
\maketitle
\flushbottom

\newpage

\section{Introduction}

In recent years, through the study of quantum information concepts within the AdS/CFT correspondence and its generalizations, particular emphasis has been placed on the fact that information and geometry are intimately related and conspire to provide new insights into gravity. Information geometry, or more specifically the study of geometry on the space of probability distributions, both in the classical and in the quantum context, is a long-standing subject that predates the AdS/CFT correspondence \cite{amari2000methods}. The relevance of information geometry for quantum field theory and the AdS/CFT correspondence was realized already in \cite{Blau:2001gj,Malek:2015hea} and explained further in 
\cite{Erdmenger:2020vmo}. The relation between information and geometry plays a particularly crucial role in the context of defining circuit complexities for quantum field theories (QFT), but a challenge lies in the infinite dimensionality of the Hilbert space.

Quantum information geometry is a manifold of quantum states (pure or mixed) equipped with a Riemannian metric \cite{petz_geometries_1996,petz_monotone_1996}. An important role is played by the \textit{Fubini--Study (FS) metric} \cite{Gibbons:1991sa,Braunstein:1994zz} which is the unique $U(N +1)$-invariant metric, up to an overall factor, 
on the finite-dimensional complex projective space $\mathbb{C}\mathrm{P}^N$, the Hilbert space of a quantum system. The Fubini--Study metric is also a natural candidate metric to consider on infinite-dimensional state spaces of QFTs as proposed in \cite{Chapman:2017rqy}. A class of such state spaces can be constructed as orbits of a reference quantum state under the action of a unitary projective representation of an infinite-dimensional symmetry group. An example is the space of Virasoro states obtained by acting with the Virasoro group \cite{Caputa:2018kdj,deBoer:2023lrd}, the symmetry group of a two-dimensional conformal field theory (CFT). In this case there is a correspondence between Virasoro states and diffeomorphisms constituting a Virasoro coadjoint orbit \cite{Witten:1987ty}, i.e.~a quotient space of the infinite-dimensional Lie group $\text{Diff}_+ S^1$.

Curves on any orbit of states can be interpreted as unitary quantum circuits, where each gate of the circuit is a unitary representation of the underlying group, and which can be assigned a circuit complexity using the original approach of Nielsen \cite{Nielsen:2005mkt}. The approach was applied to circuits on orbits of Gaussian states in free field theories in \cite{Chapman:2017rqy,Jefferson:2017sdb,Khan:2018rzm}, fermionic free field theories \cite{Hackl:2018ptj} and extended to \textit{Virasoro circuits} on the infinite-dimensional space Virasoro pure states in \cite{Caputa:2018kdj} and \cite{Flory:2020eot,Flory:2020dja}.\footnote{Virasoro circuits have also been called conformal circuits, because the corresponding unitary operators represent conformal transformations. In this paper, we prefer the name Virasoro circuit to highlight its definition using the Virasoro group.} Virasoro circuits make use of conformal transformations as computational gates. They have now also been realized as physical time-evolution in a two-dimensional CFT driven by a background metric \cite{Erdmenger:2021wzc,deBoer:2023lrd}.

In general, the definition of complexity requires the choice of a cost function which measures how costly it is to implement a particular gate, e.g.~on a quantum computer. Functions linear in the circuit generator have been used as cost functions in this context. For example, using coadjoint orbits, the associated circuit complexity may be mapped to a geometric action \cite{Erdmenger:2020sup}. Nevertheless,
it has emerged that these may be of restricted use, in particular when the reference state is an eigenstate of the gates, and the use of the Fubini--Study metric as a cost function is generically required \cite{Erdmenger:2020sup,Flory:2020dja,Flory:2020eot,Erdmenger:2021wzc}. Furthermore, the Fubini--Study metric is also natural to investigate in QFTs given its direct connection with the partition function. What differs from the original approach of Nielsen in this case is that generically, gates can be assigned different cost at different stages of the circuit. When the line element of the circuit in the Fubini--Study metric is used as the cost function (the Fubini--Study cost), the notion of complexity coincides with the Riemannian distance between points on the quantum information geometry. In this paper, we relate the Fubini--Study metric on the space of Virasoro states to the Kähler metric of a Virasoro coadjoint orbit.

Previous work considering the Fubini--Study circuit complexity on infinite-dimensional state spaces has focused on Virasoro circuits, which, however, have the shortcoming of being bound to a single Verma module (irreducible representation) of the Virasoro algebra. In realistic systems such as CFTs, the Hilbert space consists of multiple Verma modules and it is of interest to quantify complexity between them. In this paper, we take the first step in this direction and calculate the Fubini--Study cost of primary-deformed Virasoro circuits. These are circuits whose generator, in addition to containing elements of the Virasoro algebra, include modes of a general primary operator.\footnote{The linear cost function (instead of the Fubini--Study cost) for circuits generated by Kac--Moody and BMS algebras were studied in \cite{Bhattacharyya:2022ren,Bhattacharyya:2023sjr}.} They map the initial reference state to a linear combination of states in different Verma modules. Clearly, the state space explored by primary-deformed circuits is larger than the space of Virasoro pure states. We note that within quantum field theory, an example for the primary operator considered is the $J\bar{J}$ deformation \cite{Sfetsos:2013wia,Frolov:2019xzi}, whose Krylov complexity was evaluated in \cite{Chattopadhyay:2024pdj}. 

The central aim of this paper is to understand how the circuit complexity involving the Fubini--Study cost function behaves for Virasoro circuits perturbed by a primary operator. By definition, the circuit complexity of a given target state is obtained by minimizing the \textit{accumulated cost} over all circuits that connect it to the reference state, a difficult procedure we are not able to perform in general. Therefore we focus on the calculation of the accumulated cost of a single primary-deformed circuit, which provides an upper bound on the complexity of the instantaneous state of the circuit. One of our main results is a universal expression for the Fubini--Study cost up to quadratic order in the primary deformation. We neglect non-universal higher-order corrections that depend on the finer details of the primary operator algebra (spectrum and 3-point correlation function coefficients).

We consider two types of primary-deformed circuits, chiral and coupled, involving either one or two copies of the Virasoro circuit. In the primary-deformed chiral circuit, a single Virasoro circuit generator is deformed by the inclusion of a primary operator. In the coupled case, two Virasoro circuits are coupled together via a tensor product primary operator. We are able to obtain explicit results for the cost when each of the following three conditions are met: the Virasoro part of the circuit is generated only by the Virasoro zero mode $L_0$, the source of the primary deformation factorizes into a spatial and a temporal profile and the circuit starts from the vacuum state. For these circuits, all non-trivial time-dependence of the generator is encoded entirely by the primary deformation and its source function.

Our results in this case apply to arbitrary shaped spatial inhomogeneities, but simple expressions are obtained for single Fourier modes of the spatial source profile. We find that a spatially homogeneous source always has zero cost due to the fact that the corresponding primary operator mode annihilates the vacuum state. The same applies to low lying spatial Fourier modes that are below the weight of the primary operator. Therefore we find that \textit{a non-vanishing cost requires sufficiently large spatial inhomogeneity}. For sufficiently large spatial inhomogeneities, the cost is non-zero and evolves non-trivially in time. In our analysis, we prove three distinctive characteristics of the time-evolution which are universal for both chiral and coupled circuits: \textit{the cost saturates when the source becomes time-independent}, \textit{the saturation value depends on the source history} and \textit{the source returning to zero does not imply that the cost saturates to zero}. In particular, this also implies linear growth of the accumulated cost in time after the source has become time-independent. For chiral and coupled case, we showcase these characteristics in three examples. In addition to a simple constant time profile, we consider two time-dependent profiles that model global inhomogeneous quenches: as a function of time, the source is smoothly switched on from zero, after which it either remains turned on, or it is smoothly switched off again. In these examples, alongside the previously described characteristics, numerically we observe a difference given a non-trivial time profile: in the coupled circuit higher Fourier modes of the spatial source are assigned higher cost, while in the chiral circuit this is not the case.

The paper is structured as follows. First, in the remainder of the introduction, we will give a
brief summary of the setup and of the main results of the paper. Then in Section \ref{sec:conformalcircuits} we review Virasoro circuits and revisit the calculation of the Fubini--Study metric in this context. In Section \ref{sec:deformedcircuits} we introduce primary-deformed Virasoro circuits and show in full generality how its Fubini--Study cost can be calculated in perturbation theory. Finally in Section \ref{sec:trivialtimeevolution} we present our main results for the time-evolution of the Fubini--Study cost in explicit examples. We conclude with a discussion in Section \ref{sec:outlook}. Details of the calculations are relegated to the Appendices.

\subsection{Summary of results}\label{subsec:summaryofresults}

Let us now summarize our results regarding Virasoro circuits and its primary-deformed generalizations in detail.

\paragraph{Virasoro circuits.} A Virasoro circuit is a one-parameter family of quantum states, parametrized by $t$, in a Hilbert space obtained by acting on a reference state with a time-ordered exponential of Virasoro algebra generators. Explicitly
\begin{equation}
    \ket{t} = \overleftarrow{\mathcal{T}}\exp{\biggl(-i\int_0^t ds\,G(s)\biggr)}\ket{R}\,.
    \label{eq:circuitsummary}
\end{equation}
with generator of the Virasoro circuit
\begin{equation}
    G(t) = \int_0^{2\pi} d\phi\,u_t(\phi)\,T(\phi)=\sum_{n} u_{t,n}^*\,L_n \,,
\end{equation}
where $u_t(\phi) = u_t(\phi+2\pi)$ is a real periodic function with Fourier modes $u_{t,n}^* = u_{t,-n}$ and $T(\phi)$ is the stress tensor constructed from generators of the Virasoro algebra $L_n$. The Hilbert space could be for example that of a 2D conformal field theory and the reference state is commonly taken to be a highest-weight state. We also consider a tensor product circuit generated by two independent copies of the Virasoro algebra
\begin{equation}
    G(t) =\int_0^{2\pi} d\phi\,u_t(\phi)\,T(\phi)\otimes \mathbf{1}-\int_0^{2\pi} d\phi\,\overbar{u}_t(\phi)\,\mathbf{1}\otimes T(-\phi)
\end{equation}
with two independent periodic functions $u(\phi) = u(\phi+2\pi)$ and $\overbar{u}(\phi+2\pi) = \overbar{u}(\phi)$. Such circuits can be realized in 2D CFTs coupled to a time-dependent background metric \cite{Erdmenger:2021wzc,deBoer:2023lrd}.

In the first part of the paper (Section \ref{sec:conformalcircuits}) we revisit the Fubini--Study cost of Virasoro circuits. Virasoro circuits $\ket{f_t}$ are curves on the space of states corresponding to one-parameter families of diffeomorphisms $f_t$ on the manifold $\widetilde{\text{Diff}}_+S^1$. The line element of the curve in the FS metric gives the FS cost which coincides with the variance of the generator in the instantaneous state of the circuit
\begin{equation}
     \mathcal{F}_{\text{FS}}(t) = \langle \Psi(t)\vert G(t)^2\vert \Psi(t)\rangle - \langle \Psi(t)\vert G(t)\vert \Psi(t)\rangle^2\,.%\mathbb{V}_{\Psi(t)}[G(t)]
\end{equation}
We obtain the result
\begin{equation}
     \mathcal{F}_{\text{FS}}(t) = \frac{c}{12}\sum_{n=1}^{\infty} n \,\biggl(\frac{24h}{c}+n^{2}-1\biggr)\,\vert\widetilde{u}_{t,n}\vert^2\,,
\end{equation}
where $c$ is the central charge of the Virasoro algebra and $\widetilde{u}_{t,n}$ are Fourier modes of a function $\widetilde{u}(\phi)$ constructed from $u(\phi)$. To the knowledge of the authors this representation of the Fubini--Study line element has not appeared in the literature before and it uncovers a previously missed connection to the unique Kähler metric on the coadjoint orbit $\widetilde{\text{Diff}}_+S^1\slash \,\mathbb{R}$ of the Virasoro group \cite{Kirillov:1987mn}. In contrast to previous work \cite{Flory:2020dja,Flory:2020eot}, our derivation does not rely on differential regularization of the stress tensor 2-point function. To connect with previous work, we explain how the same answer can also be obtained using differential regularization.

As a simple application of our results, we consider a subset known as M\"obius circuits, which are generated by operators of the form
\begin{equation}
    G(t) = u_{t,0}^*\,L_0 + u_{t,-k}^*\,L_{-k}+u_{t,k}^*\,L_k\,,
\end{equation}
where the Virasoro generators $L_0$ and $L_{\pm k}$ (with $k=1,2,\ldots$) form an $\mathfrak{s}\mathfrak{l}(2,\mathbb{R})$ subalgebra of the Virasoro algebra. M\"obius circuits move on the manifold of M\"obius pure states for which we show that Fubini--Study metric reduces to a hyperbolic metric with curvature radius $\sqrt{h_k\slash 2}$\,,
\begin{equation}
    \mathcal{G}_{ij}^{\text{FS}}(p)\,dp^idp^j = \frac{h_k}{2}\,(d\rho^2+\sinh^2{(\rho)}\,d\chi^2)\,,\quad h_k = \frac{c}{24}\left(k-\frac{1}{k}\right)+\frac{h}{k}\,.
\end{equation}
This result matches with the metric computed in \cite{Caputa:2021sib} on the space of generalized coherent states where its volume element is related to Krylov complexity. We explain that the same metric arises, because generalized coherent states are equal to M\"obius pure states up to a phase factor cancelled in the Fubini--Study metric.

\paragraph{Primary deformation.} In the second part of the paper (Sections \ref{sec:deformedcircuits} and \ref{sec:trivialtimeevolution}), we consider primary-deformed Virasoro circuits. The (chiral) primary-deformed Virasoro circuit is given by equation \eqref{eq:circuitsummary} but with the modified generator
\begin{equation}
    G(t) =\int_0^{2\pi} d\phi\,u_t(\phi)\,T(\phi) + \lambda\int_0^{2\pi} d\phi\,J(\phi,t)\,\mathcal{O}_h(\phi)\,,
\end{equation}
where $\mathcal{O}_h(\phi)$ is a local primary operator of weight $h$ with specific commutation relations, $J(\phi,t) = J(\phi+2\pi,t)$ is the source function of the deformation and $\lambda$ is a parameter. We also consider the coupled version of the circuit consisting of a tensor product of two Virasoro circuits that are coupled together by a primary operator of weight $(h,\overbar{h})$ as
\begin{equation}
    G(t) =\int_0^{2\pi} d\phi\,u_t(\phi)\,T(\phi)\otimes \mathbf{1}-\int_0^{2\pi} d\phi\,\overbar{u}_t(\phi)\,\mathbf{1}\otimes T(-\phi) + \lambda\int_0^{2\pi} d\phi\,J(\phi,t)\,\mathcal{O}_h(\phi)\otimes \mathcal{O}_{\overbar{h}}(-\phi)\,.
\end{equation}
One of the main results of our paper is a perturbative expansion of the Fubini--Study cost of the two primary-deformed circuits up to quadratic order in the parameter $\lambda\ll 1$,
\begin{equation}
    \mathcal{F}_{\text{FS}}(t) =\mathcal{F}_{\text{FS}}^{(0)}(t) + \lambda\,\mathcal{F}_{\text{FS}}^{(1)}(t) + \lambda^2\,\mathcal{F}_{\text{FS}}^{(2)}(t) + \mathcal{O}(\lambda^3)\,.
\end{equation}
We derive an expression for each of the three terms in terms of expectation values of operators constituting the generator. In the simple case of $f_t\in \widetilde{SL}(2,\mathbb{R})$ starting from the vacuum state, we show that $\mathcal{F}_{\text{FS}}^{(0)}(t) = \mathcal{F}_{\text{FS}}^{(1)}(t) = 0 $ so that the leading correction to the cost is $\mathcal{F}_{\text{FS}}^{(2)}(t)$ for which we provide a simplified expression.

Using this result we are able to explicitly compute the time evolution of the cost when the following two conditions are met: the Virasoro circuit is generated by the zero mode $u_t(\phi) = -\overbar{u}_t(\phi) = 1$ and the deformation factorizes $J(\phi,t) = S(\phi)\,j(t)$. This is a model of an inhomogeneous global quench where $S(\phi)= S(\phi+2\pi)$ controls spatial inhomogeneities while $j(t)$ determines the quench profile. In this case, $\mathcal{F}_{\text{FS}}^{(2)}(t)$ is completely determined by the temporal source profile $j(t)$ and the real part of the vacuum 2-point function of ring operators which are defined as
\begin{equation}
    \mathcal{R}_{h}(t) \equiv \int_0^{2\pi}d\phi\, S(\phi)\,\mathcal{O}_h(\phi-t)\,,\quad \mathcal{R}_{h\overbar{h}}(t) \equiv \int_0^{2\pi}d\phi\, S(\phi)\,\mathcal{O}_h(\phi-t)\otimes \overbar{\mathcal{O}}_{\overbar{h}}(\phi+t) \,,
\end{equation}
for the chiral and coupled circuits, respectively. It is very important that only the real part contributes to the Fubini--Study cost, because the 2-point function also contains an imaginary part which, in the coupled case, has a pole at coincident times. We derive explicit expressions for the real part of the ring operator 2-point function for integer $h \geq 1$ in the chiral case and integer $h = \overbar{h}\geq 1$ in the coupled case. The results are expressed as infinite series involving Fourier modes $S_n$ of the spatial source $S(\phi) $.

We investigate, both analytically and numerically, the time-evolution $\mathcal{F}_{\text{FS}}^{(2)}(t)$ for different temporal $j(t)$ and spatial $S(\phi)$ profiles in the example of primary-deformed Virasoro zero mode evolution. We find four key characteristics (i-iv) universal to both coupled and chiral circuits. For the spatial source $S(\phi)$, we find that (i) \textit{non-vanishing cost requires sufficienly large spatial inhomogenity}: Fourier modes $S_n$ with $n < h$ and $n < 2h$ for $n\in\mathbb{N}_0$ do not contribute in the chiral and coupled circuits respectively. In particular, this implies that the spatially homogeneous global quench $S(\phi) =$ constant has vanishing cost. The reason for this is that the mode operators of the primary operator annihilate the vacuum $\mathcal{O}_{h,n}\ket{0} = 0$ when $n > -h$. We compute the non-zero cost as a function of $t$ for the case of a single spatial mode $S(\phi) = \cos{(n\phi)}$ for various time-dependent quench profiles $j(t)$. The influence of the temporal profile of the source is that (ii) \textit{the cost saturates when the source becomes time-independent}, (iii) \textit{the saturation value depends on the source history} and that (iv) \textit{source returning to zero does not imply cost saturation to zero}. Characteristic (ii) further implies linear growth at late times as a bound from above for the quantum circuit complexity of primary-deformed Virasoro circuits reaching target states that could also be reached from the vacuum via primary-deformed Virasoro zero mode evolution with source reaching a constant. All four characteristics are showcased via three explicit example circuits: $j(t) =$ constant, a switch-on profile where $j(t)$ smoothly increases from zero to a finite value, and a switch-on to switch-off profile where $j(t)$ returns to zero after the switch-on. Numerically, we furthermore observe differences in these examples in the qualitative behaviour of the cost profile and saturation values between chiral and coupled circuits.

\section{Virasoro circuits revisited}\label{sec:conformalcircuits}

A unitary quantum circuit is a one-parameter family of quantum states obtained by acting on a reference state with a sequence of successive unitary operators (gates) taken from a predetermined set. Each gate can be assigned a cost which measures the difficulty to implement it, for example by a quantum computer. The cost of a single gate can be used to assign a complexity either to the circuit or to the final (target) state reached by the circuit. The definition of the cost function is ambiguous, but for a continuous circuit there is a natural definition using quantum information geometry: the cost is an infinitesimal line element of the circuit on the Riemannian manifold of states equipped with a quantum information metric.

In this section, we will consider the Fubini--Study cost function defined using the Fubini--Study quantum information metric and the manifold of Virasoro (pure) states obtained by acting on a reference state with a unitary representation of the Virasoro group. When the reference state is a highest-weight state, the space of Virasoro pure states is a Verma module of the Virasoro algebra. Quantum circuits on such an infinite-dimensional state space are known as Virasoro (or conformal) circuits \cite{Caputa:2018kdj,Flory:2020dja,Flory:2020eot,Erdmenger:2021wzc,deBoer:2023lrd} and explicit expressions for unitary gates can be obtained due to properties of the Virasoro algebra. After defining the FS cost function and complexity, we will review the properties of unitary representations of the Virasoro group and its M\"obius subgroups. Then we will derive an explicit expression for the FS metric on Virasoro pure states which clarifies previous derivations \cite{Flory:2020dja,Flory:2020eot} regarding regularization and point out a previously missed connection to the Kähler metric of a Virasoro coadjoint orbit. On the $SL_k(2,\mathbb{R})$ M\"obius subgroup, we show that the metric reduces to the two-dimensional hyperbolic metric.

\subsection{Complexity of unitary quantum circuits}\label{subsec:complexityofcircuits}

Let us consider a Hilbert space that admits the action of a basis of Hermitian operators $\mathcal{O}_n$ that form an algebra $\mathcal{A}$ and let $\mathcal{U}$ be the set of unitary operators $e^{-i\sum_nu_n^*\mathcal{O}_n}$ (with $u_n^* = u_{-n}$) obtained by exponentiating elements of this algebra. Consider a reference state $\ket{R}$ and let $\mathbb{S}_R$ be the space of states obtained by acting with $\mathcal{U}$ on $\ket{R}$. A unitary quantum circuit is a one-parameter family of states $\mathcal{C}\colon \mathcal{U}\times (0,t)\rightarrow \mathbb{S}_R$ given by
\begin{equation}
    \ket{t} = U(t)\ket{R}\,,
    \label{eq:circuitstategeneral}
\end{equation}
where the unitary operator $U(t)$ is the solution of the equation
\begin{equation}
    \partial_t U(t) = -i\,G(t)\,U(t)\,,\quad U(0)\equiv 1\,,
    \label{eq:unitaryoperatorequation}
\end{equation}
for a Hermitian generator $G(t)$ which is a linear combination of basis operators with time-dependent coefficients
\begin{equation}
    G(t) = \sum_n u_{n,t}^*\,\mathcal{O}_n\,,\quad u_{n,t}^* = u_{-n,t}\,.
\end{equation}
The unitary operator is explicitly given by
\begin{equation}
    U(t) = \overleftarrow{\mathcal{T}}\exp{\biggl(-i\int_0^{t} ds\,G(s)\biggr)}\,,
    \label{eq:circuitunitary}
\end{equation}
where the time-ordering $\overleftarrow{\mathcal{T}}$ orders operators with larger $t$ to the left. The solution $U(t)$ can also be expressed as an ordinary exponential of a Magnus expansion involving nested commutators of the basis operators $\mathcal{O}_n$. Because the commutation relations $[\mathcal{O}_n,\mathcal{O}_m]$ close within the algebra, $U(t)$ is an ordinary exponential of a linear combination of basis operators with time-dependent coefficients.

Let us pick $N$ points $\{s_n\}_{n=0,\ldots,N-1}$ from the time-interval $(0,t)$ such that $s_0 = 0$ and $s_N = t$. The interval is divided into $N$ steps of size $\delta s_n = s_{n} - s_{n-1}$ and the circuit \eqref{eq:circuitstategeneral} decomposes to a sequence of successive unitary gates acting on the reference state
\begin{equation}
    \ket{t} = \lim_{N\rightarrow \infty}e^{-i\delta s_N\,G(s_{N-1})}\cdots e^{-i\delta s_1\,G(s_0)}\ket{R}\,.
\end{equation}
The unitary operators $e^{-i\delta s_n\,G(s_{n-1})}$ form the gates of the quantum circuit and the set of all possible gates is formed by the unitary operators $e^{-i\sum_n u_n^*\mathcal{O}_n}$.

\paragraph{Fubini--Study circuit complexity.}

Consider two one-parameter families of states $\ket{\Psi_u(\lambda)}$ and $\ket{\Psi_v(\lambda)}$ that satisfy $\ket{\Psi_u(0)} = \ket{\Psi_v(0)}\equiv \ket{R}$. The Fubini--Study (FS) metric \cite{Gibbons:1991sa,Braunstein:1994zz} at point $\ket{R}$ in the space of states, contracted with two ``tangent vectors'' $G_u$ and $G_v$ at $\ket{R}$, may be defined as the Hessian
\begin{equation}
    \mathcal{G}_{\ket{R}}^{\text{FS}}(G_u,G_v) \equiv \frac{\partial^2}{\partial \lambda_1\partial \lambda_2}\,\vert \langle \Psi_u(\lambda_1) \vert \Psi_v(\lambda_2)\rangle \vert\bigg\vert_{\lambda_1 = \lambda_2 = 0}\,,
    \label{eq:FSmetricdef}
\end{equation}
where $\vert \langle \Psi \vert \Psi'\rangle \vert \equiv \sqrt{\langle \Psi \vert \Psi'\rangle\langle \Psi \vert \Psi'\rangle^{*}}$. Defining quantum information metrics on the space of mixed states as Hessians of information divergences was first done in \cite{hasegawa1993alpha,petz_riemannian_1996} and our formula \eqref{eq:FSmetricdef} presents a special case when the states are pure. In particular, the FS metric can be written as the Hessian of the Bures distance \cite{bures1969extension} when it is restricted to pure states \cite{Braunstein:1994zz}. Therefore the analog of the FS metric for mixed states is the Bures metric.\footnote{Field-theory perturbative corrections to the Bures metric were computed in \cite{Bohra:2021zyw}.}

Let us now focus on the space of states $\ket{\Psi} = U_\Psi\ket{R}$ obtained by acting on the reference state $\ket{R}$ with a unitary operator $U_\Psi$ where $\Psi$ labels a possibly infinite set of real valued parameters. Let $U_u(\lambda)$ and $U_v(\lambda)$ be two one-parameter families of unitaries of this type obtained by taking two curves $\Psi_u(\lambda)$ and $\Psi_v(\lambda)$ on the space of parameters $\Psi$. These unitaries define two curves on the space of states of the type $U_\Psi\ket{R}$ via
\begin{equation}
    \ket{\Psi_u(\lambda)} = U_u(\lambda)\ket{R}\,,\quad \ket{\Psi_v(\lambda)} = U_v(\lambda)\ket{R}\,.
\end{equation}
Let us assume that both of these curves go through the state $\ket{\Phi} $ at $\lambda = 0$, in other words, the unitaries satisfy $U_u(0) = U_v(0) \equiv U_\Phi$. Substituting this to the FS metric, we get
\begin{equation}
    \mathcal{G}_{\ket{\Phi}}^{\text{FS}}(G_u,G_v) = \frac{1}{2}\bra{\Phi}(G_u\,G_v + G_v\,G_u)\ket{\Phi} - \bra{\Phi} G_u\ket{\Phi}\bra{\Phi} G_v\ket{\Phi}\,,
    \label{eq:FSmetricwithsymmetrization}
\end{equation}
where the tangent vectors $G_{u}$ and $G_{v}$ at the state $\ket{\Phi} = U_\Phi\ket{R}$ are defined by
\begin{equation}
    \partial_\lambda U_{u}(\lambda)\vert_{\lambda = 0} \equiv -i\,G_{u}\,U_\Phi\,,\quad \partial_\lambda U_{v}(\lambda)\vert_{\lambda = 0} \equiv -i\,G_{v}\,U_\Phi\,.
\end{equation}
Let us then consider a generic curve $\ket{\Psi(t)} = U(t)\ket{R}$ on the space of states $U_\Psi\ket{R}$ where $U(t)= U_{\Psi(t)}$ satisfies the equation \eqref{eq:unitaryoperatorequation}. The diagonal component of the FS metric \eqref{eq:FSmetricdef} at point $\ket{\Psi(t)}$ along the curve reduces to \cite{Flory:2020eot,Flory:2020dja}
\begin{equation}
\mathcal{F}_{\text{FS}}(t)\equiv\mathcal{G}_{\ket{\Psi(t)}}^{\text{FS}}(G(t),G(t)) =  \langle \Psi(t)\vert G(t)^2\vert \Psi(t)\rangle - \langle \Psi(t)\vert G(t)\vert \Psi(t)\rangle^2   \,,
\label{eq:FSmetricfinitedimensional}
\end{equation}
which is the infinitesimal line element of the curve $\ket{\Psi(t)}= U(t)\ket{R}$ in the FS metric. We call the line element \eqref{eq:FSmetricfinitedimensional} the \textit{Fubini--Study cost}, because it can be thought of measuring the cost of performing an infinitesimal unitary operation. The accumulated Fubini--Study cost (or length) of the curve is defined by
\begin{equation}
\label{eq:accumulated_cost}
    \mathcal{L}_{\text{FS}}(t) \equiv \int_{0}^tds\,\sqrt{ \mathcal{F}_{\text{FS}}(t)}\,.%\,,\quad \mathcal{S}_{\text{FS}}(t) = \int_{-\infty}^tds\,\text{Var}\,G(s) \,.
\end{equation}
Let us pick a target state $\ket{T} = U_T\ket{R}$ in the space of states $\mathbb{S}_R$. Then the FS complexity of the state $\ket{T}$, or equivalently of the unitary $U_T$ creating the state, is defined as the minimum value of the accumulated FS cost
\begin{equation}
    \mathcal{C}_{\text{FS}}[\ket{T}] = \underset{\mathcal{C}}{\text{min}}\,\mathcal{L}_{\text{FS}}(t)
    \label{eq:FScomplexity}
\end{equation}
over all curves $\mathcal{C}\colon \mathcal{U}\times (0,t)\rightarrow \mathbb{S}_R$ connecting the reference state $\ket{R}$ to the target state $\ket{T}$.

When the circuit generator is time-independent $\dot{G}(t) = 0$, the circuit unitary operator is simply an ordinary exponential:
\begin{equation}
    U(t) = e^{-itG(0)}\,,\quad \text{when}\quad G(t) = G(0)\,.
\end{equation}
Because $G(0)$ commutes with itself $[e^{-itG(0)},G(0)] = 0 $, it follows that the FS cost \eqref{eq:FSmetricfinitedimensional} is also time-independent
\begin{equation}
    \mathcal{F}_{\text{FS}}(t) = \mathcal{F}_{\text{FS}}(0) = \langle R\vert G(0)^2\vert R\rangle - \langle R\vert G(0)\vert R\rangle^2\,,
    \label{eq:timeindependentcost}
\end{equation}
and that the accumulated cost presents linear growth $ \mathcal{L}_{\text{FS}}(t) \propto t$. This does not necessarily imply that the FS complexity \eqref{eq:FScomplexity} grows linearly, because there might exist another curve between $\ket{R}$ and $e^{-itG(0)}\ket{R}$ with a time-dependent generator and a lower accumulated cost.

\paragraph{Regularization of divergences.} Equation \eqref{eq:FSmetricfinitedimensional} is the correct expression in finite dimensional Hilbert spaces. When $G(t)$ is local operator of $t$ (which will be the case below), the equal-time product $G(t)^2$ is not well defined and must be understood as a limit of a meromorphic operator valued function in the complex plane (it is a hyperfunction or a generalized distribution). More precisely, we have
\begin{equation}
    G(t)\,G(t') = \lim_{\varepsilon\rightarrow 0^+}\overleftarrow{\mathcal{I}}\{G(t+i\varepsilon)\,G(t')\}\,,\quad G(t')\,G(t) = \lim_{\varepsilon\rightarrow 0^+}\overleftarrow{\mathcal{I}}\{G(t-i\varepsilon)\,G(t')\}\,,
\end{equation}
where $\overleftarrow{\mathcal{I}}$ denotes ordering in the imaginary direction when $t \in \mathbb{C}$. Hence the product $G(t)^2$ contains both an imaginary and a real part so that the naive expression \eqref{eq:FSmetricfinitedimensional} would be complex valued and its imaginary part may also be divergent. The correct real answer is obtained by starting from the original expression \eqref{eq:FSmetricwithsymmetrization} by setting $\ket{\Phi} = \ket{\Psi(t)} =U(t)\ket{R}$, defining
\begin{equation}
    G_u = G(t)\,,\quad G_v = G(t')\,,
\end{equation}
and taking the limit $t'\rightarrow t$. This gives the expression
\begin{align}
\label{eq:FSregularization}
&\mathcal{G}_{\ket{\Psi(t)}}^{\text{FS}}(G(t),G(t))\\
&=  \lim_{t'\rightarrow t}\lim_{\varepsilon\rightarrow 0^+}\frac{1}{2}\,\langle \Psi(t)\vert\overleftarrow{\mathcal{I}}\{G(t+i\varepsilon)\,G(t') + G(t-i\varepsilon)\,G(t')\}\vert \Psi(t)\rangle - \langle \Psi(t)\vert G(t)\vert \Psi(t)\rangle^2   \,.\nonumber
\end{align}
It turns out that
\begin{equation}
    \langle \Psi(t)\vert \overleftarrow{\mathcal{I}}\{G(t+i\varepsilon)\,G(t')\}\vert \Psi(t)\rangle=\langle \Psi(t)\vert \overleftarrow{\mathcal{I}}\{G(t-i\varepsilon)\,G(t')\}\vert \Psi(t)\rangle^*
\end{equation}
so that the symmetrization picks up the real part of the Euclidean time-ordered 2-point function leading to the expression
\begin{equation}
    \mathcal{G}_{\ket{\Psi(t)}}^{\text{FS}}(G(t),G(t)) =  \lim_{t'\rightarrow t}\text{Re}\,\langle \Psi(t)\vert G(t)\,G(t')\vert \Psi(t)\rangle - \langle \Psi(t)\vert G(t)\vert \Psi(t)\rangle^2 \,,
    \label{eq:FSmetricrealpart}
\end{equation}
where we have taken the $\varepsilon\rightarrow 0^+$ limit. This expression will not be needed until in Section \ref{subsec:trivialcoupledcircuit} where we consider the coupled primary-deformed Virasoro circuit. For chiral circuits with local operators it is not needed. We emphasize that the replacement with the real part is not the regularization procedure. Regularization corresponds to the meromorphic continuation and the real part is picked up automatically by the symmetrization present in the FS metric.

\subsection{Definition of the Virasoro circuit}

A Virasoro circuit is a unitary quantum circuit where the unitary operator $U(t)$ creating the circuit coincides with a representation of the Virasoro group.

\paragraph{Virasoro group properties.} The Virasoro group $\textbf{Vir}$ is the central extension of the group $\widetilde{\text{Diff}}_+S^1$ which is the universal cover of the Lie group $\text{Diff}_+S^1$ of orientation preserving diffeomorphisms of the unit circle,
\begin{equation}
    \widetilde{\text{Diff}}_+S^1 \equiv \{f\colon \mathbb{R}\rightarrow \mathbb{R} \,\lvert\,f(\phi+2\pi) = f(\phi) +2\pi,\, f'(\phi)>0,f\text{ is smooth}\}\,.
\end{equation}
Given a diffeomorphism $f\in \widetilde{\text{Diff}}_+S^1$, its projective unitary representation $V_{f}$ satisfies the multiplication law
\begin{equation}
    V_{f}V_{h} = e^{iB(f,h)}\,V_{f\circ h}\,,
    \label{eq:virasorocomposition}
\end{equation}
where $B$ is the Thurston--Bott (TB) 2-cocycle (see \cite{Fewster:2004nj,Fewster:2018srj,Oblak:2016eij} for review)
\begin{equation}
    B(f,h) = b(f,h)+\frac{c}{48\pi}\int_{0}^{2\pi}d \phi\,\frac{h''(\phi)}{h'( \phi)}\,\log{f'(h(\phi))}
    \label{eq:2cocycle}
\end{equation}
up to the coboundary $b(f,h) = a(f\circ h) - a(f)-a(h)$ involving a functional $a(f)$. The TB 2-cocycle is the unique non-trivial group 2-cocycle of $\widetilde{\text{Diff}}_+S^1$ up to the central charge $c$ and the coboundary $b$ can be absorbed into the definition of the unitary operators by the redefinition $e^{ia(f)}\,V_f\rightarrow V_f$. By defining $V_{(f,\alpha)} \equiv e^{i\alpha}\,V_f$ with $\alpha \in \mathbb{R}$, equation \eqref{eq:virasorocomposition} can be rewritten in the form
\begin{equation}
    V_{(f,\alpha)}V_{(h,\beta)} = V_{(f,\alpha)\,\cdot\, (h,\beta)}\,,
\end{equation}
where now $(f,\alpha)\in \mathbf{Vir}$ is an element of the Virasoro group $\textbf{Vir} =\widetilde{\text{Diff}}_+S^1\ltimes \mathbb{R} $ whose group multiplication is defined as $(f,\alpha)\,\cdot\, (h,\beta)\equiv (f\circ h,\alpha+\beta+B(f,h))$. Hence $V_{(f,\alpha)}$ is an ordinary (non-projective) representation of $\textbf{Vir}$ of central charge $c$.

Consider now a one-parameter family of diffeomorphisms $f_t\in\widetilde{\text{Diff}}_+S^1$ such that $f_0 = \text{id}$ is the identity diffeomorphism $f_0(\phi) = \phi$. Starting from \eqref{eq:virasorocomposition}, it possible to write the unitary operator as a time-ordered exponential as (see \cite{deBoer:2023lrd} for proof\footnote{The definition \eqref{eq:Vft} of the unitary operator $V_{f_t}$ differs from \cite{deBoer:2023lrd} by an $f_t$ dependent phase factor. Therefore \eqref{eq:Vft} satisfies the composition law \eqref{eq:2cocycle} with this neglected phase factor appearing in the coboundary $b$.})
\begin{equation}
    V_{f_t} = \overleftarrow{\mathcal{T}}\exp{\biggl(-i\int_{0}^{t} ds\,G(f_s)\biggr)}\,,
    \label{eq:Vft}
\end{equation}
where the generator $G(f_t)$ of the Virasoro circuit is given by
\begin{equation}
    G(f_t) = \int_{0}^{2\pi} d\phi\,u_t(\phi)\,T(\phi)\,,\quad u_t(\phi) \equiv -\frac{\dot{F}_t(\phi)}{F'_t(\phi)} = (\dot{f}_t\circ f_t^{-1})(\phi)\,.
    \label{eq:genCC}
\end{equation}
Here $F_t\equiv f_t^{-1}$ is the inverse diffemorphism of $f_t$ at fixed $t$ satisfying $F_t(f_t(\phi)) = f_t(F_t(\phi)) = \phi$, a dot refers to the derivative with respect to the subscript $t$, a prime to the derivative with respect to the argument inside the brackets of $F_t(\phi)$ and $T(\phi)$ is the stress tensor operator that satisfies the Virasoro algebra \cite{Besken:2020snx}
\begin{equation}
	[T(\phi_1),T(\phi_2)] = -i\,\bigl(T(\phi_1)+T(\phi_2)\bigr)\,\delta'_{2\pi}(\phi_1-\phi_2)+\frac{ic}{24\pi}\,\delta'''_{2\pi}(\phi_1-\phi_2)\,,
 \label{eq:TTbarcommutator}
\end{equation}
where $\delta_{2\pi}(\phi) = \frac{1}{2\pi}\sum_{n=-\infty}^\infty e^{in\phi} = \sum_{n=-\infty}^\infty \delta(\phi+2\pi n)$ is the $2\pi$-periodic Dirac delta function. The Virasoro algebra $\mathfrak{v}\mathfrak{i}\mathfrak{r}$ is the Lie algebra of $\textbf{Vir}$, and in our conventions, the stress tensor has an expansion in terms of the standard Virasoro algebra generators as\footnote{The relation to the radially quantized Virasoro generators $L^{\text{rad}}_n$ is given by $L_n = L^{\text{rad}}_n -\frac{c}{24}\,\delta_{n,0} $.}
\begin{equation}
    T(\phi) = \frac{1}{2\pi}\sum_{n=-\infty}^\infty L_n\,e^{in\phi}\,,\quad [L_n,L_m] = (n-m)\,L_{n+m}+\frac{c}{12}\,n^{3}\,\delta_{n,-m}\,.
    \label{eq:stresstensorexp}
\end{equation}
We also define the complex conjugate representation $\overbar{V}_{f}$ of $\mathbf{Vir}$ via \cite{deBoer:2023lrd}
\begin{equation}
    \overbar{V}_{f_t} =\overleftarrow{\mathcal{T}}\exp{\biggl(i\int_{0}^{t} ds\,\overbar{G}(f_s)\biggr)}\,,\quad \overbar{G}(f_t) = \int_{0}^{2\pi} d\phi\,u_t(\phi)\,\overbar{T}(\phi)\,,
    \label{eq:Vbar}
\end{equation}
where $\overbar{T}(\phi)\equiv T(-\phi)$ and $u_t = \dot{f}_t\circ f_t^{-1}$ as before. Defining the reflection $Z(\phi) = -\phi$ one can show using \eqref{eq:Vft} that $\overbar{V}_{f_t} = V_{Z\circ f_t\circ Z}$. The complex conjugate representation satisfies $\overbar{V}_{f}\overbar{V}_{h} = e^{-iB(f,h)}\,\overbar{V}_{f\circ h}$ with an opposite sign for the projective phase. See Appendix \ref{subapp:conjugate_rep} for details.

A general element of $\mathfrak{v}\mathfrak{i}\mathfrak{r}$ is a Hermitian operator which is a linear combination of Virasoro generators
\begin{equation}
    G_u = \int_0^{2\pi} d\phi\,u(\phi)\,T(\phi) = \sum_{n=-\infty}^\infty u_n^*\,L_n\,,%\quad G_v = \int_0^{2\pi} d\phi\,v(\phi)\,T(\phi)\,,
    \label{eq:Gudefinition}
\end{equation}
where $u(\phi) = u(\phi+2\pi)$ is the component of a vector field on the unit circle with the Fourier expansion
\begin{equation}
    u(\phi) = \sum_{n=-\infty}^{\infty}u_n\,e^{in\phi}\,,\quad u_n^* = u_{-n}\,.
    \label{eq:ufourierexpansion}
\end{equation}
The operators $G_u$ form a centrally extended Lie algebra representation of the algebra of vector fields on the circle.

A highest-weight state $\ket{h}$ is defined by the equations $L_{n>0}\ket{h} = 0$ and $L_0 \ket{h} = (h-\frac{c}{24})\ket{h}$ and the vacuum state $\ket{0}$ satisfies $L_{\pm 1} \ket{0} = 0$ and $L_0\ket{0} = -\frac{c}{24}\ket{0}$.\footnote{The radially quantized Virasoro generators annihilate the vacuum as usual $L^{\text{rad}}_{\pm 1}\ket{0} =L^{\text{rad}}_{0}\ket{0} = 0 $.} Linear combinations of states obtained by acting on $\ket{h}$ with generators $L_{n < 0}$ spans a Verma module $\mathcal{H}_h$ of $\mathfrak{v}\mathfrak{i}\mathfrak{r}$.

\paragraph{Virasoro circuits.} Let $\ket{R}$ be a reference state in a Hilbert space which is a direct sum of Verma modules $\mathcal{H}_h$ and consider a diffeomorphism $f\in\widetilde{\text{Diff}}_+S^1$. The state defined by
\begin{equation}
    \ket{f} \equiv V_{f}\ket{R}
\end{equation}
is called a Virasoro pure state. When the reference state is a highest-weight state $\ket{R} = \ket{h}$, Virasoro states take the form
\begin{equation}
    \ket{f} = \sum_{m=0}^{\infty}\,\sum_{0<n_1\leq n_2\ldots \leq n_m}C_{n_1\ldots n_m}^{f}\ket{h;n_1,\ldots, n_m}  \,,
\end{equation}
where $C_{n_1\ldots n_m}^{f}$ are complex coefficients determined by $f$ and the descendants $\ket{h;n_1,\ldots n_m} = L_{-n_1}\cdots L_{-n_m}\ket{h}$ (with $0<n_1\leq n_2\ldots \leq n_m$) span $\mathcal{H}_h$. Therefore the space of Virasoro pure states for $\ket{R} = \ket{h}$ is a single Verma module, and in addition, there is a one-to-one map $\ket{f}\mapsto f$ to the infinite-dimensional manifold $\widetilde{\text{Diff}}_+S^1$ of diffeomorphisms.

Now a Virasoro circuit is defined to be a one-parameter family of Virasoro pure states
\begin{equation}
    \ket{f_t} \equiv V_{f_t}\ket{R}
\end{equation}
corresponding to the curve $f_t\in \widetilde{\text{Diff}}_+S^1$. This may be generalized to Virasoro circuits of the type
\begin{equation}
    \ket{f_t,\overbar{f}_t} \equiv V_{f_t,\overbar{f}_t}\ket{R}\,,
    \label{eq:tensorproductcircuit}
\end{equation}
where $(f_t,\overbar{f}_t)\in \widetilde{\text{Diff}}_+S^1\times \widetilde{\text{Diff}}_+S^1 $ is a pair of independent diffeomorphisms and we have defined the tensor product representation of $\mathbf{Vir}\times \mathbf{Vir}$ as
\begin{equation}
    V_{f_t,\overbar{f}_t} \equiv V_{f_t}\otimes \overbar{V}_{\overbar{f}_t} = \overleftarrow{\mathcal{T}}\exp{\biggl(-i\int_{0}^{t} ds\,G(f_s,\overbar{f}_s)\biggr)}
\end{equation}
with the generator
\begin{equation}
    G(f_t,\overbar{f}_t) \equiv G(f_t)\otimes \mathbf{1} - \mathbf{1}\otimes \overbar{G}(\overbar{f}_t)\,.
    \label{eq:generatorcoupledvirasoro}
\end{equation}
The use of the conjugate representation for the second tensor factor is a choice which leads to the relative minus sign in the generator. This choice is motivated by two-dimensional conformal field theory. We will call operators of the type $\mathcal{O}\otimes \mathbf{1}$ and $\mathbf{1}\otimes\mathcal{O}$ chiral and anti-chiral respectively.

\paragraph{M\"obius circuits.} 

The group $\widetilde{\text{Diff}}_+S^1$ includes subgroups of diffeomorphisms that are isomorphic to the universal cover of $SL(2,\mathbb{R})$. We call them M\"obius subgroups, because their elements can be constructed using M\"obius transformations of the complex plane as follows. The group $SL(2,\mathbb{R})$ is isomorphic to $ SU(1,1)$ which can be parametrized by matrices
\begin{equation}
    M = 
    \begin{pmatrix}
    a & b\\
    b^* & a^*
    \end{pmatrix}\,,\quad \vert a\vert^2 - \vert b\vert^2 = 1\,,\quad a,b\in \mathbb{C}\,,
\label{eq:SLmatrix}
\end{equation}
such that the group operations are multiplication $MN$ and inversion $M^{-1}$ of matrices. To each $M$ we can associate a diffeomorphism $f_M\in\widetilde{\text{Diff}}_+S^1 $ via the M\"obius transformation
\begin{equation}
   e^{i f_M(\phi)} = \frac{a e^{i\phi} +b}{b^*e^{i\phi} + a^*}\,.
    \label{eq:fM}
\end{equation}
From this definition one can check that the composition of two diffeomorphisms is given by $f_M\circ f_N = f_{MN}$ due to how two M\"obius transformations are composed and that the inverse diffeomorphism satisfies $f_M^{-1} = f_{M^{-1}}$. However, the map $M\mapsto f_M$ defined by \eqref{eq:fM} is not one-to-one, because diffeomorphisms $f_M(\phi)+2\pi m$ with $m\in \mathbb{Z}$ correspond to the same $M$. Therefore \eqref{eq:fM} is a group isomorphism from the subgroup of $\widetilde{\text{Diff}}_+S^1 $ to the universal cover $\widetilde{SL}(2,\mathbb{R})$ of $SL(2,\mathbb{R})$.

\begin{figure}[t]
\centering
\begin{subfigure}{.32\textwidth}
  \centering
  \includegraphics{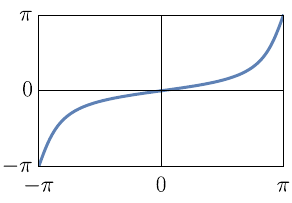}
  \caption{$k=1$}
  \label{fig:mobiusk1}
\end{subfigure}
\hfill
\begin{subfigure}{.32\textwidth}
  \centering
  \includegraphics{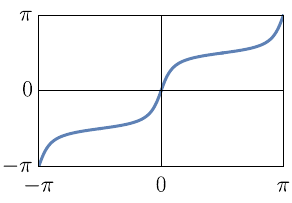}
  \caption{$k=2$}
  \label{fig:mobiusk2}
\end{subfigure}
\hfill
\begin{subfigure}{.32\textwidth}
  \centering
  \includegraphics{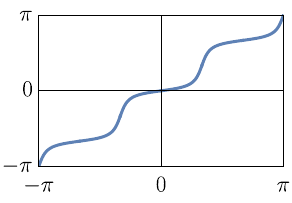}
  \caption{$k=3$}
  \label{fig:mobiusk3}
\end{subfigure}
\caption{Plot of the diffeomorphism $f_{M,k}(\phi)$ \eqref{eq:Slk2Rdiffeo} in the range $\phi\in(-\pi,\pi)$ for $(\rho,\chi,\theta) = (3\slash 2,0,0)$.}
\label{fig:mobiusdiffeos}
\end{figure}

Given three real parameters $p = (\rho,\chi,\theta)$ with ranges $\rho>0$, $\chi \in (0,2\pi)$ and $ \theta\in (0,2\pi)$, we can parametrize $SU(1,1)\cong SL(2,\mathbb{R})$ matrices as
\begin{equation}
    a = e^{i(\chi-\theta)\slash 2}\cosh{\frac{\rho}{2}}\,,\quad b = e^{i(\chi+\theta)\slash 2}\sinh{\frac{\rho}{2}}\,.
    \label{eq:SL2ab}
\end{equation}
By \eqref{eq:fM} we obtain $\widetilde{SL}(2,\mathbb{R})$ in this parametrization \cite{Fewster:2004nj}
\begin{equation}
   f_M(\phi) = \chi+2\arctan{\biggl(e^{-\rho}\tan{\frac{\phi-\theta}{2}}\biggr)}\,,\quad \phi-\theta\in(-\pi,\pi)\,,
   \label{eq:Sl2Rdiffeo}
\end{equation}
whose domain is extended to $\phi \in \mathbb{R}$ by continuity and the condition $f_M(\phi+2\pi) = f_M(\phi) + 2\pi$ (see Figure \ref{fig:mobiusdiffeos}).

There are an infinite number of subgroups $\widetilde{SL}_k(2,\mathbb{R})$ of $\widetilde{\text{Diff}}_+S^1$, labeled by a positive integer $k = 1,2,3,\ldots$, that are also isomorphic to $\widetilde{SL}(2,\mathbb{R})$, but distinct. They consist of diffeomorphisms of the form
\begin{equation}
    f_{M,k}(\phi) = \frac{1}{k}\biggl\{\chi+2\arctan{\biggl[e^{-\rho}\tan{\biggl(\frac{k\phi-\theta}{2}+\frac{\pi\,(k-2j-1)}{2}\biggr)}\biggr]}\biggr\}-\pi+\frac{2\pi j}{k}+\frac{\pi}{k}
   \label{eq:Slk2Rdiffeo}
\end{equation}
defined on the range $-\pi + \frac{2\pi j}{k} \leq \phi-\frac{\theta}{k} < -\pi + \frac{2\pi j}{k} + \frac{\pi}{k}$ where $ j = 0,1,\ldots,k-1$. This gives the definition of the diffeomorphism on the interval $\phi-\theta \in (-\pi,\pi)$ (see Figure \ref{fig:mobiusdiffeos}) which can be extended to $\phi\in \mathbb{R}$ via the condition $f_{M,k}(\phi+2\pi) = f_{M,k}(\phi) + 2\pi$. The diffeomorphisms $f_{M,k}$ are in correspondence with $SL(2,\mathbb{R})$ matrices \eqref{eq:SLmatrix} via the formula
\begin{equation}
	e^{ikf_{M,k}(\phi)} = (-1)^{k+1}\frac{ae^{ik\phi}+(-1)^{k+1}\,b}{b^{*}e^{ik\phi}+(-1)^{k+1}\,a^{*}}\,.
 \label{eq:fMSLk}
\end{equation}
From this equation it follows that $f_{M,k}\circ f_{N,k} = f_{MN,k}$ and $f_{M,k}^{-1} = f_{M^{-1},k}$ which imply $\widetilde{SL}_k(2,\mathbb{R})\cong \widetilde{SL}(2,\mathbb{R})$ as claimed.

Now let $\xi_i $ be a tangent vector in the $p^i$ direction at point $M$ defined as
\begin{equation}
    \xi_i(\phi) = (\partial_i f_{M,k}\circ f_{M,k}^{-1})(\phi) = -\frac{\partial_iF_{M,k}(\phi)}{\partial_\phi F_{M,k}(\phi)}\,,
    \label{eq:xii}
\end{equation}
where $F_{M,k} \equiv f_{M,k}^{-1}$. By substituting \eqref{eq:Slk2Rdiffeo} we obtain explicitly
\begin{gather}
    \xi_\rho(\phi) = \frac{(-1)^k}{k}\sin{(k\phi-\chi)}\,,\quad \xi_\chi(\phi) = 1\,,\nonumber\\
    \xi_\theta(\phi) = -\frac{1}{k}\,[\cosh{(\rho)}+(-1)^k\sinh{(\rho)}\cos{(k\phi-\chi)}] \,.
    \label{eq:xiexplicit}
\end{gather}
We can see that the only non-zero Fourier modes $\xi_{i,n}$ of $\xi_i(\phi)$ are $n = 0$ and $n = \pm k$. Therefore the corresponding Virasoro algebra representatives $G_{\xi_i}$ defined in \eqref{eq:Gudefinition} involve only $L_0$ and $L_{\pm k}$ which generate the $\mathfrak{s}\mathfrak{l}_k(2,\mathbb{R})$ subalgebra of the Virasoro algebra. The isomorphism $\mathfrak{s}\mathfrak{l}_k(2,\mathbb{R})\cong \mathfrak{s}\mathfrak{l}(2,\mathbb{R})$ is given by $\widetilde{L}_0 = \frac{1}{k}\,(L_0+\frac{c}{24}\,k^2)$, $\widetilde{L}_{\pm 1} = \frac{1}{k}\,L_{\pm k} $ which satisfy the same algebra as $L_0$, $L_{\pm 1}$.

Given a diffeomorphism $f_{M,k}\in \widetilde{SL}_k(2,\mathbb{R})$ of the form \eqref{eq:Slk2Rdiffeo}, a M\"obius pure state is defined as
\begin{equation}
    \ket{f_{M,k}} = V_{f_{M,k}}\ket{R}\,.
\end{equation}
A M\"obius circuit is a 1-parameter family of M\"obius pure states parametrized by a 1-parameter family of $SL(2,\mathbb{R})$ matrices $M(t)$. In Appendix \ref{app:displacementoperator} we show that (see also \cite{Fewster:2018srj} for related calculations)
\begin{equation}
    V_{f_{M,k}} = D_k(\rho e^{i\chi})\,e^{i(\theta-\chi)\slash k\, L_0}\,,\quad D_k(\zeta)\equiv e^{\frac{(-1)^k}{2k}\,(\zeta\,L_{k} -\zeta^*L_{-k})}\,,
\end{equation}
where $D_k(\zeta)$ is a generalized displacement operator. Its action on the highest weight state is given by
\begin{equation}
    V_{f_{M,k}}\ket{h} = e^{i(\theta-\chi)\,(h-\frac{c}{24})\slash k}\ket{\rho e^{i\chi},k}
    \label{eq:mobiusstatecoherentstate}
\end{equation}
where $\ket{\zeta,k} \equiv D_k(\zeta)\ket{h} $ is known as a generalized coherent state \cite{Perelomov:1986tf,Caputa:2021sib,Caputa:2022zsr}.\footnote{Similar states obtained by acting with higher-dimensional conformal groups are studied in \cite{Koch:2021tvp,Chagnet:2021uvi,Rabambi:2022jwu}.} Therefore M\"obius pure states for the reference state $\ket{R} = \ket{h}$ are equal to generalized coherent states up to a phase factor.

\subsection{Fubini--Study cost of Virasoro circuits}\label{subsec:FSmetricvirasoro}

Having introduced Fubini--Study metric and cost, we will now study them in the case of Virasoro circuits. We will derive an explicit expression for the FS metric on the infinite-dimensional space of Virasoro pure states introduced in the previous section. This problem was considered in \cite{Flory:2020dja,Flory:2020eot} which we will slightly extend while clarifying the issue of regularization. We will then restrict to M\"obius pure states where the FS metric reduces to a hyperbolic metric.

The FS metric \eqref{eq:FSmetricwithsymmetrization} at point $ V_f\ket{h}$ contracted with two tangent vectors $G_u$, $G_v$ on the space of Virasoro pure states takes the form
\begin{equation}
    \mathcal{G}_{V_f\ket{h}}^{\text{FS}}(G_u,G_v)  = \frac{1}{2}\bra{h} (\widetilde{G}_u\,\widetilde{G}_v + \widetilde{G}_v\,\widetilde{G}_u)\ket{h} - \bra{h} \widetilde{G}_u\ket{h}\bra{h} \widetilde{G}_v\ket{h}\,,
\end{equation}
where we have defined the transformed operators
\begin{equation}
    \widetilde{G}_{u} \equiv V_f^\dagger\,G_{u}\,V_f = \int_0^{2\pi}d \phi\,u(\phi)\left(F'(\phi)^2\,T(F(\phi))- \frac{c}{24\pi}\,\{F(\phi),\phi\}\right)\,.
\end{equation}
Here $F = f^{-1}$ and we have used the formula
\begin{equation}
	V_{f}^\dagger\,T(\phi)\,V_{f} = F'(\phi)^{2}\,T(F(\phi)) - \frac{c}{24\pi}\,\{F(\phi),\phi\}\,,
	\label{eq:Ttransformation_text}
\end{equation}
with the Schwarzian derivative
\begin{equation}
    \{F(\phi),\phi\} = \biggl(\frac{F''(\phi)}{F'(\phi)}\biggr)'-\frac{1}{2}\biggl(\frac{F''(\phi)}{F'(\phi)}\biggr)^2\,.
\end{equation}
By performing a change of integration variables $\phi \rightarrow F(\phi)$ and relabeling the new variable back to $\phi$, we obtain
\begin{equation}
    \widetilde{G}_u = \int_0^{2\pi}d \phi\,\widetilde{u}(\phi)\left(T(\phi)+ \frac{c}{24\pi}\,\{f(\phi),\phi\}\right)\,,
    \label{eq:tildeGu}
\end{equation}
where we have defined the transformed tangent vector
\begin{equation}
    \widetilde{u}(\phi) \equiv u(f(\phi_1))\,F'(f(\phi_1))=\frac{u(f(\phi_1))}{f'(\phi_1)}
\end{equation}
and made use of the identity $\{F(\phi),\phi\} = -F'(\phi)^{2}\,\{f(F(\phi)),F(\phi)\}$. The FS metric becomes
\begin{align}
\mathcal{G}_{V_{f}\ket{h}}^{\text{FS}}(G_u,G_v)=&\int_{0}^{2\pi}d\phi_1\int_{0}^{2\pi} d\phi_2\,\widetilde{u}(\phi_1)\,\widetilde{v}(\phi_2)\label{eq:FSdoubleintegral}\\ & \,\times\biggl(\frac{1}{2}\bra{h}T(\phi_1)\,T(\phi_2)+T(\phi_2)\,T(\phi_1)\ket{h}-\bra{h}T(\phi_1)\ket{h}\bra{h}T(\phi_2)\ket{h}\biggr)\,. \nonumber
\end{align}
Expanding $\widetilde{u}(\phi)$ and $\widetilde{v}(\phi)$ in Fourier modes $\widetilde{u}_n$ and $ \widetilde{v}_n$ as in \eqref{eq:ufourierexpansion}, this can be written as
\begin{equation}
    \mathcal{G}_{V_{f}\ket{h}}^{\text{FS}}(G_u,G_v) = \sum_{n,m=-\infty}^{\infty}\widetilde{u}_n^{*}\, \widetilde{v}_m^{*}\, \biggl(\frac{1}{2}\bra{h}L_{n}\,L_{m}+L_m\,L_n\ket{h}-\bra{h}L_n\ket{h}\bra{h}L_m\ket{h}\biggr)\,.
    \label{eq:FSmetricmodes}
\end{equation}
The 2-point function of the Virasoro generators is computed in Appendix \ref{app:primaryproperties} and the result is (see also \cite{Fitzpatrick:2015zha,Caputa:2021sib})
\begin{align}
	\bra{h}L_{n}\,L_{m}\ket{h} =  \frac{c}{12}\,n\,\biggl(\frac{24h}{c}+n^{2}-1\biggr)\,\delta_{n,-m}\,\Theta(n)+ \left(h-\frac{c}{24}\right)^{2}\delta_{n,0}\,\delta_{m,0}
 \label{eq:Virasoromode2point_text}
\end{align}
from which we obtain the anti-commutator
\begin{equation}
    \bra{h}L_{n}\,L_{m} + L_m\,L_n\ket{h} = \frac{c}{12}\,\vert n \vert\,\biggl(\frac{24h}{c}+n^{2}-1\biggr)\,\delta_{n,-m}+ 2\left(h-\frac{c}{24}\right)^{2}\delta_{n,0}\,\delta_{m,0}\,.
\label{eq:LnLmanticommutatortext}
\end{equation}
Now using equation \eqref{eq:LnLmanticommutatortext} and the 1-point function $ \bra{h}L_{n}\ket{h} = \left(h-\frac{c}{24}\right)\delta_{n,0}$, the FS metric \eqref{eq:FSmetricmodes} becomes
\begin{equation}        
\mathcal{G}_{f}^{\text{FS}}(\widetilde{u},\widetilde{v}) \equiv \mathcal{G}_{V_{f}\ket{h}}^{\text{FS}}(G_u,G_v) = \frac{c}{24}\sum_{n=1}^{\infty}n \,\biggl(\frac{24h}{c}+n^{2}-1\biggr)\,\widetilde{u}_n^{*} \widetilde{v}_n\,.
\label{eq:FSmetricdiffeos}
\end{equation}
Here $\mathcal{G}_{f}^{\text{FS}}(\widetilde{u},\widetilde{v})$ is understood as the pull-back of the FS metric to the Lie group $\widetilde{\text{Diff}}_+S^1$ and it is contracted with two vectors $\widetilde{u},\widetilde{v}$ living in the tangent space of $f$ (see \cite{deBoer:2023lrd} for a more detailed explanation). The metric vanishes in the zero mode directions $\widetilde{u}_0,\widetilde{v}_0$ so that descends to a non-degenerate metric on the quotient space $\widetilde{\text{Diff}}_+S^1\slash \,\mathbb{R}$ where $\mathbb{R}$ denotes the universal cover of the group $U(1)$ of rotations of the circle.

The expression \eqref{eq:FSmetricdiffeos} does not seem to have appeared previously in the literature in this representation where it matches exactly with the unique Kähler metric of the coadjoint orbit $\widetilde{\text{Diff}}_+S^1\slash \,\mathbb{R}$ of the Virasoro group given in \cite{Kirillov:1987mn}. This connection seems to have been missed in the literature, but similar results for higher-dimensional conformal groups have been observed in \cite{Koch:2021tvp}.

\paragraph{Differential regularization.} We may also derive the same expression for the FS metric by working directly with the double integral \eqref{eq:FSdoubleintegral}, however this requires regularization due to divergences coming from $\phi_1 = \phi_2$. Previously this has been achieved using differential regularization \cite{Flory:2020dja,Flory:2020eot}. Differential regularization exploits the fact that extracting derivatives from rational functions that are divergent as distributions renders integrals over them well-behaved \cite{Freedman:1991tk}. 
This was implemented for the energy-momentum tensor in \cite{Erdmenger:1996yc}. In the present context, the line $\phi_1 = \phi_2$ is excluded from the integration region.\footnote{Note that in \cite{Flory:2020dja,Flory:2020eot} only the diagonal component $u = v$ of the FS metric \eqref{eq:FSmetricdiffeos} is considered which is enough for the computation of the FS cost.} The integrand is then written as a total derivative of a well behaved function and integrated by parts which produces divergent terms coming from the $\phi_1 = \phi_2$ boundary of the integration region. Differential regularization and subsequent renormalization instructs to subtract these divergent boundary terms by hand leaving a finite expression. The subtraction is equivalent to including divergent counterterms in the beginning which coincides with usual renormalization \cite{Freedman:1991tk}. We will now show how the differential regularization prescription for the integral \eqref{eq:FSdoubleintegral} can be ``derived'' from the standard meromorphic continuation of two-dimensional Euclidean conformal field theories.

In meromorphic continuation (or regularization), the operator product $T(\phi)\,T(0)$ is understood as a boundary value of a meromorphic operator valued function on the complex $w$-plane at the real axis $\phi = \text{Re}\,w$. This boundary value is then to be integrated against a test function on the real axis defining a so called hyperfunction or a generalized distribution. The meromorphic function in question is the Euclidean time-ordered operator product $\overleftarrow{\mathcal{I}}\{T(w)\,T(0)\}$ whose form is dictated by the operator product expansion (OPE)
\begin{equation}
    \overleftarrow{\mathcal{I}}\{T(w_1)\,T(w_2)\} = \frac{1}{(2\pi)^{2}}\biggl[\frac{c\slash 2}{16\sin^{4}(\frac{w_1-w_2}{2})}-\frac{2\pi\,T(w_1)+c\slash 24}{2\sin^{2}(\frac{w_1-w_2}{2})}+\frac{2\pi\,\partial_{w_1}T(w_1)}{2\tan(\frac{w_1-w_2}{2})}\biggr] + \ldots\,,
    \label{eq:TTOPE_text}
\end{equation}
where $w_1,w_2\in \mathbb{C}$, $\overleftarrow{\mathcal{I}}$ denotes ordering in $\text{Im}\,w_{1,2}$ and ellipsis denote a series of operators whose coefficients are finite in the $w_2\rightarrow w_1$ limit (see Appendix \ref{app:primaryproperties} for details). In our conventions, the operator products are defined as
\begin{equation}
    T(\phi_1)\,T(\phi_2) = \lim_{\varepsilon\rightarrow 0^+}\overleftarrow{\mathcal{I}}\{T(\phi_1 + i\varepsilon)\,T(\phi_2)\}\,,\quad T(\phi_2)\,T(\phi_1) = \lim_{\varepsilon\rightarrow 0^+}\overleftarrow{\mathcal{I}}\{T(\phi_1 - i\varepsilon)\,T(\phi_2)\}\,,
\end{equation}
so that the two operator orderings are obtained by taking the boundary value by approaching from two sides of the real axis. It follows that any expectation value of the product $T(\phi_1)\,T(\phi_2)$ contains in addition to the real part an imaginary part which is identified with expectation value of the commutator $[T(\phi_1),T(\phi_2)]$ \eqref{eq:TTbarcommutator} as proven in Appendix \ref{app:stresstensormodecorrelators}. The commutator is completely determined by the divergent terms in the OPE and contains only contact terms involving derivatives of the delta function. On the other hand, the real part coincides with the expectation value of the anti-commutator $[T(\phi_1),T(\phi_2)]_+ \equiv T(\phi_1)\,T(\phi_2)+T(\phi_2)\,T(\phi_1)$ which contains contributions also from the ellipsis in the OPE. It is the expectation value of this anti-commutator that appears in the FS metric \eqref{eq:FSdoubleintegral}.

To compute $\bra{h}T(\phi_1)\,T(\phi_2)+T(\phi_2)\,T(\phi_1)\ket{h}$, we use that $\bra{h}\overleftarrow{\mathcal{I}}\{T(w_1)\,T(w_2)\}\ket{h}$ is fixed by the OPE between the stress tensor and the primary operator creating the state $\ket{h}$ \cite{Datta:2019jeo}. As shown in Appendix \ref{app:stresstensormodecorrelators}, this gives
\begin{equation}
    \bra{h}T(\phi_1)\,T(\phi_2)\ket{h} = \lim_{\varepsilon\rightarrow 0^+}\frac{1}{(2\pi)^{2}}\biggl[\frac{c\slash 2}{16\sin^{4}(\frac{\phi_1-\phi_2+i\varepsilon}{2})}-\frac{h}{2\sin^{2}(\frac{\phi_1-\phi_2+i\varepsilon}{2})}+\left(h-\frac{c}{24}\right)^{2}\biggr]\,.
    \label{eq:stress2point_text}
\end{equation}
The expectation value of the anti-commutator is the real part of this expression which gives the Hadamard finite part distribution (also known as the Hadamard 2-point function)
\begin{equation}
    \bra{h}[T(\phi_1)\,T(\phi_2)]_+\ket{h} = \frac{2}{(2\pi)^2}\biggl[\,\mathcal{H}\,\biggl(\frac{c}{32\sin^{4}(\frac{\phi_1-\phi_2}{2})}-\frac{h}{2\sin^{2}(\frac{\phi_1-\phi_2}{2})}\biggr)+\left(h-\frac{c}{24}\right)^{2}\delta_{n,0}\,\delta_{m,0}\biggr]\,,
    \label{eq:TTanticommutatortext}
\end{equation}
which is defined in terms of the Cauchy principal value distribution as (see Appendix \ref{app:stresstensormodecorrelators})
\begin{equation}
    \mathcal{H}\,\biggl(\frac{c}{32\sin^{4}(\frac{\phi_1-\phi_2}{2})}-\frac{h}{2\sin^{2}(\frac{\phi_1-\phi_2}{2})}\biggr) = \frac{c}{12}\,\partial_{\phi}\,\biggl(-\partial_{\phi}^2-1+\frac{24h}{c}\biggr)\,\mathcal{P}\left(\frac{1}{2\tan{(\frac{\phi}{2}})}\right)\bigg\vert_{\phi = \phi_1-\phi_2}\,.
\end{equation}
By integrating \eqref{eq:TTanticommutatortext} against $e^{-in\phi_1-im\phi_2}$ gives equation \eqref{eq:LnLmanticommutatortext} as shown in Appendix \ref{app:stresstensormodecorrelators} so that the FS metric \eqref{eq:FSmetricdiffeos} is reproduced. In addition, the calculation is exactly the same as performing differential regularization. Hence we have provided a derivation of the differential regularization prescription from the meromorphic continuation (OPE) \eqref{eq:stress2point_text} of the operator product.

\paragraph{Virasoro circuits.} Let us now consider Virasoro circuits. A Virasoro circuit $f_t$ has the tangent vector $u_t(\phi) = (\dot{f}_t\circ f_t^{-1})(\phi) = -\frac{\dot{F}_t(\phi)}{F_t'(\phi)}$ at point $f_t$. Given that, the transformed tangent vector $\widetilde{u}_t(\phi)$ takes the form 
\begin{equation}
    \widetilde{u}_t(\phi)=\frac{u_t(f_t(\phi))}{f_t'(\phi)}=\frac{\dot{f}_t(\phi)}{f_t'(\phi)}
\end{equation}
and it has Fourier modes $\widetilde{u}_{t,n}$. The infinitesimal length element in the FS metric of this curve becomes
\begin{equation}
     \mathcal{G}_{f_t}^{\text{FS}}(u_t,u_t) = \frac{c}{12}\sum_{n=1}^{\infty} n \,\biggl(\frac{24h}{c}+n^{2}-1\biggr)\,\vert \widetilde{u}_{t,n}\vert^2\,,
    \label{eq:FSmetricconformalcircuit}
\end{equation}
which is the FS cost of the Virasoro circuit. For a tensor product Virasoro circuit \eqref{eq:tensorproductcircuit} with $(f_t,\overbar{f}_t)$, the full FS cost is the sum of the individual costs for $f_t$ and $\overbar{f}_t$ separately. We will now study the expression \eqref{eq:FSmetricconformalcircuit} in simple cases.

\paragraph{Virasoro zero mode circuits.}

Let us investigate the case of evolution where $G(t)=L_0$ is the Virasoro zero mode. This is implemented by a Virasoro circuit with $f_t(\phi)=\phi+t\equiv r_t(\phi)$ being a rotation of the circle. The tangent vectors for this circuit are simply $\widetilde{u}_t(\phi)=u_t(\phi)=1$. Hence, the only nonvanishing Fourier mode contribution to the FS metric is the zero mode $\widetilde{u}_{t,0}=1$. From this it is clear that the derived formula correctly reproduces the expected vanishing of the FS metric 
\begin{align}
    ds^2=\mathcal{G}_{V_{r_t}\ket{h}}^{\text{FS}}(G_{1},G_{1}) = 0\,.
\end{align}
The metric is expected to vanish because evolution generated by $G(t)=L_0$ only modifies the reference state by a phase and the FS metric is insensitive to phase modifications.

\paragraph{M\"obius circuits.} 

The $SL(2,\mathbb{R})$ matrices $M = M(p)$ are parametrized by three real coordinates $p = (\rho,\chi,\theta)$ via equation \eqref{eq:SL2ab}. Therefore the Fubini--Study metric at point $f_{M,k}$ on $\widetilde{SL}_k(2,\mathbb{R})$ can be written as
\begin{equation}
    ds^2 = \mathcal{G}_{ij}^{\text{FS}}(p)\,dp^idp^j\,,\quad \mathcal{G}_{ij}^{\text{FS}}(p) \equiv \mathcal{G}_{f_{M(p),k}}^{\text{FS}}(\xi_i,\xi_j)\,.
\end{equation}
To calculate the FS metric, we need the Fourier modes $\widetilde{\xi}_{i,n}$ of $\widetilde{\xi}_i(\phi)  = \frac{\partial_if_{M,k}(\phi)}{\partial_\phi f_{M,k}(\phi)}$ which are obtained using \eqref{eq:Slk2Rdiffeo}. By substituting to \eqref{eq:FSmetricdiffeos}, we obtain
\begin{equation}
    \mathcal{G}_{ij}^{\text{FS}}(p)\,dp^idp^j = \frac{h_k}{2}\,(d\rho^2+\sinh^2{(\rho)}\,d\chi^2)\,,\quad h_k = \frac{c}{24}\left(k-\frac{1}{k}\right)+\frac{h}{k}\,,
    \label{eq:hyperbolicmetricSLk}
\end{equation}
which is a hyperbolic metric of curvature radius $\sqrt{h_k\slash 2}$ in the $(\rho,\chi)$ directions and which has vanishing components in the $\theta$-direction.

The hyperbolic metric \eqref{eq:hyperbolicmetricSLk} with $k = 1$ was shown in \cite{Caputa:2021sib} to be equal to the FS metric \eqref{eq:FSmetricdef} on the space of generalized coherent states $\ket{\zeta} = D_1(\zeta) \ket{h} =e^{\xi L_{-1} - \xi^*L_{1}}\ket{h}  $ with parameter $\zeta = 2\xi^* = \rho e^{-i\chi} $ (see also \cite{Chagnet:2021uvi}). The fact that the FS metric is the same on both coherent and M\"obius states follows from the fact that they are equal up to a phase \eqref{eq:mobiusstatecoherentstate} which cancels in the absolute value appearing in the FS metric \eqref{eq:FSmetricdef}.

\section{Primary-deformed Virasoro circuits}\label{sec:deformedcircuits}

In the previous section, we studied Virasoro circuits evolving on the space of Virasoro pure states which is a single highest-weight representation of the Virasoro group (a Verma module). We will now generalize the setup to circuits that move between different Verma modules. To this end, we consider a Hilbert space which is a direct sum of the vacuum module and multiple modules of non-zero weights; tensor product representations described in Section \ref{sec:conformalcircuits} will also be considered. For the circuit to move between the modules,  we supplement the generator of the circuit with a coupling to a primary operator determined by a time-dependent source function $J(\phi,t)$. We derive a general formula for the Fubini--Study cost function up to second order in a perturbative expansion in the primary deformation.

\subsection{Definition and information geometry}\label{subsec:definition_info_geometry} 

Virasoro circuits have the shortcoming that they are confined to the Verma module (irreducible representation of $\mathbf{Vir}$) of the reference state $\ket{R}$. We would like to consider circuits producing linear combinations of states of multiple Verma modules with different weights. To this end, we introduce a local primary operator of integer weight $h =1,2,3,\ldots$ defined by the mode expansion (see \cite{Fewster:2018srj} for example)
\begin{equation}
    \mathcal{O}_h(\phi) = \frac{1}{2\pi}\sum_{n=-\infty}^{\infty}\mathcal{O}_{h,n}\,e^{in\phi}\,,\quad [L_n,\mathcal{O}_{h,m}] = [(h-1)\,n-m]\,\mathcal{O}_{h,n+m}\,,
    \label{eq:primarydefinition}
\end{equation}
where the action of the mode operators on the vacuum is defined to be
\begin{equation}
\label{eq:primarymodesactonvacuum}
    \mathcal{O}_{h,n}\ket{0} = 0\,,\quad n>-h\,.
\end{equation}
A conjugate operator is defined as $\overbar{\mathcal{O}}(\phi) \equiv \mathcal{O}(-\phi)$. It follows that the highest weight state is obtained as $\ket{h} = \mathcal{O}_{h,-h}\ket{0} $ with normalization assumed to be $\braket{h} = 1$.\footnote{We may check that $\mathcal{O}_{h,-h}\ket{0}$ satisfies the properties of a highest weight state by using the commutation relations \eqref{eq:primarydefinition}.} At the level of local operators, the commutator can be written as
\begin{equation}
[T(\phi_1),\mathcal{O}_h(\phi_2)] = -i\,(h-1)\,\mathcal{O}_h'(\phi_1)\,\delta_{2\pi}(\phi_1-\phi_2)-i\,h\,\mathcal{O}_h(\phi_1)\,\delta'_{2\pi}(\phi_1-\phi_2)\,,
\label{eq:TOcommutator}
\end{equation}
which implies that under the adjoint action of the Virasoro group, the primary operator transforms as (note that $F = f^{-1}$)
\begin{equation}
V_{f}^\dagger\,\mathcal{O}_h(\phi)\,V_{f}= F'(\phi)^{h}\,\mathcal{O}_h(F(\phi))\,.
\label{eq:primarytransformation}
\end{equation} 
The algebra of primaries $\{\mathcal{O}_{h_i,n}\}_{i\in W}$ can include multiple primaries with equal or different weights. In general, the commutation relations between the primaries are \cite{Caputa:2021sib}
\begin{equation}
	[\mathcal{O}_{h_i,n},\mathcal{O}_{h_j,m}] = D_{ij}\,\binom{n+h_i-1}{2h_i-1}\,\delta_{n,-m} + \sum_{k\in W}C_{ij}^k\,p_{ij}^k(n,m)\,\mathcal{O}_{h_k,\,n+m}\,,
 \label{eq:primaryalgebra}
\end{equation}
where $D_{ij}$ is the normalization of the 2-point function of two primary operators of weights $h_i,h_j$ (we may set $D_{ij} = \delta_{ij}$), $C_{ij}^k$ is the coefficient of the 3-point function of three primaries and $p_{ij}^k(n,m)$ are polynomials of $n,m$. The first term is derived in Appendix \ref{app:primaryproperties} and it is the same for all primary operator algebras, because it follows from the transformation law \eqref{eq:primarytransformation}. The remaining terms are non-universal and depend on the OPE coefficients of the primary operator algebra in question.

We may define two types of deformations of a Virasoro circuit that we refer to as \textit{chiral} and \textit{coupled} circuits.
The chiral circuit is obtained by including a primary operator $ \mathcal{O}_h(\phi)$ into the generator \eqref{eq:genCC} of the Virasoro circuit as
\begin{equation}
    \textbf{Chiral circuit:}\quad G(t) = G(f_t) + \int_0^{2\pi} d\phi\,J(\phi,t)\,\mathcal{O}_h(\phi)\,,
    \label{eq:chiraldeformationCC}
\end{equation}
where $J(\phi,t) = J(\phi+2\pi,t)$ is the source. Alternatively, instead of the deformation \eqref{eq:chiraldeformationCC} of a single Virasoro circuit, we may also consider a deformation of the tensor product circuit \eqref{eq:tensorproductcircuit} given by
\begin{equation}
    \textbf{Coupled circuit:}\quad G(t) = G(f_t,\overbar{f}_t) + \int_0^{2\pi} d\phi\,J(\phi,t)\,\mathcal{O}_h(\phi)\otimes \overbar{\mathcal{O}}_{\overbar{h}}(\phi)\,,
    \label{eq:coupledcircuitgen}
\end{equation}
where $\overbar{\mathcal{O}}_{h}(\phi) = \mathcal{O}_{h}(-\phi)$ (see Appendix \ref{subapp:conjugate_rep} for properties of the conjugate representation). The circuit couples the two Virasoro circuits to each other and $\overbar{h} \neq h$ in general. We will only consider a deformation with a single primary for simplicity.

The quantum information geometry explored by primary-deformed Virasoro circuits is larger than the space of Virasoro pure states. The primary mode operators $\mathcal{O}_{h,n}$ are vectors pointing in directions orthogonal to the Virasoro algebra that generate a flow in an infinite number of new dimensions. The larger manifold consists of states that are linear combinations of descendants of the vacuum $\ket{0}$ and of descendants of all highest-weight states $h_{i\in W}$ present in the algebra. Therefore Hilbert spaces explored by primary-deformed circuits are direct sums
\begin{equation}
    \mathcal{H} =
    \begin{dcases}
         \bigoplus_{i\in  W}\mathcal{H}_{h_i}\,,\quad &\text{chiral circuit}\\
         \bigoplus_{i,j\in  W}\mathcal{H}_{h_i}\otimes \mathcal{H}_{h_j}\,,\quad &\text{coupled circuit}\\
    \end{dcases}
\end{equation}
of multiple Verma modules (which we take to include the vacuum module) of the Virasoro algebra. The coefficients of linear combinations of descendants of the different modules parametrize the larger manifold.

For Virasoro circuits, we are able to parametrize the instantaneous state $\ket{t} = U(t)\ket{R}$ of the circuit as a Virasoro pure state $\ket{f_t} = V_{f_t}\ket{R}$ where $f_t$ is a curve on the manifold infinite-dimensional manifold $\widetilde{\text{Diff}}_+S^1$. In other words, given a time-dependent generator $G(t)$, we have $U(t) = V_{f_t}$ where $f_t$ is the solution of the first order differential equation $\dot{f}_t = u_t\circ f_t$ with the boundary condition $f_0 = \text{id}$. This is possible, because the Virasoro algebra is the Lie algebra of a known Lie group, the Virasoro group, which is the central extension of $\widetilde{\text{Diff}}_+S^1$. When $G(t)$ contains a primary deformation, the unitary operator $U(t)$ involves also the commutator \eqref{eq:primaryalgebra}. This does not form the Lie algebra of a known Lie group due to the unknown OPE coefficients $C_{ij}^k$. Therefore it is not possible to parametrize the space of states explored by primary-deformed Virasoro circuits as an infinite-dimensional manifold by writing $U(t)$ as a representation of some group. To evade this problem in what follows, we compute $U(t)$ only to quadratic order in a perturbative expansion in the primary deformation, which does not require the knowledge of $C_{ij}^k$.

\subsection{Perturbative expansion of the FS metric}

Let us start to investigate the FS metric of a circuit with generator
\begin{equation}
    G(t) = C(t) + \lambda P(t)
\end{equation}
where $C(t)$ is the Virasoro circuit generator and $P(t)$ the primary deformation involving the source $J(\phi,t)$. $C(t)$ and $P(t)$ refer to the first and second summand in \eqref{eq:chiraldeformationCC} and \eqref{eq:coupledcircuitgen} for chiral and coupled cases respectively. Additionally, $\lambda\ll1$ is a small parameter that enables us to employ perturbation theory to find corrections to the Fubini--Study cost originating from the primary deformation. To this end we consider the curve $\ket{\Psi(t)} = U(t)\ket{h'}$ on the space of states where
\begin{equation}
    U(t) =\overleftarrow{\mathcal{T}}\exp{\biggl(-i\int_{0}^{t} ds\, G(s)\biggr)}\,,
\end{equation}
and where the reference state $\ket{h'}$\footnote{For the coupled case, in a slight abuse of notation, let us denote the (spinless) highest weight state of weight $h'$ by the same symbol $\ket{h'}\equiv\ket{h'}\otimes \ket{h'}$. In particular, the coupled case vacuum is $\ket{0}\equiv\ket{0}\otimes\ket{0}$.} is a primary state whose weight does not necessarily coincide with the weight of the primary $h$ used to deform the generator. As shown in Appendix \ref{sec:appendix_splittingtheunitary}, this circuit unitary can be factorized as
\begin{equation}
    U(t) = V(t)\;U_P(t)
    \label{eq:Utfactorization}
\end{equation}
where
\begin{equation}
    V(t)  = \overleftarrow{\mathcal{T}}\exp{\biggl(-i\int_{0}^{t} ds\, C(s)\biggr)}\,,\quad U_P(t) = \overleftarrow{\mathcal{T}}\exp{\biggl(-i\int_0^t ds\, V(s)^\dagger\,\lambda P(s)\,V(s)\biggr)}\,,
    \label{eq:unitaryfactors}
\end{equation}
This split is useful, because $V(t) = V_{f_t}$ (or $V(t) = V_{f_t,\overbar{f}_t}$ in the coupled case) has a known action on operators as given in \eqref{eq:Ttransformation_text} and \eqref{eq:primarytransformation}. 

Now given the decomposition \eqref{eq:Utfactorization}, the FS cost \eqref{eq:FSmetricfinitedimensional} of the curve $\ket{\Psi(t)} = U(t)\ket{h'}$ reduces to
\begin{equation}
    \mathcal{F}_{\text{FS}}(t) = \bra{h'}U_P(t)^\dagger\,\widetilde{G}(t)^2\,U_P(t)\ket{h'}- \bra{h'}U_P(t)^\dagger\,\widetilde{G}(t)\,U_P(t)\ket{h'}^2\,,
    \label{eq:variancestartingpoint}
\end{equation}
where we have defined the generator transformed by the Virasoro circuit $V(t)$ as
\begin{equation}
   \widetilde{G}(t) =\widetilde{C}(t) + \lambda\widetilde{P}(t) =V(t)^\dagger\,C(t)\,V(t)+V(t)^\dagger\,\lambda P(t)\,V(t) \,.
\end{equation}
At this point, we notice the difficulty in dealing with this deformed Virasoro circuit. Previously, in the case without deformation, the FS metric could be calculated analytically because the action of the Virasoro circuit on the operators is known. Because the primary deformation does not span a Lie algebra of a known Lie group, the FS metric is hard to calculate for a general deformation. However, for any operator $A(t)$ we can obtain the transformed result up to second order in $ \lambda$ via (see Appendix \ref{subapp:perturbative_expansion})
\begin{equation}
\label{eq:perturbative_expansion_maintext}
        U_P(t)^{\dagger}\,A(t)\,U_P(t)\\
= A(t)+ \lambda\, I_P(A(t))+\lambda^2\,I_{PP}(A(t)) + \mathcal{O}(\lambda^3)\,,
\end{equation}
where the first two corrections are given by
\begin{equation}
\label{eq:corrections_trafolawunderprimary}
    I_P(A(t))\equiv i\,\int_{0}^{t}ds\,[\widetilde{P}(s),A(t)]\,,\quad I_{PP}(A(t))\equiv-\int_{0}^{t} ds_1\int_{0}^{s_1} ds_2\,[\widetilde{P}(s_2),[\widetilde{P}(s_1),A(t)]]\,.
\end{equation}
Here $\widetilde{P}(t)$ appears, because it appears in the exponent of $U_P(t)$. Substituting \eqref{eq:perturbative_expansion_maintext} to \eqref{eq:variancestartingpoint} gives the expansion
\begin{equation}
    \mathcal{F}_{\text{FS}}(t) =\mathcal{F}_{\text{FS}}^{(0)}(t) + \lambda\,\mathcal{F}_{\text{FS}}^{(1)}(t) + \lambda^2\,\mathcal{F}_{\text{FS}}^{(2)}(t) + \mathcal{O}(\lambda^3)\,,
\end{equation}
where the zeroth order term is independent of the source and coincides with the line element of the Virasoro circuit
\begin{equation}
\label{eq:zerothorderFS}
    \mathcal{F}_{\text{FS}}^{(0)}(t) = \langle\widetilde{C}(t)^2\rangle-\langle\widetilde{C}(t)\rangle^2\,.
\end{equation}
At leading order in the source, we obtain (let us omit writing the time dependence)
\begin{equation}
\label{eq:firstorderFS}
    \mathcal{F}_{\text{FS}}^{(1)}(t) = \langle\widetilde{C}\widetilde{P}+\widetilde{P}\widetilde{C}\rangle-2\langle\widetilde{P}\rangle\langle\widetilde{C}\rangle+\langle \widetilde{C}I_P(\widetilde{C})+I_P(\widetilde{C})\,\widetilde{C}\rangle-2\langle I_P(\widetilde{C})\rangle\langle\widetilde{C}\rangle\,.
\end{equation}
where we have introduced the shorthand $\langle \,\cdots\,\rangle \equiv \bra{h'}\cdots\ket{h'}$. At second order in the source, the result is
\begin{align}
    \mathcal{F}_{\text{FS}}^{(2)}(t) &=\langle \widetilde{P}^2\rangle-\langle\widetilde{P}\rangle^2+\langle I_{P}(\widetilde{C})^2\rangle-\langle I_{P}(\widetilde{C})\rangle^2+\langle \widetilde{C}I_P(\widetilde{P})+I_P(\widetilde{P})\,\widetilde{C}\rangle-2\langle I_P(\widetilde{P})\rangle\langle\widetilde{C}\rangle\nonumber\\
    &+\langle \widetilde{C}I_{PP}(\widetilde{C})+I_{PP}(\widetilde{C})\,\widetilde{C}\rangle-2\langle I_{PP}(\widetilde{C})\rangle\langle \widetilde{C}\rangle+\langle \widetilde{P}I_{P}(\widetilde{C})+I_{P}(\widetilde{C})\,\widetilde{P}\rangle-2\langle I_{P}(\widetilde{C})\rangle\langle \widetilde{P}\rangle.\label{eq:secondorderFS}
\end{align}
At this point, let us highlight that this equation is fully general for any primary-deformed Virasoro circuit. The corrections to the cost of the Virasoro circuit depend on the commutator of the stress tensor with the primaries \eqref{eq:TOcommutator} as well as the commutator of the primaries with themselves \eqref{eq:primaryalgebra} which both contribute in \eqref{eq:corrections_trafolawunderprimary}. Given the many terms that contribute, we will move forward by identifying a subclass of the general case where we can make analytic progress.

\subsection{Cost of primary-deformed M\"obius circuits}

The expressions for the cost \eqref{eq:zerothorderFS} - \eqref{eq:secondorderFS} simplify when the reference state $\ket{h'}$ is an eigenstate of $\widetilde{C}(t)$. This is only possible when (a) $\widetilde{C}(t) = L_0$ or when (b) $\vert h'\rangle = \vert 0\rangle$ and $\widetilde{C}(t)$ is the generator of a M\"obius circuit implementing $f_t\in \widetilde{SL}(2,\mathbb{R})$. In both cases, the zeroth and first order terms vanish $\mathcal{F}_{\text{FS}}^{(0)}(t) = \mathcal{F}_{\text{FS}}^{(1)}(t) = 0$ and the variance is only given by the second order term \eqref{eq:secondorderFS} which simplifies further due to the fact that $\ket{h'}$ is an eigenstate of $\widetilde{C}(t) $. The result for the full variance \eqref{eq:variancestartingpoint} up to quadratic order is thus in these two cases
\begin{equation}
    \mathcal{F}_{\text{FS}}(t) = \lambda^2\mathcal{F}_{\text{FS}}^{(2)}(t) + \mathcal{O}(\lambda^3)
\end{equation}
with the leading term given by
\begin{equation}
    \mathcal{F}_{\text{FS}}^{(2)}(t) = \langle \widetilde{P}^2\rangle-\langle\widetilde{P}\rangle^2+\langle I_{P}(\widetilde{C})^2\rangle-\langle I_{P}(\widetilde{C})\rangle^2+\langle \widetilde{P}I_{P}(\widetilde{C})+I_{P}(\widetilde{C})\widetilde{P}\rangle-2\langle I_{P}(\widetilde{C})\rangle\langle \widetilde{P}\rangle \,.
    \label{eq:variancesimplequadratic}
\end{equation}
Moreover, in case (b), the two terms involving $\langle \widetilde{P}\rangle$ vanish due to the fact that the primary operator vacuum expectation value vanishes. This case is the main focus for the remainder of this paper. Although the restriction to M\"obius circuits $f_t\in \widetilde{SL}(2,\mathbb{R})$ is a strong restriction, it has been of particular interest in recent investigations \cite{deBoer:2023lrd,Jiang:2024hgt,Das:2024vqe,Malvimat:2024vhr}. So far we have neglected regularization of local operators in the expression for the cost which is discussed at the end of Section \ref{subsec:complexityofcircuits}. We will give a proper treatment in Section \ref{subsec:trivialcoupledcircuit} where it is needed to obtain finite results. 

For simplicity, in what follows we will consider sources that factorize into space and time profiles as
\begin{align}
\label{eq:sourcefactorization}
    J(\phi,t)=  S(\phi)\,j(t)\,,
\end{align}
where $S(\phi)$ is the spatial modulation and $j(t)$ encodes time-dependence of the source. Factorized sources are enough for our applications, but the calculations can be generalized to arbitrary sources as well.

\paragraph{Chiral circuit.} Let us assume that we work with a chiral primary insertion into the chiral sector of a general Virasoro circuit
\begin{align}
    C(t)=\int_0^{2\pi}d \phi\,u_t(\phi)\,T(\phi)\,,\quad\label{eq:chiral_C(t)_generalform}\\ P(t)=\int_0^{2\pi} d\phi\,J(\phi,t)\,\mathcal{O}_h(\phi)\,,
\end{align}
from which we obtain the transformed generators (see derivation of \eqref{eq:tildeGu} above)
\begin{equation}
    \widetilde{C}(t) = \int_0^{2\pi}d \phi\,\widetilde{u}_t(\phi)\left(T(\phi)+ \frac{c}{24\pi}\,\{f_t(\phi),\phi\}\right)\,,\quad \widetilde{P}(t)=\int_0^{2\pi} d\phi\,\widetilde{J}_h(\phi,t) \,\mathcal{O}_h(F_t(\phi))\,,
\end{equation}
where we have defined the transformed source $\widetilde{J}_h(\phi,t) \equiv J(\phi,t)\,F_t'(\phi)^h$. It follows that
\begin{equation}
   [\widetilde{P}(s),\widetilde{C}(t)] = -\int_0^{2\pi}d\phi_1\int_0^{2\pi}d\phi_2\, \widetilde{u}_t(\phi_1)\,\widetilde{J}_{h}(\phi_2,s)\,[T(\phi_1),\mathcal{O}(F_s(\phi_2))]\,.
\end{equation}
By substituting the commutator \eqref{eq:TOcommutator}, we obtain after integrations by parts
\begin{equation}
    I_{P}(\widetilde{C})=\int_{0}^{t}ds\int_0^{2\pi}d\phi_2\, \widetilde{J}_h(\phi_2,s)\left(\widetilde{u}_t(\phi_1)\,\partial_{\phi_1}+h\,\widetilde{u}_t'(\phi_1)\right)\mathcal{O}_h(\phi_1)\bigg\lvert_{\phi_1 = F_s(\phi_2)}
\end{equation}
For notational convenience, let us define the differential operator
\begin{align}
\label{eq:diffop}
    \mathcal{D}_{t}(\phi)\equiv\widetilde{u}_t(\phi)\,\partial_{\phi}+h\,\widetilde{u}_t'(\phi)\,.
\end{align}
Assuming the factorization of the source \eqref{eq:sourcefactorization}, it is useful to define the two \textit{ring operators} as
\begin{align}
    \mathcal{R}_{h}(t)&\equiv\int_0^{2\pi}d\phi\, \widetilde{S}_{h}(\phi)\,\mathcal{O}_h(F_t(\phi))\,,\nonumber\\
\mathcal{D}\mathcal{R}_h(t)&\equiv\int_0^{2\pi}d\phi\, \widetilde{S}_{h}(\phi) \left(\mathcal{D}_{t}(\phi_1)\,\mathcal{O}_h(\phi_1)\right)\bigg\lvert_{\phi_1 = F_t(\phi)}\,,\label{eq:chiralringoperators}
\end{align}
where we have defined the transformed spatial source $\widetilde{S}_{h}(\phi)\equiv S(\phi)\,F_t'(\phi)^h$. We call them ring operators, because they are localized in time, but smeared over the spatial circle. With these definitions, we get simply
\begin{equation}
    \widetilde{P}(t)= j(t)\,\mathcal{R}_{h}(t)\,,\quad I_{P}(\widetilde{C})= \int_{0}^{t}ds\,j(s)\,\mathcal{D}\mathcal{R}_h(s)\,.
\end{equation}
Substituting to \eqref{eq:variancesimplequadratic} and using that $\langle \mathcal{O}_h(\phi) \rangle  = 0$ gives
\begin{align}
\label{eq:variance_chiral}
    \mathcal{F}_{\text{FS}}^{(2)}(t) =&\,j(t)^2\,\langle\mathcal{R}_{h}(t)^2\rangle+j(t)\int_{0}^{t}ds\,j(s) \,\langle \mathcal{R}_{h}(t)\,\mathcal{D}\mathcal{R}_{h}(s)\rangle\nonumber\\&+\int_{0}^{t}ds_1\int_{0}^{t}ds_2\,j(s_1)\,j(s_2)\,\langle \mathcal{D}\mathcal{R}_{h}(s_1)\,\mathcal{D}\mathcal{R}_{h}(s_2)\rangle\,.
\end{align}

\paragraph{Coupled circuit.} Let us investigate the same object for the coupled case where
\begin{align}
    &C(t)=\int_0^{2\pi}d \phi\,u_t(\phi)\,T(\phi)\otimes\mathbf{1}-\int_0^{2\pi}d \phi\,\overbar{u}_t(\phi)\,\mathbf{1}\otimes\overbar{T}(\phi)\label{eq:coupled_C(t)_generalform}\\
    &P(t)=\int_0^{2\pi} d\phi\,J(\phi,t)\,\mathcal{O}_h(\phi)\otimes \overbar{\mathcal{O}}_{\overbar{h}}(\phi)
\end{align}
Similarly as above, we obtain
\begin{align}
   \widetilde{C}(t)&=\int_0^{2\pi}d \phi\,\widetilde{u}_t(\phi)\,(T(\phi)\otimes\mathbf{1})-\int_0^{2\pi}d \phi\,\widetilde{\overbar{u}}_t(\phi)\,(\mathbf{1}\otimes \overbar{T}(\phi)) + (\text{Schwarzians})\,,\\
   \widetilde{P}(t)&=\int_0^{2\pi} d\phi\,\widetilde{J}_{h\overbar{h}}(\phi,t)\,\mathcal{O}_h(F_t(\phi))\otimes \overbar{\mathcal{O}}_{\overbar{h}}(\overbar{F}_t(\phi))\,,
\end{align}
where we have defined $\widetilde{J}_{h\overbar{h}}(\phi,t)\equiv J(\phi,t)\,F_t'(\phi)^h\,\overbar{F}_t'(\phi)^{\overbar{h}}$. The Schwarzian terms cancel in the calculation below because $\widetilde{C}$ only appears inside commutators. To evaluate $I_{P}(\widetilde{C})$ in this case, we need to calculate the commutator
\begin{align}
    [\widetilde{P}(s),\widetilde{C}(t)] = &-\int_0^{2\pi}d\phi_1\int_0^{2\pi}d\phi_2\, \widetilde{u}_t(\phi_1)\,\widetilde{J}_{h\overbar{h}}(\phi_2,s)\,[T(\phi_1)\otimes \mathbf{1},\mathcal{O}_h(F_s(\phi_2))\otimes \overbar{\mathcal{O}}_{\overbar{h}}(\overbar{F}_s(\phi_2))]\nonumber\\
    &+\int_0^{2\pi}d\phi_1\int_0^{2\pi}d\phi_2\, \widetilde{\overbar{u}}_t(\phi_1)\,\widetilde{J}_{h\overbar{h}}(\phi_2,s)\,[\mathbf{1}\otimes\overbar{T}(\phi_1),\mathcal{O}_h(F_s(\phi_2))\otimes \overbar{\mathcal{O}}_{\overbar{h}}(\overbar{F}_s(\phi_2))]
\end{align}
where the commutators of tensor product operators can be evaluated via the commutator \eqref{eq:TOcommutator} in each sector. This makes it clear that the chiral half of the stress tensor leaves the antichiral part of the primary unaffected and vice versa. Hence, the result for the coupled case goes through very similarly as the chiral case above and we obtain
\begin{align}
    I_{P}(\widetilde{C})&=\int_{0}^{t}ds\int_0^{2\pi}d\phi\, \widetilde{J}_{h\overbar{h}}(\phi,s)\left[\left(\widetilde{u}_t(\phi_1)\,\partial_{\phi_1}+h\,\widetilde{u}_t'(\phi_1)\right)\mathcal{O}_h(\phi_1)\right]\otimes\overbar{\mathcal{O}}_{\overbar{h}}(\phi_1)\bigg\lvert_{\phi_1 = F_s(\phi)}\nonumber\\
    &+\int_{0}^{t}ds\int_0^{2\pi}d\phi\, \widetilde{J}_{h\overbar{h}}(\phi,s)\,\mathcal{O}_{h}(\phi_1)\otimes\left[\left(\widetilde{\overbar{u}}_t(\phi_1)\,\partial_{\phi_1}+\overbar{h}\,\widetilde{\overbar{u}}_t'(\phi_1)\right)\mathcal{\overbar{O}}_{\overbar{h}}(\phi_1)\right]\bigg\lvert_{\phi_1 = \overbar{F}_s(\phi)}
\end{align}
where the difference in sign of the chiral and antichiral conformal generator has cancelled with the different sign for the imaginary unit in the $T\mathcal{O}$ and $\overbar{T}\overbar{\mathcal{O}}$ commutator (see Appendix \ref{subapp:conjugate_rep}). Analogously as in the chiral case, we introduce two ring operators involving the coupled primary
\begin{align}
    \mathcal{R}_{h\overbar{h}}(t)&\equiv\int_0^{2\pi}d\phi\, \widetilde{S}_{h\overbar{h}}(\phi)\,\mathcal{O}_h(F_t(\phi))\otimes \overbar{\mathcal{O}}_{\overbar{h}}(\overbar{F}_t(\phi))\,,\label{eq:coupledringoperators}\\
    \mathcal{D}\mathcal{R}_{h\overbar{h}}(t)&\equiv\int_0^{2\pi}d\phi\, \widetilde{S}_{h\overbar{h}}(\phi)\, (\mathcal{D}_{t}(\phi_1)+\overbar{\mathcal{D}}_{t}(\phi_2))(\mathcal{O}_h(\phi_1)\otimes \overbar{\mathcal{O}}_{\overbar{h}}(\phi_2))\bigg\lvert_{\phi_1 = F_t(\phi),\phi_2 = \overbar{F}_t(\phi)}\,,\nonumber
\end{align}
where we have defined the source $\widetilde{S}_{h\overbar{h}}(\phi)\equiv S(\phi)\,F_t'(\phi)^h \overbar{F}_t'(\phi)^{\overbar{h}}$ (assuming factorization \eqref{eq:sourcefactorization}) and the differential operator $\overbar{\mathcal{D}}_{t}(\phi)\equiv\widetilde{\overbar{u}}_t(\phi)\,\partial_{\phi}+\overbar{h}\,\widetilde{\overbar{u}}_t'(\phi)$. Therefore we get in the coupled case
\begin{equation}
    \widetilde{P}(t)= j(t)\,\mathcal{R}_{h\overbar{h}}(t)\,,\quad I_{P}(\widetilde{C})= \int_{0}^{t}ds\,j(s)\,\mathcal{D}\mathcal{R}_{h\overbar{h}}(s)\,.
\end{equation}
Substituting to \eqref{eq:variancesimplequadratic} and using that $\langle \mathcal{O}_h(\phi) \rangle = \langle \overbar{\mathcal{O}}_{\overbar{h}}(\phi) \rangle = 0$ gives
\begin{align}
\label{eq:variance_coupled}
    \mathcal{F}_{\text{FS}}^{(2)}(t) =j(t)^2\,\langle\mathcal{R}_{h\overbar{h}}(t)^2\rangle &+j(t)\int_{0}^{t}ds\,j(s)\, \langle\mathcal{R}_{h\overbar{h}}(t)\,\mathcal{D}\mathcal{R}_{h\overbar{h}}(s)+\mathcal{D}\mathcal{R}_{h\overbar{h}}(s)\,\mathcal{R}_{h\overbar{h}}(t)\rangle\nonumber\\&+\int_{0}^{t}ds_1\int_{0}^{t}ds_2\,j(s_1)\,j(s_2)\,\langle\mathcal{D}\mathcal{R}_{h\overbar{h}}(s_1)\,\mathcal{D}\mathcal{R}_{h\overbar{h}}(s_2)\rangle\,.
\end{align}
At this point, let us highlight an important simplification in the sum of the derivative operators in the ring operator. By expressing the derivatives in terms of $t$ and $\phi$ via (see for example \cite{deBoer:2023lrd})
\begin{equation}
	\partial_{F_t(\phi)} =\frac{1}{F_t'(\phi)}\frac{\partial_t + \overbar{u}_t(\phi)\, \partial_\phi}{\overbar{u}_t(\phi)-u_t(\phi)}\,, \quad \partial_{\overbar{F}_t(\phi)} = -\frac{1}{\overbar{F}_t'(\phi)}\frac{\partial_t + u_t(\phi)\, \partial_\phi}{\overbar{u}_t(\phi)-u_t(\phi)}\,,
	\label{eq:tildepartial}
\end{equation}
and by employing $u_t(\phi)=-\dot{F}_t(\phi)/F_t'(\phi)$, $\widetilde{u}_t(F_t(\phi))=u_t(\phi)\,F_t'(\phi)$, we find
\begin{align}
\label{eq:derivativesimplification_coupled}
     \mathcal{D}_{t}(F_t(\phi))+\overbar{\mathcal{D}}_{t}(\overbar{F}_t(\phi))=-\partial_t+h\,\widetilde{u}_t'(F_t(\phi))+\overbar{h}\,\widetilde{\overbar{u}}_t'(\overbar{F}_t(\phi))\,,
\end{align}
where all spatial derivative contributions fully cancel. The identity \eqref{eq:derivativesimplification_coupled} allows to simplify the operator $\mathcal{D}\mathcal{R}_{h\overbar{h}}(t)$ and will be used below.

In this subsection, we have derived a formula for the leading correction, originating from inclusion of a primary operator in the circuit generator, to the (vanishing) Fubini--Study cost of a Virasoro circuit when the reference state is the vacuum state and the Virasoro part $C(t)$ implements M\"obius transformations $f_t\in \widetilde{SL}(2,\mathbb{R})$. The result depends on vacuum 2-point functions of previously defined ring operators $\mathcal{R}$ and $\mathcal{DR}$ in both the chiral and the coupled case. In order to gain intuition for the obtained analytic results, let us now focus on a specific instance where the Virasoro sector is as simple as possible, while the primary-deformation is kept as general as possible.

\section{Primary-deformed Virasoro zero mode circuit}\label{sec:trivialtimeevolution}

In this section, we consider the most simple example of a primary-deformed Virasoro circuit which captures the essential properties of the deformation. Let us choose the vacuum state as reference state $\ket{R} = \ket{0}$. The main simplifying assumption is now that we assume the Virasoro circuit generator to be the Virasoro zero mode, referring to $C(t) = L_0$ in the chiral case and to $C(t) = L_0\otimes \mathbf{1} + \mathbf{1}\otimes L_0$ in the coupled case. While we choose the most simple instance of the Virasoro circuit generator $C(t)$, we do not require additional assumptions on the source of the primary other than the previously stated factorization into spatial and temporal parts $J(\phi,t) =  S(\phi)\,j(t)$. This simple example is sufficient to capture non-trivial properties of the circuit due to the presence of the primary deformation and showcase the formulae derived in the previous section. To set up the examples, we will investigate the following spatial and temporal time profiles.

\paragraph{Choice of spatial and temporal source profiles.}

\begin{figure}[t]
\begin{subfigure}{.45\textwidth}
  \centering
  \includegraphics[width=\linewidth]{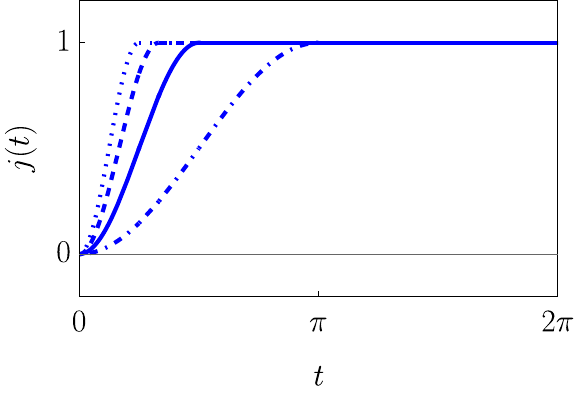}
  \caption{Sinusoidal switch-on with $k=1$ (dot-dashed), $k=2$ (solid), $k=3$ (dashed) and $k=4$ (dotted).}
  \label{fig:chiral_source_on}
\end{subfigure}
\hfill
\begin{subfigure}{.45\textwidth}
  \centering
  \includegraphics[width=\linewidth]{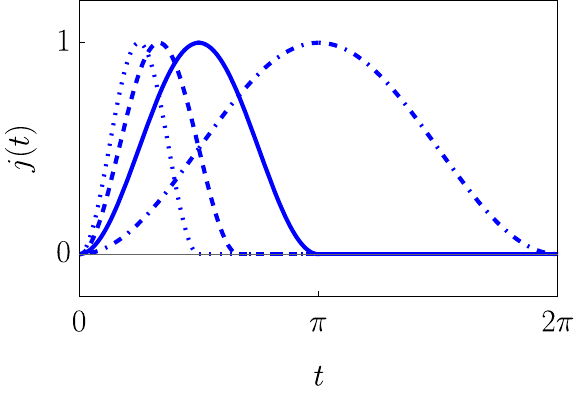}
  \caption{Sinusoidal switch-on to switch-off with $k=1$ (dot-dashed), $k=2$ (solid), $k=3$ (dashed) and $k=4$ (dotted).}
  \label{fig:chiral_source_onoff}
\end{subfigure}
\caption{Temporal source profiles $j(t)$ of interest for different positive integer $k$: (a) sinusoidal switch-on \eqref{eq:source_on} and (b) sinusoidal switch-on to switch-off \eqref{eq:source_onoff}. Both profiles are smooth at every point and reach a constant value at a characteristic time.}
\label{fig:sources}
\end{figure}

For all following cases we keep the primary weight $h$ as a parameter and assume spatial sources of the form 
\begin{equation}
\label{eq:spatial_source}
    S(\phi)=\cos{(n\phi)}\,,\quad n\in\mathbb{N}_0
\end{equation}
In terms of Fourier modes, this means that only the $n$-th Fourier mode and its conjugate $S_{-n}=S_n^*$ contribute with $S_{\pm n} = 1\slash 2$ while all other Fourier modes vanish. Having specified the spatial profile of the source, let us in the following examples investigate three distinct time profiles. We choose to investigate a source which is constant in time
\begin{equation}
\label{eq:source_const}
    j(t)=1\,,
\end{equation}
a smooth sinusoidal switch-on procedure
\begin{equation}
\label{eq:source_on}
    j(t)=\begin{dcases}
        \frac{1-\cos(kx)}{2}&t<\pi/k\,,\\
        1 &t\geq \pi/k\,,
    \end{dcases}
\end{equation}
with mode $k\in \mathbb{N}$, as well as a smooth sinusoidal switch-on to switch-off
\begin{equation}
\label{eq:source_onoff}
    j(t)=\begin{dcases}
        \frac{1-\cos(kx)}{2}&t<2\pi/k\,,\\
        0 &t\geq 2\pi/k\,,
    \end{dcases}
\end{equation}
with mode $k\in \mathbb{N}$. The two latter time profiles are visualized in Figure \ref{fig:sources}. These three source profiles constitute a set of sources that cover three processes of interest: a circuit that runs with constant addition of the primary term, a smooth switch-on of the primary term as well as the case where the primary operator is only switched on for a limited time before getting switched off again. Let us now move towards explicit investigation in the chiral circuit.

\subsection{Chiral circuit}

Zero mode time evolution in the Virasoro sector in the chiral case corresponds to the choice that the Virasoro circuit implements the path
\begin{align}
    f_t(\phi)=\phi+t\,.
\end{align}
or, in terms of the generator of the Virasoro circuit \eqref{eq:chiral_C(t)_generalform}, it takes the form 
\begin{align}
    C(t)=L_0\,.
\end{align}
The inverse diffeomorphism is $F_t(\phi)=\phi-t$ correspondingly. The generator of this circuit has the tangent vector $u_t(\phi)=1$ which in this special case is identical to the transformed tangent vector $\widetilde{u}_t(\phi)=1$. Because the tangent vectors are constant, their spatial derivatives vanish such that $\widetilde{u}_t'(\phi)=0$. Hence, the ring operators \eqref{eq:chiralringoperators} take the simple form
\begin{equation}
    \mathcal{R}_{h}(t) = \int_0^{2\pi}d\phi\, S(\phi)\,\mathcal{O}_h(\phi-t)\,,\quad \mathcal{D}\mathcal{R}_h(t) = -\partial_t \mathcal{R}_{h}(t)\,.
\end{equation}
where we have used $F_t'(\phi) = 1$ so that $\widetilde{S}_h(\phi) = S(\phi)$ and that
\begin{equation}
    \mathcal{D}_t(\phi_1)\,\mathcal{O}(\phi_1)\vert_{\phi_1 = F_t(\phi)} = \mathcal{O}_h'(\phi-t) = -\partial_t \mathcal{O}_h(\phi-t)\,.
\end{equation}
Importantly, we see that the derivative ring operator is just a pure time derivative of a ring operator without the derivative. Given this, the leading correction to the FS cost is
\begin{align}
    \mathcal{F}_{\text{FS}}^{(2)}(t) =j(t)^2\,\langle\mathcal{R}_{h}(t)^2\rangle &-j(t)\int_{0}^{t}ds\,j(s)\, \partial_s \langle\mathcal{R}_{h}(t)\,\mathcal{R}_{h}(s)+\mathcal{R}_{h}(s)\,\mathcal{R}_{h}(t)\rangle\nonumber\\&+\int_{0}^{t}ds_1\int_{0}^{t}ds_2\,j(s_1)\,j(s_2)\,\partial_{s_1}\partial_{s_2}\langle\mathcal{R}_{h}(s_1)\,\mathcal{R}_{h}(s_2)\rangle\,.
    \label{eq:finalexpression0}
\end{align}
We can integrate by parts in $s$ which gives\footnote{Integrating by parts the second term in \eqref{eq:finalexpression0} gives $-2\,j(t)^2\,\langle\mathcal{R}_{h}(t)^2\rangle+j(t)\,j(0)\,\langle[\mathcal{R}_{h}(t)\,\mathcal{R}_{h}(0)]_+\rangle+j(t)\int_0^t ds\,\partial_s j(s)\,\langle[\mathcal{R}_{h}(s)\,\mathcal{R}_{h}(t)]_+\rangle $. On the other hand, the boundary terms coming from the third term are $j(t)^2\,\langle\mathcal{R}_{h}(t)^2\rangle + j(0)^2\,\langle\mathcal{R}_{h}(0)^2\rangle -j(t)\,j(0)\,\langle[\mathcal{R}_{h}(t)\,\mathcal{R}_{h}(0)]_+\rangle-\int_0^t ds\,\partial_s j(s)\,[j(t)\,\langle[\mathcal{R}_{h}(s)\,\mathcal{R}_{h}(t)]_+\rangle-j(0)\,\langle[\mathcal{R}_{h}(s)\,\mathcal{R}_{h}(0)]_+\rangle] $. The sum of these is given by $j(0)^2\,\langle\mathcal{R}_{h}(0)^2\rangle-j(t)^2\,\langle\mathcal{R}_{h}(t)^2\rangle+j(0)\int_0^t ds\,\partial_s j(s)\,\langle[\mathcal{R}_{h}(s)\,\mathcal{R}_{h}(0)]_+\rangle$ which leads to \eqref{eq:finalexpression}. Here $[\,\cdot\,,\cdot\,]_+$ is the anticommutator.}
\begin{align}
    \mathcal{F}_{\text{FS}}^{(2)}(t) =j(0)^2\,\langle\mathcal{R}_{h}(0)^2\rangle &+j(0)\int_{0}^{t}ds\,\partial_sj(s)\, \langle\mathcal{R}_{h}(t)\,\mathcal{R}_{h}(s)+\mathcal{R}_{h}(s)\,\mathcal{R}_{h}(t)\rangle\nonumber\\&+\int_{0}^{t}ds_1\int_{0}^{t}ds_2\,\partial_{s_1}j(s_1)\,\partial_{s_2}j(s_2)\,\langle\mathcal{R}_{h}(s_1)\,\mathcal{R}_{h}(s_2)\rangle\,,
    \label{eq:finalexpression}
\end{align}
where there has been a cancellation of boundary terms supported at $s = 0$ and $s = t$ due to the important minus sign in the second term of \eqref{eq:finalexpression0}. The 2-point function of ring operators in the vacuum state has been computed in Appendix \ref{app:2pt_chiral} and the result is
\begin{equation}
     \bra{0} \mathcal{R}_h(t_1)\,\mathcal{R}_h(t_2)\ket{0} = \sum_{n=h}^\infty\, \vert S_n\vert^2\;\binom{h+n-1}{n-h}\, e^{-in(t_1-t_2)}\,,
\end{equation}
where $t_1$ is understood to have a small imaginary part $-i\varepsilon$ with $\varepsilon>0$ to ensure convergence of the series for general modes of the spatial source $S_n$.\footnote{This prescription comes from the definition of the operator product as a limit $\mathcal{O}_h(\phi_1)\,\mathcal{O}_h(\phi_2) = \lim_{\varepsilon \rightarrow 0^+}\overleftarrow{\mathcal{I}}\{\mathcal{O}_h(\phi_1+i\varepsilon)\,\mathcal{O}_h(\phi_2)\}$ corresponding to a prescription $\phi_1\rightarrow \phi_1 + i\varepsilon$. Since the argument of the primary operator in the ring operator is $\phi-t$ this translates to the prescription $t_1\rightarrow t_1 - i\varepsilon$.} The 2-point function is clearly complex valued, however, the imaginary part vanishes in the coincidence limit $t_1 = t_2$ and therefore does not contribute to the first term in \eqref{eq:finalexpression}. In addition, the imaginary part is an odd function of $t_1-t_2$ so it does not contribute to the other two terms of \eqref{eq:finalexpression} either, because they are symmetric under the exchange $t_1\leftrightarrow t_2$ of the arguments of the 2-point function. As a result, only the symmetric real part contributes which gives
\begin{align}
    \mathcal{F}_{\text{FS}}^{(2)}(t) =j(0)^2\,\text{Re}\,\langle\mathcal{R}_{h}(0)^2\rangle &+2\,j(0)\int_{0}^{t}ds\,\partial_sj(s)\, \text{Re}\,\langle\mathcal{R}_{h}(t)\,\mathcal{R}_{h}(s)\rangle\label{eq:finalexpressionrealpart}\\&+\int_{0}^{t}ds_1\int_{0}^{t}ds_2\,\partial_{s_1}j(s_1)\,\partial_{s_2}j(s_2)\,\text{Re}\,\langle\mathcal{R}_{h}(s_1)\,\mathcal{R}_{h}(s_2)\rangle\,\nonumber
\end{align}
with
\begin{equation}
    \text{Re}\bra{0} \mathcal{R}_h(t_1)\,\mathcal{R}_h(t_2)\ket{0} =\sum_{n=h}^\infty\, \vert S_n\vert^2\;\binom{h+n-1}{n-h} \cos{(n(t_1-t_2))}\,.
\end{equation}
The choice of spatial source \eqref{eq:spatial_source} leads to the simplified real part
\begin{equation}
\text{Re}\bra{0}\mathcal{R}_h(t_1)\,\mathcal{R}_h(t_2)\ket{0} = \begin{dcases}
        \frac{1}{4}\binom{h+n-1}{n-h} \cos{(n(t_1-t_2))} & n\geq h\,,\\
        0 & n<h\,.
    \end{dcases} 
\end{equation}
As all three terms in the cost \eqref{eq:finalexpressionrealpart} depend on the real part of the ring operator vacuum 2-point function, it is clear that the cost vanishes for a vanishing real part, so for all $n<h$.\footnote{In particular, this observation directly implies that the cost for $n=0$ always vanishes for all $h\in\mathbb{N}$.} This shows the key characteristic that (i) \textit{non-vanishing cost requires sufficienly large spatial inhomogenity} by which we refer to the necessity of having $n\geq h$ to get a non-vanishing result. We can see that this is consistent and expected by looking at the mode expansion of the ring operator
\begin{equation}
    \mathcal{R}_h(t) =\sum_{m=-\infty}^\infty S_m\,\mathcal{O}_{h,m}\,e^{-im t}=\frac{1}{2}\,\mathcal{O}_{h,n}\,e^{-in t}+\frac{1}{2}\,\mathcal{O}_{h,-n}\,e^{in t}
\end{equation}
and notice the fact that, because  $\mathcal{O}_{h,n}\ket{0} = 0$ for $n>-h$ as stated in \eqref{eq:primarymodesactonvacuum}, the ring operator annihilates the vacuum for the case $n<h$ given $n\in\mathbb{N}_0$ because in this case both of its primary modes $\mathcal{O}_{h,n}$ and $\mathcal{O}_{h,-n}$ annihilate the vacuum. Hence, it is clear that we  only get nonvanishing cost for the case where the ring operators act nontrivially on the vacuum.

\paragraph{Saturation and circuit history dependence.}

Let us elaborate on three key features concerning the influence of the temporal part of the source on the cost \eqref{eq:finalexpressionrealpart}, namely the previously stated features that (ii) \textit{the cost saturates when the source becomes time-independent}, (iii) \textit{the saturation value depends on the source history} and that (iv) \textit{source returning to zero does not imply cost saturation to zero}. It is important to highlight that time appears as an upper integral boundary only. What governs the specific time-dependence of the result is the derivative of the source $\partial_t j(t)$, not the source itself. This is consistent with the fact that for a time-independent source we get
\begin{equation}
    \mathcal{F}_{\text{FS}}^{(2)}(t) = \mathcal{F}_{\text{FS}}^{(2)}(0)  = j(0)^2\,\langle\mathcal{R}_{h}(0)^2\rangle\,,\quad \partial_tj(t) = 0\,,
\end{equation}
which matches with the expected time-independent result \eqref{eq:timeindependentcost}. Let us illuminate the fact that the derived cost saturates for sources reaching a constant value by investigating such a profile explicitely, given by
\begin{align}
\label{eq:saturatingsource}
    j(t)=\begin{dcases}
        j_{\text{pre}}(t)&t<t_c\,,\\
        j_s=j_{\text{pre}}(t_c)&t\geq t_c\,,
    \end{dcases}
\end{align}
where $j_{\text{pre}}(t)$ is a general profile pre-constant, $j_s$ is a constant and $t_c$ is the point in time where the source reaches the constant value. Plugging this time profile into \eqref{eq:finalexpressionrealpart}, we find for times after the source has reached a constant $t>t_c$
\begin{align}
     \mathcal{F}_{\text{FS}}^{(2)}(t) &=j_{\text{pre}}(0)^2\,\text{Re}\langle\mathcal{R}_{h}(0)^2\rangle +2j_{\text{pre}}(0)\int_{0}^{t_c}ds\,\partial_sj_{\text{pre}}(s)\, \text{Re}\langle\mathcal{R}_{h}(t)\,\mathcal{R}_{h}(s)\rangle\nonumber\\&\quad\quad+\int_{0}^{t_c}ds_1\int_{0}^{t_c}ds_2\,\partial_{s_1}j_{\text{pre}}(s_1)\,\partial_{s_2}j_{\text{pre}}(s_2)\,\text{Re}\langle\mathcal{R}_{h}(s_1)\,\mathcal{R}_{h}(s_2)\rangle\,\nonumber\\
     &=\text{constant,}
\end{align}
because the second and third term vanish in the regime of constant source. This shows that for a source that becomes time-independent at $t_c$, the derived Fubini--Study cost does so accordingly. Importantly, the saturation value does (nontrivially) depend on the history of the circuit trough the derivative of the temporal source $j_{\text{pre}}(t)$. Because of this history dependence, the constant value reached by the source does not alone govern the saturation value but the whole source profile is important. These observations prove analytically the three lessons (ii), (iii) and (iv) in the setup of chiral primary-deformed Virasoro zero mode evolution.

Let us make some further comments before moving on: given a general profile $j(t)$ with $j(0)=0$ ensuring smooth switch-on of the source from zero, only the last of the three terms remains in \eqref{eq:finalexpressionrealpart}. Hence, for all time profiles with $j(0)=0$, this term alone captures the cost of such circuits completely. Let us further highlight that the structure of the double integral term is also consistent with the saturation value we expect for a source profile which is constant and hence given by the first term in \eqref{eq:finalexpressionrealpart}. We can see the consistency between first and third term by plugging in a step function source $j(t)=\Theta(t-\varepsilon)$ and sending $\varepsilon\to0$. It is clear that while the first and the second term vanish in this case because this source satisfies $j(0)=0$, the third term recovers what is given in the constant source case by the first term because the derivatives of the heavyside step function are delta distributions which pick exactly $\text{Re}\,\langle\mathcal{R}_{h}(0)^2\rangle$ as the result of the double integral in the $\varepsilon\to0$ limit.  Let us now showcase the lessons of this paragraph explicitely in three examples.

\paragraph{Three simple chiral circuits.}

We begin our concrete investigation of examples in the most simple case, namely, a deformation of the  chiral Virasoro zero mode circuit with a temporal source profile which is constant in time \eqref{eq:source_const}. The spatial source is again \eqref{eq:spatial_source}. As discussed previously, because of vanishing derivative of the source, the only contribution to the final result \eqref{eq:finalexpressionrealpart} comes from the first term. The final expression in case of constant temporal source reads therefore
\begin{align}
\label{eq:chiral_costresult_const}
    \mathcal{F}_{\text{FS}}^{(2)}(t) =\text{Re}\,\langle\mathcal{R}_{h}(0)^2\rangle=\begin{dcases}
       \frac{1}{4}\binom{h+n-1}{n-h}&n\geq h\,,\\
       0&n<h\,.
    \end{dcases}
\end{align}
As already explained before, a time-independent source leads to a constant cost. For the case of a spatial inhomogeneity with mode $n$ less than the chiral primary weight $h$, there is no cost because the spatial inhomogeneity picks up no primary mode that does not annihilate the vacuum. This includes the case of homogenous spatial excitation $n=0$. One common feature that we find is that higher spatial inhomogeneities have larger cost. In other words, short wavelength spatial inhomogeneities are more costly than long wavelength ones. For larger weights $h$ of the primary, this effect gets more pronounced as we can read off the prefactor in the final result which introduces a higher scaling of the result in powers of $n$ for larger weights, i.e.
\begin{align}
    h=1\colon\;\;\,\mathcal{F}_{\text{FS}}^{(2)}(t)\propto n\,,\qquad h=2\colon\;\; \,\mathcal{F}_{\text{FS}}^{(2)}(t)\propto (n^3-n)\,,\qquad\ldots\,.
\end{align}
Let us highlight here that the observed scaling with $n$ in the $h=2$ case agrees with the $n$-scaling of the FS cost for the Virasoro circuit without deformation \eqref{eq:FSmetricdiffeos}. This is fully consistent with the fact that the stress tensor, the circuit generator in the pure Virasoro circuit case, is a quasi-primary of weight $h=2$.

\begin{figure}[t]
\begin{subfigure}{.45\textwidth}
  \centering
  \includegraphics[width=\linewidth]{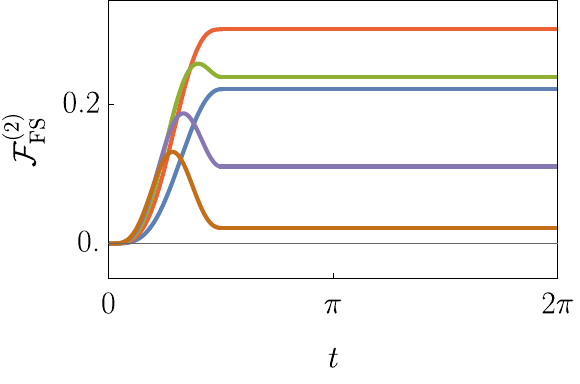}
  \caption{$h=1$, $k=2$ and variable modes $n=1$ (blue), $n=2$ (red), $n=3$ (green), $n=4$ (purple), $n=5$ (brown)}
  \label{fig:Chiral_Cost_On_k2_nvariable_h1}
\end{subfigure}
\hfill
\begin{subfigure}{.45\textwidth}
  \centering
  \includegraphics[width=\linewidth]{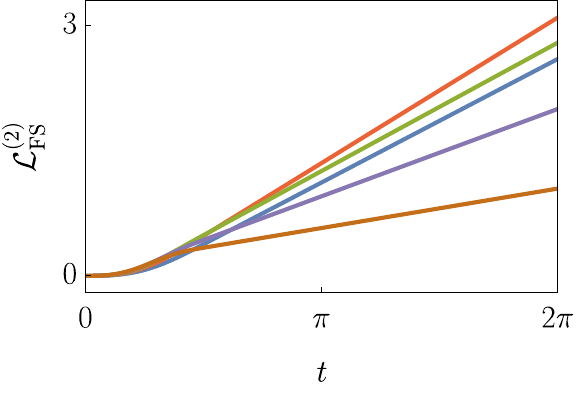}
  \caption{$h=1$, $k=2$ and variable modes $n=1$ (blue), $n=2$ (red), $n=3$ (green), $n=4$ (purple), $n=5$ (brown)}
  \label{fig:Chiral_AccumCost_On_k2_nvariable_h1}
\end{subfigure}
\caption{Time-evolution of (a) the leading quadratic correction $\mathcal{F}^{(2)}_{\text{FS}}(t)$ to the Fubini--Study cost function and (b) the corresponding accumulated FS cost $\mathcal{L}^{(2)}_{\text{FS}}(t)$ in the primary-deformed chiral circuit when the source profile $j(t)$ is a sinusoidal switch-on \eqref{eq:source_on}. The primary weight $h$, the temporal source mode $k$ and the spatial source modes $n$ are fixed parameters. The cost saturates and the accumulated cost grows linearly after the source reaches a constant value.}
\label{fig:chiral_On_1}
\end{figure}

The second example we are interested in is a gradual sinusoidal switch-on in the temporal part of the source, given in \eqref{eq:source_on}. Because this source satisfies $j(0)=0$, only the double integral contribution in \eqref{eq:finalexpressionrealpart} is relevant. Before this source has reached a constant, the relevant source contribution to the integral is the derivative
\begin{align}
\label{eq:source_on_derivative}
    \partial_t j(t)=\frac{k}{2}\sin(kt)\,.
\end{align}
Given this derivative, we can put this into the relevant expression \eqref{eq:finalexpressionrealpart} and solve the double integration exactly. Again, we find vanishing cost for $n<h$ and for $n\geq h$ the result reads
\begin{equation}
\label{eq:chiral_costresult_on}
    \mathcal{F}_{\text{FS}}^{(2)}(t)=\frac{1}{4}\binom{h+n-1}{n-h}\begin{dcases}
         \scriptstyle\frac{k^2 \left(k^2+n^2 \sin^2{(k t)}+k (k \cos (k t) (\cos (k t)-2 \cos (n t))-2 n \sin (k t) \sin (n t))\right)}{4 \left(k^2-n^2\right)^2} &t<\frac{\pi}{k}\,,\\
        \frac{k^4 \cos ^2\left(\frac{\pi  n}{2 k}\right)}{\left(k^2-n^2\right)^2}&t\geq\frac{\pi}{k}\,.
    \end{dcases}    
\end{equation}
An example visualization of this cost as well as the associated accumulated cost, introduced in \eqref{eq:accumulated_cost}, is given in Figure \ref{fig:chiral_On_1} for various  modes $n$ of the spatial source.

Let us highlight key features of this result. First of all, this result is well defined for all $k\in\mathbb{N},n\in\mathbb{N}_0$ as the limit $k\to n$ of this expression is finite. We see that the result saturates at $t=\frac{\pi}{k}$. The influence of the primary weight on the result is again specified by the appearance of $h$ in the binomial prefactor. We also notice that the intuition from the constant case that higher inhomogeneities $n$ always have higher cost, does not fully translate. In this example, the behaviour of the cost with $n$ does depend on $h$. For $h=1$ and $h=2$, for fixed $k$, the cost reaches a maximum for some mode $n'$ before going back down to zero. For higher conformal weights, on the other side, the cost increases with larger $n$.

The main difference to the constant case is that the saturation value is modulated by a factor which depends on the mode $k$ of the switch-on. This is how the switch-on history enters the saturation value. There is one more interesting observation: for $\frac{\pi n}{2k}=\frac{\pi}{2}+m\pi$ where $m$ is integer, the saturation value is identically zero because the exact combination of $n$ and $k$ has picked a zero of the cosine. This is the case when the source as well as the real part of the vacuum two point function of ring operators combine in such a way that the double integral over the real part and the two source derivatives gives zero because of their symmetries and periodicity. For completeness, let us also mention that in the $k\to\infty$ limit, the limit of infinitely fast switch-on, the cost of the circuit with constant source \eqref{eq:chiral_costresult_const} is recovered, showcasing again the consistency between terms one and three in \eqref{eq:finalexpressionrealpart} that was discussed at the end of the previous paragraph.

\begin{figure}[t]
\begin{subfigure}{.45\textwidth}
  \centering
  \includegraphics[width=\linewidth]{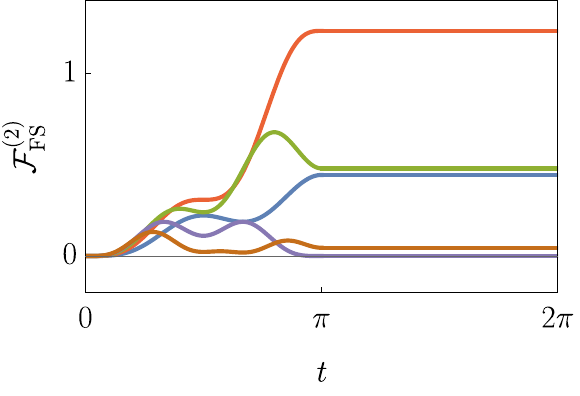}
  \caption{$h=1$, $k=2$ and variable modes $n=1$ (blue), $n=2$ (red), $n=3$ (green), $n=4$ (purple), $n=5$ (brown)}
  \label{fig:Chiral_Cost_OnOff_k2_nvariable_h1}
\end{subfigure}
\hfill
\begin{subfigure}{.45\textwidth}
  \centering
  \includegraphics[width=\linewidth]{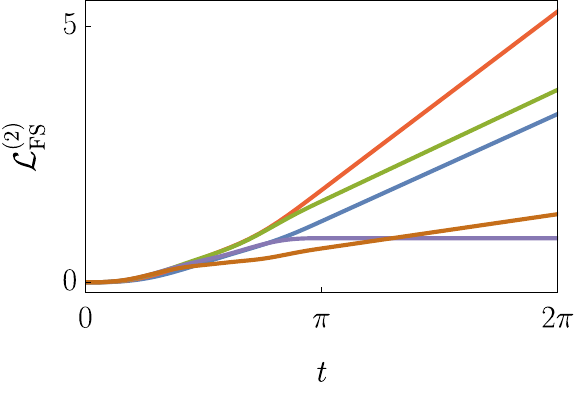}
  \caption{$h=1$, $k=2$ and variable modes $n=1$ (blue), $n=2$ (red), $n=3$ (green), $n=4$ (purple), $n=5$ (brown)}
  \label{fig:Chiral_AccumCost_OnOff_k2_nvariable_h1}
\end{subfigure}
\caption{Time-evolution of (a) the leading quadratic correction $\mathcal{F}^{(2)}_{\text{FS}}(t)$ to the Fubini--Study cost function and (b) the corresponding accumulated FS cost $\mathcal{L}^{(2)}_{\text{FS}}(t)$ in the primary-deformed chiral circuit when the source profile $j(t)$ is a sinusoidal switch-on to switch-off \eqref{eq:source_onoff}.  The primary weight $h$, the temporal source mode $k$ and the spatial source modes $n$ are fixed parameters. The cost saturates and the accumulated cost grows linearly after the temporal source reaches a constant value.}
\label{fig:chiral_OnOff_1}
\end{figure}

The third and final example of interest is the sinusoidal switch-on to switch-off \eqref{eq:source_onoff}, a peaked excitation of the primary where the temporal source returns to zero after a one peak excitation. This source also starts from zero $j(0)=0$ and the derivative $\partial_tj(t)$ is again \eqref{eq:source_on_derivative} for times before the source has returned to zero. The difference to the previous example is that this source only goes to a constant at time $t=2\pi/k$ compared to $t=\pi/k$ in the previous case. Inserting this into \eqref{eq:finalexpression} can again be solved analytically. For spatial excitations $n<h$ this once again vanishes, while for $n\geq h$ we obtain
\begin{equation}
\label{eq:chiral_costresult_onoff}
    \mathcal{F}_{\text{FS}}^{(2)}(t)=\frac{1}{4}\binom{h+n-1}{n-h}\begin{dcases}
        \scriptscriptstyle \frac{k^2 \left(k^2+n^2 \sin ^2(k t)+k (k \cos (k t) (\cos (k t)-2 \cos (n t))-2 n \sin (k t) \sin (n t))\right)}{4 \left(k^2-n^2\right)^2} &t<\frac{2\pi}{k}\\
        \frac{2 k^4 \sin ^2\left(\frac{\pi  n}{k}\right)}{\left(k^2-n^2\right)^2}&t\geq\frac{2\pi}{k}
    \end{dcases}    
\end{equation}
An example visualization of this cost as well as the accumulated cost is given in Figure \ref{fig:chiral_OnOff_1}. 

This result coincides with the previous switch-on result for $t\leq\pi/k$ by construction. This is meaningful since this is exactly the region where the temporal source profiles coincide. While the second example circuit saturates at $t=\pi/k$, the third example circuit continues to dynamically evolve until $t=2\pi/k$. The scaling of the resulting cost with $h,k$ and $n$ is similar to the previous case. Let us highlight the standout feature of the result: the cost saturates to a nonzero value although the temporal source returns to zero. As already expected from the previous analytical investigation, the history of the source, rather than just the saturation value of the source profile, determines the saturation value of the cost in this example. The physical interpretation of this third example is that the performed quench of finite duration changes the state in a significant way such that the evolution with the Virasoro zero mode $L_0$ is costly after the quench. This result showcases explicitly that the presented quantum circuit model achieves what we aspired, namely accessing a non-vacuum Verma module. After the source has returned to zero, evolution with $L_0$ gets assigned nonzero cost. This is only possible if the circuit states at that time do not lie in the Verma module of the vacuum state.

\subsection{Coupled circuit}\label{subsec:trivialcoupledcircuit}

Let us now turn our attention to the coupled circuit and investigate the same examples. The tensor product Virasoro zero mode circuit corresponds to
\begin{align}
    f_t(\phi)=\phi+t\,,\quad\quad\overbar{f}_t(\phi)=\phi-t\,.
    \label{eq:trivialtimediffeos}
\end{align}
Because of the relative sign in the definition of the generator \eqref{eq:coupled_C(t)_generalform} of the tensor product Virasoro circuit, the opposite sign of $t$ in the chiral and anti-chiral sectors ensures the generator takes the form\footnote{In a two-dimensional conformal field theory, the diffeomorphisms \eqref{eq:trivialtimediffeos} are related to light-ray coordinates $x^{\pm} = \phi \pm t$ on a Lorentzian cylinder and $C(t)$ coincides with the Hamiltonian of the cylinder on a constant-$t$ slice. The relative sign in the generator \eqref{eq:generatorcoupledvirasoro} ensures the CFT Hamiltonian takes the form \eqref{eq:trivialCt} which is bounded from below.}
\begin{align}
    C(t)=L_0\otimes\mathbf{1}+\mathbf{1}\otimes L_0\,.
    \label{eq:trivialCt}
\end{align}
The chiral tangent vectors behave as pointed out in the previous Subsection. The important difference in the anti-chiral sector is the sign of the time derivative of $\overbar{f}_t(\phi)$ such that the tangent vectors are $\overbar{u}_t(\phi)=\widetilde{\overbar{u}}_t(\phi)=-1$. Given the simplification of the derivative operator in the coupled case \eqref{eq:derivativesimplification_coupled}, it is clear that $\mathcal{D}_t(\phi+t)+\overbar{\mathcal{D}}_t(\phi-t)=-\partial_t$ so that the ring operators \eqref{eq:coupledringoperators} take the form
\begin{equation}
    \mathcal{R}_{h\overbar{h}}(t)=\int_0^{2\pi}d\phi\, S(\phi)\,\mathcal{O}_h(\phi-t)\otimes \overbar{\mathcal{O}}_{\overbar{h}}(\phi+t)\,,\quad \mathcal{D}\mathcal{R}_{h\overbar{h}}(t) = -\partial_t \mathcal{R}_{h\overbar{h}}(t)\,.
\end{equation}
The FS cost \eqref{eq:variance_coupled} for this setup simplifies to 
\begin{align}
    \mathcal{F}_{\text{FS}}^{(2)}(t) =j(t)^2\,\langle\mathcal{R}_{h\overbar{h}}(t)^2\rangle &-j(t)\int_{0}^{t}ds\,j(s)\, \partial_s \langle\mathcal{R}_{h\overbar{h}}(t)\,\mathcal{R}_{h\overbar{h}}(s)+\mathcal{R}_{h\overbar{h}}(s)\,\mathcal{R}_{h\overbar{h}}(t)\rangle\nonumber\\&+\int_{0}^{t}ds_1\int_{0}^{t}ds_2\,j(s_1)\,j(s_2)\,\partial_{s_1}\partial_{s_2}\langle\mathcal{R}_{h\overbar{h}}(s_1)\,\mathcal{R}_{h\overbar{h}}(s_2)\rangle\,.
    \label{eq:finalexpression0_coupled}
\end{align}
We can once again integrate by parts in $s$ which gives
\begin{align}
    \mathcal{F}_{\text{FS}}^{(2)}(t) =j(0)^2\,\langle\mathcal{R}_{h\overbar{h}}(0)^2\rangle &+j(0)\int_{0}^{t}ds\,\partial_sj(s)\, \langle\mathcal{R}_{h\overbar{h}}(t)\,\mathcal{R}_{h\overbar{h}}(s)+\mathcal{R}_{h\overbar{h}}(s)\,\mathcal{R}_{h\overbar{h}}(t)\rangle\nonumber\\&+\int_{0}^{t}ds_1\int_{0}^{t}ds_2\,\partial_{s_1}j(s_1)\,\partial_{s_2}j(s_2)\,\langle\mathcal{R}_{h\overbar{h}}(s_1)\,\mathcal{R}_{h\overbar{h}}(s_2)\rangle\,.
    \label{eq:finalexpression_coupled}
\end{align}
For a time-independent source, this produces the correct result \eqref{eq:timeindependentcost} due to the minus sign in the second term of \eqref{eq:finalexpression0_coupled} analogously to the chiral circuit. From this point onwards, we will restrict to spinless primaries $\overbar{h} = h$ for simplicity.

Up to this point we have neglected the divergence issues associated with 2-point functions of local operators at coincident insertions appearing in the formula \eqref{eq:finalexpression}. We can see the problem explicitly from the 2-point function of ring operators computed in Appendix \ref{app:2pt_coupled},
\begin{align}
\label{eq:Coupled_RingOperatorVevGeneral_text}
    &\bra{0} \mathcal{R}_{hh}(t_1)\,\mathcal{R}_{hh}(t_2)\ket{0}\\
    &=\frac{1}{(2\pi)^2}\sum_{n=-\infty}^\infty\vert S_{n}\vert^2\,\frac{ (2 h+\vert n\vert-1)! }{(2 h-1)!\,\vert n\vert!}\,e^{-i(2h+\vert n\vert)(t_1-t_2)}\, _2F_1\left(2 h,2 h+\vert n\vert;1+\vert n\vert;e^{-2 i (t_1-t_2)}\right)\,,\nonumber
\end{align}
where again $t_1\rightarrow t_1 - i\varepsilon$ with $\varepsilon\rightarrow 0^+$ is understood to ensure convergence for $t_1\neq t_2$. This 2-point function is complex valued in general which can be seen explicitly for example in the simple $h = 1$ expression,
\begin{equation}
    \bra{0} \mathcal{R}_{11}(t_1)\,\mathcal{R}_{11}(t_2)\ket{0}=\frac{1}{4}\frac{1}{(2\pi)^2}\sum_{n=-\infty}^\infty\vert S_{n}\vert^2\,e^{-i\vert n\vert\Delta t}\,(-\vert n\vert+i\cot{\Delta t})\csc^2{\Delta t}\,,\label{eq:ring2point_marginal}
\end{equation}
where we have defined $\Delta t\equiv t_1-t_2$. For general $h$, the real and imaginary parts of the 2-point function of ring operators can be extracted using identities for the hypergeometric function. The real part can be written as (see Appendix \ref{app:2pt_coupled})
\begin{align}
	&\text{Re}\bra{0} \mathcal{R}_{hh}(t_1)\,\mathcal{R}_{hh}(t_2)\ket{0}\label{eq:REtwopoint}\\
	&=\frac{1}{(2\pi)^2}\sum_{m=h}^\infty \frac{(-1)^{h+m}\,(m+h)!}{(2h-1)!\,(m-h)!}\,\biggl[\frac{\vert S_{2m}\vert^2}{2\,(m+h)}\, _2F_1\biggl(h-m,h-m;\frac{1}{2};\cos^{2}{(\Delta t)}\biggr)\biggr.\nonumber\\
	&\hspace{5cm}\biggl.+\vert S_{2m+1}\vert^2\,\cos{(\Delta t)}\, _2F_1\biggl(h-m+1,h-m;\frac{3}{2};\cos^{2}{(\Delta t)}\biggr)\biggr]\,,\nonumber
\end{align}
where we again use the shorthand notation $\Delta t\equiv t_1-t_2$. The real part is completely finite in the coincidence limit $t_2 \rightarrow t_1$, because the hypergeometric functions are order $m-h$ polynomials of $\cos^2{(\Delta t)} $, while the imaginary part is divergent as explained in Appendix \ref{app:2pt_coupled}. To again give an example, the divergent imaginary part in the $h = 1$ expression \eqref{eq:ring2point_marginal} evidently corresponds to the $\csc^3{(\Delta t)}$ pole. Given this analysis, the FS cost \eqref{eq:finalexpression_coupled} appears to be both complex and divergent. The resolution of this problem was discussed already at the end of Section \ref{subsec:complexityofcircuits}: starting from the FS metric involving the symmetrized product of $G_u$, $G_v$ and understanding that the operator products are limits of a meromorphic function in the complex plane, one ends up with the expression for the variance where only the real part of the 2-point function appears\footnote{This has not been relevant for the primary-deformed chiral circuit, because the imaginary part of the 2-point function vanishes at $t_1 = t_2$ and never contributes to the first term in the cost.}
\begin{align}
    \mathcal{F}_{\text{FS}}^{(2)}(t) =j(0)^2\,\text{Re}\,\langle\mathcal{R}_{hh}(0)^2\rangle &+2\,j(0)\int_{0}^{t}ds\,\partial_sj(s)\, \text{Re}\,\langle\mathcal{R}_{hh}(t)\,\mathcal{R}_{hh}(s)\rangle\label{eq:finalexpression_coupled_real}\\&+\int_{0}^{t}ds_1\int_{0}^{t}ds_2\,\partial_{s_1}j(s_1)\,\partial_{s_2}j(s_2)\,\text{Re}\,\langle\mathcal{R}_{hh}(s_1)\,\mathcal{R}_{hh}(s_2)\rangle\,,\nonumber
\end{align}
where the real part of the vacuum expectation value of two ring operators associated to a spinless primary with $\overbar{h}=h$ is given by equation \eqref{eq:REtwopoint}. In the last two terms, the divergent imaginary part is automatically cancelled by symmetricity in the arguments of the 2-point function. Equation \eqref{eq:finalexpression_coupled_real} is the correct finite expression for the cost of the coupled circuit which is one of the important technical results of this paper.

Let us highlight one key fact: this expression structurally agrees, up to the difference of having a different real part as a kernel weighting the contribution of the derivative of the temporal source (which is not important for the following argument), with the expression for the cost in the chiral case \eqref{eq:finalexpressionrealpart}. Hence, all previously discussed main features of the cost with respect to the temporal source profile $j(t)$ in the chiral case are also present in the coupled case, i.e.~(ii) \textit{the cost saturates when the source becomes time-independent}, (iii) \textit{the saturation value depends on the source history} and that (iv) \textit{source returning to zero does not imply cost saturation to zero}.

Let us investigate the obtained cost \eqref{eq:finalexpression_coupled_real} with respect to the same examples of spatial and temporal source that we have previously analyzed for the chiral circuit. The choice of $S(\phi)=\cos(n\phi),\,n\in\mathbb{N}_0$ will again amount to picking a single Fourier mode with $|S_n|^2=1/4$ and all other modes zero, simplifying the real part of the ring operator expectation value to
\ref{app:2pt_coupled})
\begin{align}
\label{eq:CoupledRETwoPt1Spatialmode}
	&\text{Re}\bra{0} \mathcal{R}_{hh}(t_1)\,\mathcal{R}_{hh}(t_2)\ket{0}=\frac{1}{16\pi^2}\frac{(-1)^{h}}{(2h-1)!}\\
	&\times \begin{dcases}
	    0&n<2h\\
     \frac{(-1)^{\tfrac{n}{2}}\,(\tfrac{n}{2}+h)!}{(\tfrac{n}{2}-h)!}\,\frac{1}{2h+n}\, _2F_1\biggl(h+\tfrac{n}{2},h-\tfrac{n}{2};\frac{1}{2};\cos^{2}{(\Delta t)}\biggr)&n\geq2h\text{ even}\\
     \frac{(-1)^{\tfrac{n-1}{2}}\,(\tfrac{n-1}{2}+h)!}{(\tfrac{n-1}{2}-h)!}\,\cos{(\Delta t)}\, _2F_1\biggl(h+\tfrac{n-1}{2}+1,h-\tfrac{n-1}{2};\frac{3}{2};\cos^{2}{(\Delta t)}\biggr)&n\geq2h\text{ odd}
	\end{dcases}\nonumber
\end{align}
This directly shows that the cost for $n<2h$ always vanishes, highlighting (i) \textit{non-vanishing cost requires sufficienly large spatial inhomogenity} also in the coupled case. We will now again investigate the three examples of a constant source time profile \eqref{eq:source_const}, a sinusoidal switch-on with mode with mode $k\in\mathbb{N}$ \eqref{eq:source_on} and a sinusoidal switch-on to switch-off with mode $k\in\mathbb{N}$ \eqref{eq:source_onoff}.

\paragraph{Three simple coupled circuits.}

Let us begin the investigation of examples of the coupled primary-deformed Virasoro zero mode circuit by assuming the constant source \eqref{eq:source_const}. The only contributing term will be the first term in \eqref{eq:finalexpression_coupled}. The coincidence limit of the real part of the vacuum expectation value of ring operator in the coupled case is derived in the appendix and given in \eqref{eq:coupled_re_coincidencelimit_app}. Given this, the result for the cost function reads
\begin{align}
\label{eq:coupled_costresult_const}
    \mathcal{F}_{\text{FS}}^{(2)}(t) =\lim_{t_2\to t_1}\text{Re}\bra{0} \mathcal{R}_{hh}(t_1)\mathcal{R}_{hh}(t_2)\ket{0}=\begin{dcases}
       \frac{1}{4}\frac{1}{(2\pi)^2}\binom{2h+n-1}{n-2h}&n\geq 2h\\
       0&n<2h
    \end{dcases}\,.
\end{align}
Consistently, again, we find a constant cost for a circuit with constant temporal source. Interestingly, this result agrees with the chiral sector result \eqref{eq:chiral_costresult_const} under the replacement $h\to2h$ up to a factor of $(2\pi)^{-2}$ which originates from the fact that we coupled two primaries but still only have one spatial integration in the generator such that one factor of $(2\pi)^{-1}$ from the normalization of the primary mode expansion \eqref{eq:primarydefinition} is additional in each ring operator in the coupled case.

\begin{figure}[t]
\begin{subfigure}{.45\textwidth}
  \centering
  \includegraphics[width=\linewidth]{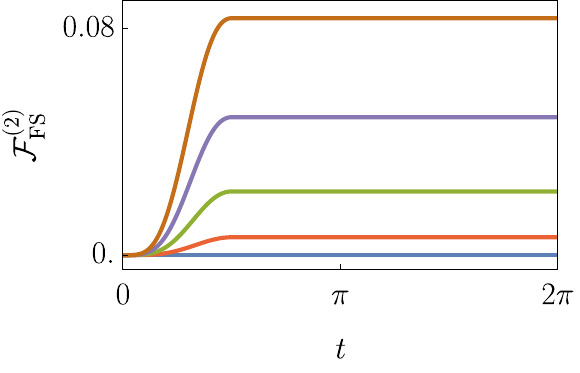}
  \caption{$h=1$, $k=2$ and variable modes $n=1$ (blue), $n=2$ (red), $n=3$ (green), $n=4$ (purple), $n=5$ (brown)}
  \label{fig:Coupled_Cost_On_k2_nvariable_h1}
\end{subfigure}
\hfill
\begin{subfigure}{.45\textwidth}
  \centering
  \includegraphics[width=\linewidth]{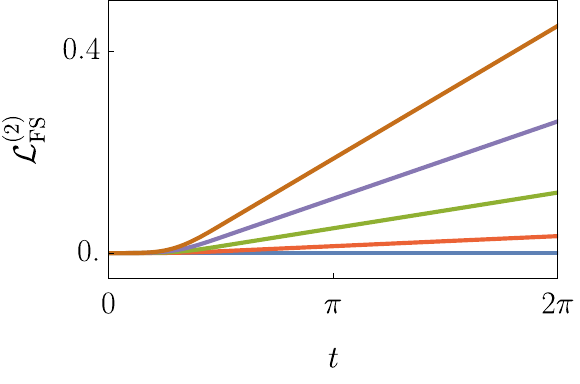}
  \caption{$h=1$, $k=2$ and variable modes $n=1$ (blue), $n=2$ (red), $n=3$ (green), $n=4$ (purple), $n=5$ (brown)}
  \label{fig:Coupled_AccumCost_On_k2_nvariable_h1}
\end{subfigure}
\caption{Time-evolution of (a) the leading quadratic correction $\mathcal{F}^{(2)}_{\text{FS}}(t)$ to the Fubini--Study cost function and (b) the corresponding accumulated FS cost $\mathcal{L}^{(2)}_{\text{FS}}(t)$ in the primary-deformed coupled circuit when the source profile $j(t)$ is a sinusoidal switch-on \eqref{eq:source_on}.  The primary weight $h$, the temporal source mode $k$ and the spatial source modes $n$ are fixed parameters. The cost saturates and the accumulated cost grows linearly after the temporal source reaches a constant value.}
\label{fig:coupled_On_1}
\end{figure}

The second example is a again a sinusoidal switch-on \eqref{eq:source_on}. Because the intricate and more complicated structure of the real part of the ring operator vacuum 2-point function \eqref{eq:CoupledRETwoPt1Spatialmode}, we resort to numerical integration to analyze the time evolution of the cost for the sinusoidal switch-on in the coupled case. The outcome is visualized in Figure \ref{fig:coupled_On_1}. In general, the coupled circuit behaves similarly to the chiral circuit but differences are also visible. As in the chiral circuit, the clear imprint of the source gives rise to a cost function that increases dynamically before saturating. In the coupled case, higher spatial inhomogenities always have higher cost, in contrast to the chiral case where the saturation value can also reach a maximum for a specific mode $n'$ before going down for higher $n$, given specific choices of $h$.

\begin{figure}[t]
\begin{subfigure}{.45\textwidth}
  \centering
  \includegraphics[width=\linewidth]{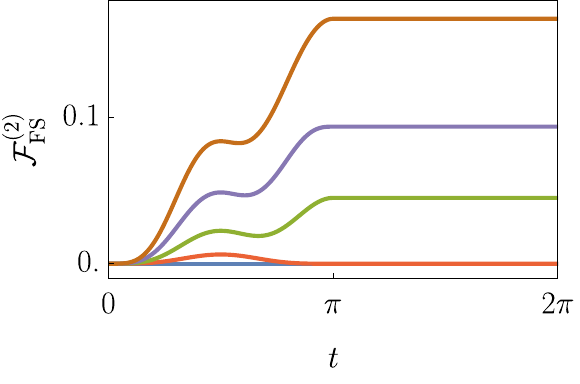}
  \caption{$h=1$, $k=2$ and variable modes $n=1$ (blue), $n=2$ (red), $n=3$ (green), $n=4$ (purple), $n=5$ (brown)}
  \label{fig:Coupled_Cost_OnOff_k2_nvariable_h1}
\end{subfigure}
\hfill
\begin{subfigure}{.45\textwidth}
  \centering
  \includegraphics[width=\linewidth]{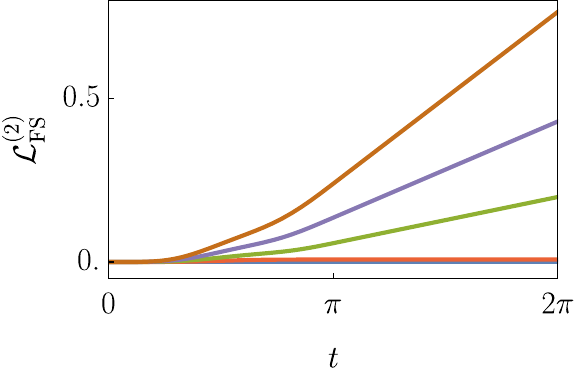}
  \caption{$h=1$, $k=2$ and variable modes $n=1$ (blue), $n=2$ (red), $n=3$ (green), $n=4$ (purple), $n=5$ (brown)}
  \label{fig:Coupled_AccumCost_OnOff_k2_nvariable_h1}
\end{subfigure}
\caption{Time-evolution of (a) the leading quadratic correction $\mathcal{F}^{(2)}_{\text{FS}}(t)$ to the Fubini--Study cost function and (b) the corresponding accumulated FS cost $\mathcal{L}^{(2)}_{\text{FS}}(t)$ in the primary-deformed coupled circuit when the source profile $j(t)$ is a sinusoidal switch-on to switch-off \eqref{eq:source_onoff}.  The primary weight $h$, the temporal source mode $k$ and the spatial source modes $n$ are fixed parameters. The cost saturates and the accumulated cost grows linearly after the temporal source reaches a constant value.}
\label{fig:coupled_OnOff_1}
\end{figure}

The last investigated example for the coupled circuit is the sinusoidal switch-on to switch-off given in \eqref{eq:source_onoff}. This is again treated numerically. The results are visualized in Figure \ref{fig:coupled_OnOff_1}. The observations are similar to the previous example: while a similar behaviour compared to the chiral case occurs, once again there is no falloff in the saturation value for excitations with larger $n$. The demonstration of non-vanishing cost after saturation has again a physical interpretation: Because evolution with purely the Virasoro zero mode is assigned a non-vanishing cost, the primary-deformed circuit has accessed a non-vacuum Verma module.

\paragraph{Results and implications for complexity bounds.}
In this section we have investigated three explicit realizations of simple chiral and coupled primary-deformed Virasoro zero mode circuits. The three examples were a constant source deformation, a circuit with sinusoidal switch-on of the deformation and one with sinusoidal switch-on to switch-off. Both chiral and the coupled realizations exhibit the same key features: concerning the spatial source $S(\phi)$, it is clear that (i) a \textit{non-vanishing cost requires sufficiently large spatial inhomogeneity}. In more detail, the cost vanishes for spatial inhomogeneities $S(\phi)=\cos{(n\phi)}$ with $n<h$ and $n<2h$ in the chiral and coupled case, respectively. The reason for this characteristic is that these spatial profiles excite modes of the primary operator that annihilate the vacuum. Furthermore, investigating the influence of the temporal part of the source $j(t)$ on the cost, we show that (ii) \textit{the cost saturates when the source becomes time-independent}, (iii) \textit{the saturation value depends on the source history} and that (iv) \textit{source returning to zero does not imply cost saturation to zero}. The investigated examples explicitly showcase the features (i-iv). The fact that primary-deformed circuits access non-vacuum Verma modules is clearly demonstrated by the switch-on to switch-off deformation: after the source has been switched off and the remaining evolution of the circuit is generated only by Virasoro zero mode, the cost is still non-vanishing. A numerically observed difference between the chiral and the coupled circuits is that, while in the chiral circuit higher spatial inhomogeneities get assigned higher cost for general primary weight only in the example of the constant temporal source, in the coupled circuit this feature is observed in all three examples.

To conclude, let us highlight the interpretation of the results bounding quantum circuit complexity of primary-deformed Virasoro circuits from above. The fact that (ii) \textit{source reaching a constant value implies saturation of cost} can be rephrased in the language of the accumulated cost to (ii*) \textit{source reaching a constant value implies linear growth of accumulated cost}. Recall that complexity is defined as the minimal value of the accumulated cost over all curves connecting the reference state to the target state. Let us consider the class of target states that can be reached by a primary-deformed Virasoro zero mode circuit with source reaching a constant value. Hence, for all primary-deformed Virasoro circuits that reach such a target state, the quantum circuit complexity is bounded from above by linear growth at late times.

\section{Discussion and outlook}\label{sec:outlook}

We have studied the Fubini--Study cost of a Virasoro circuit and its generalization where a deformation by a primary operator is included to the circuit generator. For Virasoro circuits, we related the differential regularization prescription used in the literature in the context of Virasoro circuits to meromorphic continuation of the operator product, derived explicit expressions for the Fubini--Study metric on the space of Virasoro and M\"obius pure states and uncovered a connection to the Kähler metric of a Virasoro coadjoint orbit. For primary-deformed Virasoro circuits, one of our main results is an explicit expression for the FS cost of general primary-deformed Virasoro circuits up to second order in the primary deformation. Applying the formula in a simple case, we have proved four universal characteristics of the evolution. A complete summary of results is given in Section \ref{subsec:summaryofresults}.

Let us outline a few extensions and future directions based on our work.

\paragraph{Explicit solutions for other generators.} We only computed the FS cost explicitly for primary-deformed circuits where the Virasoro circuit part of the generator is the Virasoro zero mode $C(t) = L_0$ or $C(t) = L_0\otimes \mathbf{1} + \mathbf{1}\otimes L_0$ in the chiral and coupled case respectively. This example is enough to study the effects of the primary deformation on the FS cost, however, it does not allow to study the interplay of the Virasoro and primary parts of the circuit. The simplest extension is to take $C(t)$ to be an element of the $\mathfrak{s}\mathfrak{l}(2,\mathbb{R})$ subalgebra of the Virasoro algebra generating a curve $f_t\in \widetilde{SL}(2,\mathbb{R})$ parametrized by three time-dependent real numbers of a corresponding $SL(2,\mathbb{R})$ matrix. By the results of Section \ref{subsec:FSmetricvirasoro}, the FS cost at zeroth order in the primary deformation is a line element in hyperbolic space. Preliminary investigations suggest that the calculation of the FS cost up to second order in the primary deformation $J(\phi,t)$ could be possible. It will be interesting to see if the extra direction parametrized by the source $J(\phi,t)$ allows for paths with smaller FS cost than paths confined to the hyperbolic space.

A further interesting case is the generalization to periodic temporal source profiles in the primary deformation part of the full generator. This models a system with Floquet dynamics whose information geometry has been of interest for example in the context of driven CFTs in \cite{Wen:2018agb,Lapierre:2019rwj,deBoer:2023lrd}. In the case of a periodic deformation, it is expected to find a growing cost as we can think of periodic driving as performing multiple successive switch-on to switch-off procedures, accumulating non-vanishing addition to the cost each time.

\paragraph{Non-perturbative extensions.} Our results for the FS cost of the primary-deformed circuit are perturbative in nature and valid to second order in the primary deformation. The result is universal and fixed completely by the fact that the deformation operator is a Virasoro primary. An extension to higher orders would be welcomed, however, universality of the result is lost already at third order where the three-point function of the primary operator enters: the three-point function involves an undetermined OPE coefficient which is not fixed by the transformation properties of the primary under the adjoint action of the Virasoro group. The result becomes even more non-universal at higher orders which are sensitive to the exact fusion rules of the various operators assumed to be present in the algebra. In particular, the non-universality is reflected in the non-trivial OPE (and commutation relation) of $\mathcal{O}_h$ with itself.

Non-perturbative information could be extracted in the specific case of a Virasoro--Kac--Moody algebra. This would correspond to a chiral primary $\mathcal{O}_1$ of weight $h = 1$ whose commutator with itself is fixed and involves the Kac--Moody central charge. Time-ordered exponentials $U_P(t)$ (see equation \eqref{eq:unitaryfactors}) of the generators are completely fixed by the algebra and would allow the computation of the FS cost non-perturbatively in the source similarly to the Virasoro circuit case. This algebra has already been considered in \cite{Bhattacharyya:2022ren}, however, the Fubini--Study cost function was not computed and the focus there was on a cost function linear in the generator. A similar algebra is the BMS algebra studied in \cite{Bhattacharyya:2023sjr}. Another situation where non-perturbative results could possibly be obtained is when $\mathcal{O}_h$ is a generalized free field whose higher-point functions decompose to sums of products of 2-point functions according to Wick's theorem.

Our perturbative calculations reveal a regime of linear growth of accumulated Fubini--Study cost after the quench. It will be interesting if non-perturbative corrections reveal a plateau and Poincaré recurrence at very late times, as discussed in this context in \cite{Erdmenger:2022lov}. We leave the analysis of higher-order corrections and non-perturbative effects to future work.

\paragraph{Quantum information geometry and geodesics.}

In this paper, we have answered how the addition of a primary operator with a given source $J(\phi,t)$ to the generator of a Virasoro zero mode circuit modifies its Fubini--Study cost. Firstly, an important future direction is to leverage these insights to study the quantum information geometry explored by general primary-deformed Virasoro circuits such as its negative curvature regions. As discussed at the end of Section \ref{subsec:definition_info_geometry}, the basic difficulty here lies in the fact that we do not know the Lie group generated by the Lie algebra of $L_n$ and $\mathcal{O}_n$. The final point reached by a Virasoro circuit is labeled by the Lie group element $(f,\overbar{f})$, but for a primary-deformed circuit, we do not know how to label the final point in an analogous manner since we do not have the Lie group in this case. Put another way, we are not able to write down a coordinate system in the general case.

Secondly, it would be crucial to find geodesics on the manifold explored by primary-deformed circuits to study the existence of conjugate points and to evaluate circuit complexities of states. So far we have only focused on non-geodesic curves and their associated accumulated costs. To this end, it is necessary to explicitly derive the cost function for a general circuit generated by arbitrary ``vector fields'' $(u_t(\phi),\overbar{u}_t(\phi),J(\phi,t))$ and not only for the zero mode subclass. Assuming we are able to write down a good coordinate system, this would give the cost functional for a general curve on the space of states generated by the algebra $L_n,\mathcal{O}_n$ which we are currently missing. From the resulting functional the geodesic equation could be derived via extremization over all curves.

Immediate progress in the above directions could be possible using the results of this paper if one restricts to the submanifold explored by primary-deformed Virasoro zero mode circuits. This restriction can be thought of as assigning infinite penalty factors to all Virasoro generators $L_n$ other than the zero mode. We leave the analysis of curvature, geodesics and complexity to future work.

\paragraph{Realization in a driven CFT.} The tools used in this paper are completely algebraic and relied solely on the Virasoro algebra and the transformation property of the primary under the Virasoro group. Combined with the definition of a highest-weight state, all necessary 2-point functions for calculations of the FS cost are fixed up to a choice of normalization of the operators. However, these algebras appear naturally in CFTs so that the circuits studied in this paper should also have a CFT interpretation.

Conformal (tensor product) circuits have already been realized as physical unitary time-evolution in a driven CFT on a Lorentzian cylinder coupled to a time-dependent background metric \cite{Erdmenger:2021wzc,deBoer:2023lrd}. The Hamiltonian of the CFT on a constant-time slice coincides with the generator $C(t)$ of the circuit and the components of the background metric determine the tangent vectors $u_t(\phi)$ and $\overbar{u}_t(\phi)$ of the curve $(f_t,\overbar{f}_t)$. The coupled primary-deformed Virasoro circuit also has a similar CFT realization where the CFT is driven, in addition to the metric, by a non-trivial background source for a scalar field $\mathcal{O}(\phi,t)$. The scalar field coincides with the coupled primary operator schematically as $\mathcal{O} = \mathcal{O}_h\otimes \mathcal{O}_h$ and its scaling dimension $\Delta = 2h$ is determined by the primary weight.

This is a model of a global quench driven by the scalar operator which has been studied in CFTs and other quantum systems in \cite{Dymarsky:2017awt,Das:2017sgp,DiGiulio:2021oal}. We wish to report on the CFT realization of the primary-deformed circuit in the nearby future.

\paragraph{Holographic interpretation.} If the primary-deformed circuit is realized in a driven CFT that has a holographic dual via the AdS\slash CFT correspondence, one naturally obtains a gravitational realization of the circuit. The gravity dual of a Virasoro circuit was analyzed in \cite{Erdmenger:2021wzc,deBoer:2023lrd} and it corresponds to a large diffeomorphism of a solution of three-dimensional pure gravity that does not include any non-trivial dynamics (see also \cite{Jiang:2024hgt}). To describe the primary-deformed circuit holographically, pure gravity must be supplemented with a dynamical bulk scalar field $\Phi$ which is dual to the deformation $\mathcal{O} = \mathcal{O}_h\otimes \mathcal{O}_h$: the mass of $\Phi$ is fixed by $h$ and its boundary value by the source function $J(\phi,t)$. Holographic duals of global quenches have been studied in higher dimensions for example in \cite{Buchel:2013lla,Buchel:2013gba}.

In general, $\Phi$ backreacts to the bulk metric, leading to shockwaves and possible black hole formation modelled by a time-dependent Vaidya geometry \cite{Anous:2016kss}. Previous work has considered complexity equals action and volume proposals in such situations in higher dimensions \cite{Moosa:2017yvt,Chapman:2018lsv,Chapman:2018dem,Fan:2018xwf}. Furthermore, shockwaves are a prime candidate for studies of the switchback effect \cite{Stanford:2014jda} for which our approach offers potential new insights from the boundary point of view. Similarly, it will be interesting to understand whether the Fubini--Study cost of primary-deformed circuits studied in this paper contain  signatures of black hole collapse in the dual gravity theory. Detecting such signatures most likely requires analyses beyond perturbation theory and the use of  large-$c$ methods similar to those in \cite{Anous:2016kss}. It is conceivable that quantum information geometry concepts such as the Fubini--Study metric will offer further insights into holography, in particular also in the context of black hole collapse.

\acknowledgments

We thank Anna-Lena Weigel for collaboration during the early stages of this work and Michal P. Heller, Stefan Kehrein, Arnab Kundu, Ren\'e Meyer and Dominik Neuenfeld for useful discussions. J.~E.~and J.~K. are supported by the Deutsche Forschungsgemeinschaft (DFG, German Research Foundation) through the German-Israeli Project Cooperation (DIP) grant ‘Holography and the Swampland’, as well as under Germany’s Excellence Strategy through the W\"{u}rzburg-Dresden Cluster of Excellence on Complexity and Topology in Quantum Matter - ct.qmat (EXC 2147, project-id 390858490). The project was also supported by  DFG individual grant ER 301/8-1. T.S.~would like to thank ICTS Bengaluru and the organizers of the program 'Quantum Information, Quantum Field Theory, and Gravity' (ICTS/qftg2024/08) for their hospitality during the final stages of this project.

\begin{appendix}

\section{Displacement operator of $SL_k(2,\mathbb{R})$ as a Virasoro unitary}\label{app:displacementoperator}

The Virasoro group element $V_{f_{M,k}}$ corresponding to $f_{M,k}$ defined in \eqref{eq:Slk2Rdiffeo} can be obtained as a time-ordered exponential involving a curve $h_\rho\in \widetilde{SL}_k(2,\mathbb{R})$ with circuit evolution parameter $\rho \in (0,\infty)$ such that $h_0 = \text{id}$ and $h_{\rho} = f_{M,k}$. One can see that $f_{M,k}(\phi)\vert_{\rho = 0} = \phi + \frac{1}{k}(\chi-\theta) = r_{(\chi-\theta)\slash k}(\phi)$ where we have defined the rotation of the circle $r_t(\phi) \equiv \phi + t$. Thus we can define the 1-parameter family of diffeomorphisms
\begin{equation}
    h_\rho \equiv  f_{M,k}\circ r_{(\theta-\chi)\slash k}\,,
\end{equation}
which satisfies $h_0 = \text{id}$. Using the formula \eqref{eq:Vft}, we obtain
\begin{equation}
    V_{h_\rho} = \overleftarrow{\mathcal{T}}\exp{\biggl(-i\int_0^\rho ds\int_0^{2\pi}d\phi\,(\partial_s h_s\circ h_s^{-1})(\phi)\,T(\phi)\biggr)}\,.
    \label{eq:Vhrho}
\end{equation}
Since $\partial_\rho h_\rho = (\partial_\rho f_{M,k})\circ r_{(\theta-\chi)\slash k}$, we get by using \eqref{eq:xiexplicit} explicitly
\begin{equation}
    (\partial_\rho h_\rho\circ h_\rho^{-1})(\phi) = (\partial_\rho f_{M,k}\circ f_{M,k}^{-1})(\phi)  = \frac{(-1)^k}{k}\sin{(k\phi-\chi)}\,.
\end{equation}
The integrand in \eqref{eq:Vhrho} is independent of $s$ so that the time-ordered exponential can be replaced with an ordinary exponential with $\int_0^\rho ds = \rho$ in the exponent. By substituting into \eqref{eq:Vhrho} and using $L_n = \int_{0}^{2\pi} d\phi\,e^{-in \phi}\,T(\phi)$, we obtain
\begin{equation}
    V_{h_\rho} = \exp{\biggl[\frac{(-1)^k}{2k}\,\rho\,(e^{i\chi}L_{k} -e^{-i\chi}L_{-k})\biggr]} = e^{\frac{(-1)^k}{2k}\,(\zeta\,L_{k} -\zeta^*L_{-k})}\equiv D_k(\zeta)\,,
\end{equation}
where $\zeta = \rho e^{i\chi}$. Using the composition law \eqref{eq:virasorocomposition}, we have (the 2-cocycle vanishes since $r_t''(\phi) = 0$)
\begin{equation}
    V_{h_\rho} = V_{f_{M,k}}V_{r_{(\theta-\chi)\slash k}}\quad \Rightarrow\quad V_{f_{M,k}} = V_{h_\rho}V_{r_{(\chi-\theta)\slash k}}\,,
\end{equation}
where $V_{r_{t}} = e^{-itL_0}$ since $(\dot{r}_{t}\circ r_{-t})(\phi) = 1$. Thus we get the formula
\begin{equation}
    V_{f_{M,k}} = D_k(\rho e^{i\chi})\,e^{i(\theta-\chi)\slash k\, L_0}\,.
    \label{eq:Virasoro_unitary_displacement_app}
\end{equation}

\paragraph{Matrix representation.} We can check the formula \eqref{eq:Virasoro_unitary_displacement_app} by using a two by two matrix representation of the Virasoro algebra generators. Let us define the operators
\begin{equation}
    \widetilde{L}_0 = \frac{1}{k}\,\biggl(L_0+\frac{c}{24}\,k^2\biggr)\,,\quad \widetilde{L}_{\pm 1} = \frac{1}{k}\,L_{\pm k}\,,
    \label{eq:Ltilde_generators}
\end{equation}
which, by the fact that $L_n$ obeys the Virasoro algebra \eqref{eq:stresstensorexp}, satisfy the $SL(2,\mathbb{R})$ algebra in the form
\begin{equation}
    [\widetilde{L}_n,\widetilde{L}_m] = (n-m)\, \widetilde{L}_{n+m}\,,\quad n,m = 0,\pm 1\,.
\end{equation}
Let us consider a two by two matrix representation of this $SL(2,\mathbb{R})$ algebra $\widetilde{L}_n \mapsto \widetilde{L}_n^{2\times 2}$ (mapping products and inverses of operators to products and inverses of matrices) given by
\begin{equation}
    \widetilde{L}_0^{2\times 2} = \frac{1}{2}\begin{pmatrix}
        -1 & 0\\
        0 & 1
    \end{pmatrix}\,,\quad
    \widetilde{L}_{-1}^{2\times 2} = \begin{pmatrix}
        0 & 0\\
        1 & 0
    \end{pmatrix}\,,\quad
    \widetilde{L}_{1}^{2\times 2} = \begin{pmatrix}
        0 & -1\\
        0 & 0
    \end{pmatrix}\,.
\end{equation}
By solving from \eqref{eq:Ltilde_generators} we obtain a two by two matrix representation for $L_0$, $L_{\pm k}$ as
\begin{equation}
    L_0^{2\times 2} = k\widetilde{L}_0^{2\times 2}-\frac{c}{24}\,k^2\,,\quad
    L_{-k}^{2\times 2} = k\widetilde{L}_{-k}^{2\times 2}\,,\quad
    L_{k}^{2\times 2} = k\widetilde{L}_{k}^{2\times 2}\,.
\end{equation}
Substituting this to \eqref{eq:Virasoro_unitary_displacement_app} gives the matrix representation of the M\"obius unitary
\begin{equation}
    V_{f_{M,k}}^{2\times 2} =  e^{i\frac{ck}{24}(\chi-\theta)} \,\exp{\biggl[\frac{(-1)^k}{2}\,\rho\,(e^{i\chi}\widetilde{L}_{1}^{2\times 2} -e^{-i\chi}\,\widetilde{L}_{-1}^{2\times 2})\biggr]}\,e^{i(\theta-\chi)\, \widetilde{L}_0^{2\times 2}}\,,
\end{equation}
which can be explicitly computed in this matrix representation to be
\begin{equation}
    V_{f_{M,k}}^{2\times 2} = e^{i\frac{ck}{24}(\chi-\theta)}\begin{pmatrix}
         e^{i(\chi-\theta)\slash 2}\cosh{\frac{\rho}{2}} & (-1)^{k+1}\,e^{i(\chi+\theta)\slash 2}\sinh{\frac{\rho}{2}}\\
        (-1)^{k+1}\,e^{-i(\chi+\theta)\slash 2}\sinh{\frac{\rho}{2}} & e^{-i(\chi-\theta)\slash 2}\cosh{\frac{\rho}{2}}
    \end{pmatrix}\,.
\end{equation}
This result is proportional to a $SU(1,1)\cong SL(2,\mathbb{R})$ matrix which, in terms of the parametrization \eqref{eq:SL2ab}, can be written as
\begin{equation}
     V_{f_{M,k}}^{2\times 2} = e^{i\frac{ck}{24}(\chi-\theta)}\begin{pmatrix}
         a & (-1)^{k+1}\,b\\
         (-1)^{k+1}\,b^* & a^*
     \end{pmatrix}\,.
\end{equation}
Because of the phase factor, this is a $ GL(2,\mathbb{R})$ matrix whose determinant is non-zero, but not equal to unity. Any invertible $ GL(2,\mathbb{R})$ matrix defines a diffeomorphism $f_k \in \widetilde{\text{Diff}}_+S^1$ via the homomorphism
\begin{equation}
    \begin{pmatrix}
        m_1 & m_2\\
        m_3 & m_4
    \end{pmatrix}\mapsto f_k\,,\quad   e^{ikf_k(\phi)} = \frac{m_1 e^{ik\phi} + m_2}{m_3e^{ik\phi} + m_4}\,,\quad m_1 m_4-m_2m_3 \neq 0\,,
    \label{eq:GL_to_diffeo}
\end{equation}
which orrectly maps products of matrices to compositions of diffeomorphisms. Because of invariance under rescaling of the whole $ GL(2,\mathbb{R})$ matrix $m_i\rightarrow \lambda m_i$, $\forall i =1,2,3,4$ and $\lambda\in \mathbb{C}$, the definition of $f_k$ is independent of the phase. By equation \eqref{eq:fMSLk}, we see that $V_{f_{M,k}}^{2\times 2} \mapsto f_{M,k} $ under the mapping \eqref{eq:GL_to_diffeo}. Therefore the product of two unitaries \eqref{eq:Virasoro_unitary_displacement_app} amounts to the composition of the corresponding diffeomorphisms as required.

\section{Correlation functions}\label{app:primaryproperties}

Let us first review basics of operator products in two dimensions. Let $w_1,w_2\in \mathbb{C}$ be two points on the complex plane and let $A(w), B(w)$ be two local operator. We define the product ordered in the imaginary direction as
\begin{equation}
    \overleftarrow{\mathcal{I}}\{A(w_1)\,B(w_2)\} =A(w_1)\,B(w_2)\;\Theta(\text{Im}\,w_1 - \text{Im}\,w_2)+B(w_2)\,A(w_1)\; \Theta(\text{Im}\,w_2 - \text{Im}\,w_1)\,,
    \label{eq:radialordering}
\end{equation}
where the Heaviside step function $\Theta(x) = 1$ when $x >0$ and zero otherwise. Here $\overleftarrow{\mathcal{I}}$ orders operators with larger imaginary parts to the left of operators with smaller imaginary parts. We assume that it is a meromorphic operator valued function on $\mathbb{C}^2$ and define the product of two local operators on the real line as
\begin{equation}
    A(\phi_1)\,B(\phi_2) \equiv \lim_{\varepsilon \rightarrow 0^+}\overleftarrow{\mathcal{I}}\{A(\phi_1+i\varepsilon)\,B(\phi_2)\}\,,\quad B(\phi_2)\,A(\phi_1)\equiv \lim_{\varepsilon \rightarrow 0^+}\overleftarrow{\mathcal{I}}\{A(\phi_1-i\varepsilon)\,B(\phi_2)\}\,.
    \label{eq:operatorproduct}
\end{equation}
Notice that these are not well defined function when $\phi_1 = \phi_2$ and should be understood as an operator valued hyperfunction on the real line obtained as the $\varepsilon\rightarrow 0^+$ limit of the meromorphic operator valued function \eqref{eq:radialordering}. Using these we can define the real and imaginary parts as (valid inside expectation values so that the complex conjugation does not act on the components of the operators themselves)
\begin{align}
    \frac{1}{2i}\,[A(\phi_1),B(\phi_2)]\equiv \lim_{\varepsilon \rightarrow 0^+}\text{Im}\,\overleftarrow{\mathcal{I}}\{A(\phi_1+i\varepsilon)\,B(\phi_2)\}\label{eq:commutatorlimit}\\
    \frac{1}{2}\,[A(\phi_1),B(\phi_2)]_+\equiv \lim_{\varepsilon \rightarrow 0^+}\text{Re}\,\overleftarrow{\mathcal{I}}\{A(\phi_1+i\varepsilon)\,B(\phi_2)\}\label{eq:anticommutatorlimit}
\end{align}
When $\phi_1\neq \phi_2$ they reduce to the ordinary commutator $[A(\phi_1),B(\phi_2)] = A(\phi_1)\,B(\phi_2) - B(\phi_2)\,A(\phi_1)$ and the anti-commutator $[A(\phi_1),B(\phi_2)]_+ =A(\phi_1)\,B(\phi_2) + B(\phi_2)\,A(\phi_1) $ respectively.

\subsection{Stress tensor operator}\label{app:stresstensormodecorrelators}

The adjoint action of the Virasoro group on the stress tensor is given by
\begin{equation}
	V_{f}\,T(\phi)\,V_{f}^\dagger = f'(\phi)^{2}\,T(f(\phi)) - \frac{c}{24\pi}\,\{f(\phi),\phi\}\,,
	\label{eq:Ttransformation}
\end{equation}
which implies the commutation relation
\begin{equation}
	[T(\phi_1),T(\phi_2)] = -i\,\bigl(T(\phi_1)+T(\phi_2)\bigr)\,\delta'_{2\pi}(\phi_1-\phi_2)+\frac{ic}{24\pi}\,\delta'''_{2\pi}(\phi_1-\phi_2)\,.
 \label{eq:TTbarcommutatorapp}
\end{equation}
We will now show that this commutator follows from the stress tensor OPE
\begin{equation}
    \overleftarrow{\mathcal{I}}\{T(w_1)\,T(w_2)\} = \frac{1}{(2\pi)^{2}}\biggl[\frac{c\slash 2}{16\sin^{4}(\frac{w_1-w_2}{2})}-\frac{2\pi\,T(w_1)+c\slash 24}{2\sin^{2}(\frac{w_1-w_2}{2})}+\frac{2\pi\,\partial_{w_1}T(w_1)}{2\tan(\frac{w_1-w_2}{2})}\biggr] + \ldots\,,
\label{eq:TTOPE}
\end{equation}
using the formula \eqref{eq:commutatorlimit}. In this equation ellipsis denote a series of local operators at $w_1$ with coefficients that are finite in the $w_2\rightarrow w_1$ limit. Using the Mittag-Leffler theorem we can write
\begin{equation}
    \frac{1}{2\tan{(\frac{\phi+i\varepsilon}{2}})} = \frac{1}{\phi+i\varepsilon} + \sum_{n\neq 0}\biggl(\frac{1}{\phi+i\varepsilon + 2\pi n} + \frac{1}{2\pi n}\biggr)\,.
\end{equation}
By using the identity
\begin{equation}
    \lim_{\varepsilon\to 0^+}\frac{1}{\phi+i\varepsilon} = -i\pi\,\delta(\phi) + \mathcal{P}\,\frac{1}{\phi}\,,
\end{equation}
we get
\begin{equation}
    \lim_{\varepsilon\to 0^+}\frac{1}{2\tan{(\frac{\phi+i\varepsilon}{2}})} = -i\pi\,\delta_{2\pi}(\phi) + \mathcal{P}\left(\frac{1}{2\tan{(\frac{\phi}{2}})}\right)\,.
    \label{eq:limitinversetan}
\end{equation}
where the $2\pi$-periodic delta function $\delta_{2\pi}(\phi) = \frac{1}{2\pi}\sum_{n=-\infty}^\infty e^{in\phi}$. Here the Cauchy principal value $\mathcal{P}$ of a function $\frac{1}{\mathcal{F}(\phi)}$ with the Taylor expansion $\mathcal{F}(\phi) = \phi + \mathcal{O}(\phi^2)$ at $\phi = 0$ is defined by
\begin{equation}    \int_0^{2\pi}\,d\phi\,\varphi(\phi)\,\mathcal{P}\left(\frac{1}{\mathcal{F}(\phi)}\right) = \lim_{\epsilon\rightarrow 0^+}\int_{\epsilon}^{2\pi-\epsilon}\,d\phi\,\frac{\varphi(\phi)}{\mathcal{F}(\phi)}\,,
\end{equation}
where $\varphi(\phi)$ is a test function. The formula \eqref{eq:limitinversetan} allows to compute the $\varepsilon\rightarrow 0^+$ limit of the third term in the OPE. The limits of the other terms in the OPE can be obtained by using the identities
\begin{equation}
    \frac{1}{4\sin^{2}(\frac{\phi+i\varepsilon}{2})} = -\partial_\phi\,\frac{1}{2\tan{(\frac{\phi+i\varepsilon}{2}})}\,,\quad 
          \frac{1}{16\sin^{4}(\frac{\phi+i\varepsilon}{2})} = -\frac{1}{6}\,(\partial_{\phi}^3 + \partial_\phi)\,\frac{1}{2\tan{(\frac{\phi+i\varepsilon}{2}})}
\end{equation}
which yield the result
\begin{align}
    \lim_{\varepsilon\rightarrow 0^+}\frac{1}{4\sin^{2}(\frac{\phi+i\varepsilon}{2})} &= i\pi\,\delta'_{2\pi}(\phi)+\mathcal{H}\left(\frac{1}{4\sin^{2}(\frac{\phi}{2})}\right)\label{eq:Hidentity1}\\
    \lim_{\varepsilon\rightarrow 0^+}\frac{1}{16\sin^{4}(\frac{\phi+i\varepsilon}{2})}&=\frac{i\pi}{6}\,(\delta_{2\pi}'''(\phi)+\delta_{2\pi}'(\phi))+\mathcal{H}\left(\frac{1}{16\sin^{4}(\frac{\phi}{2})}\right)\label{eq:Hidentity2}
\end{align}
where we have defined the Hadamard finite part
\begin{equation}
    \mathcal{H}\left(\partial_\phi^k\,\frac{1}{\mathcal{F}(\phi)}\right) \equiv (-1)^k\,\partial_\phi^k\,\mathcal{P}\,\left(\frac{1}{\mathcal{F}(\phi)}\right)\,,
\end{equation}
whose integral against a test function $\varphi(\phi)$ is given by
\begin{equation}    \int_0^{2\pi}\,d\phi\,\varphi(\phi)\,\mathcal{H}\left(\partial_\phi^k\,\frac{1}{\mathcal{F}(\phi)}\right) = \lim_{\epsilon\rightarrow 0^+}\int_{\epsilon}^{2\pi-\epsilon}\,d\phi\,\frac{\partial_\phi^k\varphi(\phi)}{\mathcal{F}(\phi)}\,.
\end{equation}
This follows from the definition of a the Cauchy principal value and the derivative of a distribution: the derivative of a distribution integrated against a test function equals (up to minus signs) the distribution integrated against derivatives of the test function.\footnote{This rule for distributions can be derived using residue theorem and contour integration from the OPE.} Due to the Taylor expansion $\mathcal{F}(\phi) = \phi + \mathcal{O}(\phi^2)$, integrals of the Hadamard finite part distribution are finite. 

Using these formulae, we can compute the commutator which is encoded in the imaginary part of the OPE via \eqref{eq:commutatorlimit}. The result is
\begin{equation}
[T(\phi_1),T(\phi_2)] = -2i\,T(\phi_1)\,\delta'_{2\pi}(\phi_1-\phi_2)-i\,T'(\phi_1)\,\delta_{2\pi}(\phi_1-\phi_2)+\frac{ic}{24\pi}\,\delta'''_{2\pi}(\phi_1-\phi_2)\,,
\label{eq:firstcommutator}
\end{equation}
where all contributions from the terms denoted by ellipsis in the OPE vanish. Relabeling $ \phi_1 \leftrightarrow \phi_2 $, subtracting from \eqref{eq:firstcommutator} and dividing by two gives \eqref{eq:TTbarcommutatorapp}.
\begin{figure}
    \centering
    \includegraphics[scale=1]{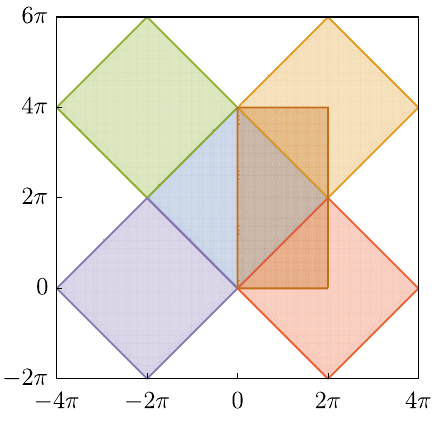}
    \caption{The $(\phi_-,\phi_+)$-plane where the horizontal axis parametrizes $\phi_- = \phi_1-\phi_2$ while the vertical axis $\phi_+ = \phi_1+\phi_2$. The blue diamond in the center is the integration region $(\phi_1,\phi_2)\in [0,2\pi]^2$ of the double integral \eqref{eq:doubleintidentity} in terms of $(\phi_-,\phi_+)$. Diamonds shaded with other colors correspond to displacements of the center diamond by $\pm 2\pi$ in either $\phi_1$ or $\phi_2$ direction. The dark brown shaded rectangle is the region $(\phi_-,\phi_+)\in [0,2\pi]\times [0,4\pi]$. The identity \eqref{eq:doubleintidentity} states that the double integral of a function $\mathcal{F}(\phi_1,\phi_2)  = \mathcal{F}(\phi_1+2\pi,\phi_2)=\mathcal{F}(\phi_1,\phi_2+2\pi) $ over the blue shaded diamond is equal to the double integral over the dark brown shaded rectangle.}
    \label{fig:intregion}
\end{figure}

\paragraph{2-point functions.} When the OPE \eqref{eq:TTOPE} is evaluated inside an expectation value, it simplifies to \cite{Datta:2019jeo}
\begin{equation}
    \bra{h}\overleftarrow{\mathcal{I}}\{T(w_1)\,T(w_2)\}\ket{h} = \lim_{\varepsilon\rightarrow 0^+}\frac{1}{(2\pi)^{2}}\biggl[\frac{c\slash 2}{16\sin^{4}(\frac{w_1-w_2}{2})}-\frac{h}{2\sin^{2}(\frac{w_1-w_2}{2})}+\left(h-\frac{c}{24}\right)^{2}\biggr]
    \label{eq:OPEvev}
\end{equation}
from which we obtain
\begin{equation}
    \bra{h}T(\phi_1)\,T(\phi_2)\ket{h} = \lim_{\varepsilon\rightarrow 0^+}\frac{1}{(2\pi)^{2}}\biggl[\frac{c\slash 2}{16\sin^{4}(\frac{\phi_1-\phi_2+i\varepsilon}{2})}-\frac{h}{2\sin^{2}(\frac{\phi_1-\phi_2+i\varepsilon}{2})}+\left(h-\frac{c}{24}\right)^{2}\biggr]\,.
    \label{eq:stress2point}
\end{equation}
Due to the Fourier expansion \eqref{eq:stresstensorexp}, we can use this to calculate the 2-point functions of Virasoro generators via
\begin{equation}
	\bra{h}L_{n}\,L_{m}\ket{h} = \int_{0}^{2\pi}d\phi_1 \int_{0}^{2\pi}d\phi_2\,e^{-in\phi_1-im\phi_2}\,\bra{h}T(\phi_1)\,T(\phi_2)\ket{h} \,.
 \label{eq:virasoromode2pointapp1}
\end{equation}
Substituting \eqref{eq:stress2point} gives explicitly
\begin{align}
	\bra{h}L_{n}\,L_{m}\ket{h} &=\frac{1}{(2\pi)^{2}}\int_{0}^{2\pi}d\phi_1 \int_{0}^{2\pi}d\phi_2\,e^{-in\phi_1-im\phi_2}\,\biggl[\frac{c}{32\sin^{4}(\frac{\phi_1-\phi_2+i\varepsilon}{2})}-\frac{h}{2\sin^{2}(\frac{\phi_1-\phi_2+i\varepsilon}{2})}\biggr]\nonumber\\
	&+\left(h-\frac{c}{24}\right)^{2}\delta_{n,0}\,\delta_{m,0}
\label{eq:stresstensormidstep}
\end{align}
For any function $\mathcal{F}(\phi_1,\phi_2)$ which is $2\pi$-periodic in both arguments $\mathcal{F}(\phi_1+2\pi,\phi_2)=\mathcal{F}(\phi_1,\phi_2+2\pi) = \mathcal{F}(\phi_1,\phi_2) $, we have the identity
\begin{equation}
    \int_{0}^{2\pi}d\phi_1 \int_{0}^{2\pi}d\phi_2\,\mathcal{F}(\phi_1,\phi_2) = \frac{1}{2} \int_{0}^{2\pi}d\phi_{-}\int_{0}^{4\pi}d\phi_{+}\,\mathcal{F}(\tfrac{1}{2}(\phi_++\phi_-),\tfrac{1}{2}(\phi_+-\phi_-))\,,
    \label{eq:doubleintidentity}
\end{equation}
where we have defined $\phi_\pm = \phi_1\pm \phi_2$ and used $d\phi_1d\phi_2 = \frac{1}{2}\,d\phi_{+}d\phi_{-}$. This identity can be explained using Figure \ref{fig:intregion} where the integration region on the left-hand side of \eqref{eq:doubleintidentity} is the blue shaded diamond in the center while the integration region on the right is the dark brown shaded rectangle. Let us denote the overlap of the blue diamond and the brown rectangle by $O$. Denote the complement of $O$ within the blue diamond by $O^{\text{c}}_\text{blue}$ and the complement within the brown rectangle by $O^{\text{c}}_\text{brown}$. The integral over $O$ accounts for half of the identity \eqref{eq:doubleintidentity}. Due to $2\pi$-periodicity of $\mathcal{F}(\phi_1,\phi_2)$, the integral over $O^{\text{c}}_\text{blue}$ and $O^{\text{c}}_\text{brown}$ are equal which accounts for the remaining half.

Now given any periodic function $\mathcal{F}(\phi+2\pi) = \mathcal{F}(\phi)$, \eqref{eq:doubleintidentity} implies
\begin{align}
&\int_{0}^{2\pi}d\phi_1 \int_{0}^{2\pi}d\phi_2\,e^{-in\phi_1-im\phi_2}\,\mathcal{F}(\phi_1-\phi_2)\nonumber\\
&= \frac{1}{2} \int_{0}^{4\pi}d\phi_{+}e^{-i(m+n)\phi_{+}\slash 2}\int_{0}^{2\pi}d\phi_{-}\,e^{i(m-n)\,\frac{\phi_{-}}{2}}\,\mathcal{F}(\phi_-)\,,
\end{align}
Then by using 
\begin{equation}
    \int_{0}^{4\pi}d\phi_{+}e^{-i(m+n)\,\phi^{+}\slash 2} = 4\pi\,\delta_{n,-m}\,,
\end{equation}
we get
\begin{equation}
    \int_{0}^{2\pi}d\phi_1 \int_{0}^{2\pi}d\phi_2\,e^{-in\phi_1-im\phi_2}\,\mathcal{F}(\phi_1-\phi_2) = 2\pi\,\delta_{n,-m}\int_{0}^{2\pi}d\phi\,e^{-in\phi}\,\mathcal{F}(\phi)\,.
\label{eq:curlyFintegral}
\end{equation}
Applying this identity in \eqref{eq:stresstensormidstep} and defining the complex variable $z = e^{i\phi}$, we get
\begin{equation}
	\bra{h}L_{n}\,L_{m}\ket{h} = \frac{1}{2\pi}\,\delta_{n,-m}\oint_{S^{1}}dz\,\biggl[\frac{c}{2iz^{n-1}(z-e^{\varepsilon})^{4}}+\frac{2h}{iz^{n}(z-e^{\varepsilon})^{2}}\biggr]+ \left(h-\frac{c}{24}\right)^{2}\delta_{n,0}\,\delta_{m,0}
\end{equation}
with the integration running counter-clockwise. The residues of the poles are given by
\begin{align}
	&2\pi i\,\text{Res}_{z = z_0}\,\biggl[\frac{c}{2iz^{n-1}(z-e^{\varepsilon})^{4}}+\frac{2h}{iz^{n}(z-e^{\varepsilon})^{2}}\biggr]\\
 &= \frac{\pi c}{6}\,n\,\biggl(\frac{24h}{c}+n^{2}-1\biggr)\times
	\begin{dcases}
		-1,\quad &z_0 = e^{\varepsilon}\\
		\Theta(n),\quad &z_0 = 0\\
		-\Theta(-n),\quad &z_0 = \infty
	\end{dcases}\,,
\end{align}
where $\Theta(n)$ is the Heaviside step function.  
Since $\varepsilon>0$ only the residue of the pole at $z = 0$ is picked up which gives the result
\begin{align}
	\bra{h}L_{n}\,L_{m}\ket{h} =  \frac{c}{12}\,n\,\biggl(\frac{24h}{c}+n^{2}-1\biggr)\,\delta_{n,-m}\,\Theta(n)+ \left(h-\frac{c}{24}\right)^{2}\delta_{n,0}\,\delta_{m,0}\,.
 \label{eq:Virasoromode2pointapp}
\end{align}
This result is consistent with our convention that highest weight states satisfy $L_{n>0}\ket{h} = 0$.
From \eqref{eq:Virasoromode2pointapp} it follows that the commutator becomes
\begin{equation}
	\bra{h}[L_{n},L_{m}]\ket{h} = \frac{c}{12}\,n\,\biggl(1-n^{2}-\frac{24h}{c}\biggr)\,\delta_{n,-m}\,.
 \label{eq:commutatorfromcontoursapp}
\end{equation}
We can compare this with the expectation value of the commutator \eqref{eq:stresstensorexp} given by
\begin{equation}
    \bra{h}[L_{n},L_{m}]\ket{h} = (n-m)\,\bra{h}L_{n+m}\ket{h}+\frac{c}{12}n^{3}\,\delta_{n,-m}\,.
\end{equation}
By substituting $\bra{h}L_{n+m}\ket{h} = \left(h-\frac{c}{24}\right)\,\delta_{n,-m}$ this matches with \eqref{eq:commutatorfromcontoursapp}. For the anti-commutator we get
\begin{equation}
    \bra{h}L_{n}\,L_{m} + L_m\,L_n\ket{h} = \frac{c}{12}\,\vert n \vert\,\biggl(\frac{24h}{c}+n^{2}-1\biggr)\,\delta_{n,-m}+ 2\left(h-\frac{c}{24}\right)^{2}\delta_{n,0}\,\delta_{m,0}\,,
\label{eq:LnLmanticommutator}
\end{equation}
where we have used $n\,(\Theta(n) - \Theta(-n)) = n\,\text{sgn}\,{(n)} = \vert n \vert$.

\paragraph{Alternative calculation.} Above we calculated $\bra{h}L_{n}\,L_{m}\ket{h}$ using the standard $i\varepsilon$ prescription which gives an unambiguous finite answer \eqref{eq:LnLmanticommutator} for the anti-commutator $\bra{h}[L_n,L_m]_+\ket{h}$ defined by $[L_n,L_m]_+ \equiv L_n\,L_m + L_m\,L_n$. However, we can also obtain the same answer by using differential regularization\slash renormalization as follows.

Two times the real part of \eqref{eq:stress2point} gives via the formula \eqref{eq:anticommutatorlimit} and the identities \eqref{eq:Hidentity1} - \eqref{eq:Hidentity2} the anti-commutator
\begin{equation}
    \bra{h}[T(\phi_1)\,T(\phi_2)]_+\ket{h} = \frac{2}{(2\pi)^2}\biggl[\,\mathcal{H}\,\biggl(\frac{c}{32\sin^{4}(\frac{\phi_1-\phi_2}{2})}-\frac{h}{2\sin^{2}(\frac{\phi_1-\phi_2}{2})}\biggr)+\left(h-\frac{c}{24}\right)^{2}\delta_{n,0}\,\delta_{m,0}\biggr]\,,
    \label{eq:TTanticommutatorapp}
\end{equation}
where the Hadamard finite part distribution is given by
\begin{equation}
    \mathcal{H}\,\biggl(\frac{c}{32\sin^{4}(\frac{\phi_1-\phi_2}{2})}-\frac{h}{2\sin^{2}(\frac{\phi_1-\phi_2}{2})}\biggr) = \frac{c}{12}\,\partial_{\phi}\,\biggl(-\partial_{\phi}^2-1+\frac{24h}{c}\biggr)\,\mathcal{P}\left(\frac{1}{2\tan{(\frac{\phi}{2}})}\right)\bigg\vert_{\phi = \phi_1-\phi_2}\,.
\end{equation}
We have
\begin{equation}
    \bra{h}[L_n,L_m]_+\ket{h} = \int_{0}^{2\pi}d\phi_1 \int_{0}^{2\pi}d\phi_2\,e^{-in\phi_1-im\phi_2}\,\bra{h}[T(\phi_1),T(\phi_2)]_+\ket{h} 
 \label{eq:virasoromode2pointapp2}
\end{equation}
so that by the identity \eqref{eq:curlyFintegral} we get
\begin{align}
    &\bra{h}[L_n,L_m]_+\ket{h}\nonumber\\
    &= \frac{2}{2\pi}\,\delta_{n,-m}\int_{0}^{2\pi}d\phi\,e^{-in\phi}\,\frac{c}{12}\,\partial_{\phi}\,\biggl(-\partial_{\phi}^2-1+\frac{24h}{c}\biggr)\,\mathcal{P}\left(\frac{1}{2\tan{(\frac{\phi}{2}})}\right)+2\left(h-\frac{c}{24}\right)^{2}\delta_{n,0}\,\delta_{m,0}\,,
    \label{eq:cauchyintegral1}
\end{align}
Since the derivative of a distribution integrated against a test function equals (up to minus signs) the distribution integrated against derivatives of the test function, we get 
\begin{align}
    &\bra{h}[L_n,L_m]_+\ket{h}\nonumber\\
    &= \frac{2}{2\pi}\,\delta_{n,-m}\lim_{\epsilon\rightarrow 0^+}\int_{\epsilon}^{2\pi-\epsilon}d\phi\,\frac{1}{2\tan{(\frac{\phi}{2}})}\,\frac{c}{12}\,(-\partial_{\phi})\,\biggl(-\partial_{\phi}^2-1+\frac{24h}{c}\biggr)\,e^{-in\phi}\nonumber\\
    &+2\left(h-\frac{c}{24}\right)^{2}\delta_{n,0}\,\delta_{m,0}\,,
    \label{eq:cauchyintegral2}
\end{align}
where $\epsilon$ is the regulator defining the Cauchy principal value which should not be confused with $i\varepsilon$. Calculating the derivatives gives
\begin{align}
    &\bra{h}[L_n,L_m]_+\ket{h}\nonumber\\
    &= \frac{2}{2\pi}\frac{c}{12}\,in\,\biggl(\frac{24h}{c}+n^2-1\biggr)\,\delta_{n,-m}\lim_{\epsilon\rightarrow 0^+}\int_{\epsilon}^{2\pi-\epsilon}d\phi\,\frac{e^{-in\phi}}{2\tan{(\frac{\phi}{2}})}+2\left(h-\frac{c}{24}\right)^{2}\delta_{n,0}\,\delta_{m,0}\,.
\end{align}
The remaining integral is given by
\begin{equation}
    \lim_{\epsilon\rightarrow 0^+}\int_{\epsilon}^{2\pi-\epsilon}d\phi\,\frac{e^{-in\phi}}{2\tan{(\frac{\phi}{2}})} =
\begin{dcases}
-i\pi\,\text{sgn}\,(n)\,,\quad &n\in \mathbb{Z}\,\backslash \{0\}\\
0\,,\quad &n = 0
\end{dcases}
\end{equation}
Thus we get
\begin{equation}
    \bra{h}[L_n,L_m]_+\ket{h} = \frac{c}{12}\,\vert n\vert\,\biggl(\frac{24h}{c}+n^2-1\biggr)\,\delta_{n,-m}
	+2\left(h-\frac{c}{24}\right)^{2}\delta_{n,0}\,\delta_{m,0}\,,
\end{equation}
which matches with \eqref{eq:LnLmanticommutator} obtained using the $i\varepsilon$ prescription.

This calculation using the Cauchy principal value is equivalent to a calculation using differential regularization. The reason is that the Cauchy principal value \eqref{eq:cauchyintegral1} is defined as the integrated by parts expression \eqref{eq:cauchyintegral2} without total derivative terms which integrate to divergent boundary terms at $\phi = \epsilon$ and $\phi = 2\pi - \epsilon$. These are exactly the same boundary terms that are subtracted by hand in differential regularization: in that case, one starts with \eqref{eq:cauchyintegral1} without the Cauchy principal value.

\subsection{Primary operators}

The properties of a primary operator $\mathcal{O}_h(\phi)$ of weight $h$ are completely determined by its transformation property under the adjoint action of $\mathbf{Vir}$ given by
\begin{equation}
V_{f}\,\mathcal{O}_h(\phi)\,V_{f}^{\dagger} = f'(\phi)^{h}\,\mathcal{O}_h(f(\phi))\,,
\label{primarytransformation}
\end{equation}
which implies the commutator
\begin{equation}
[T(\phi_1),\mathcal{O}_h(\phi_2)] = -i\,(h-1)\,\mathcal{O}_h'(\phi_1)\,\delta_{2\pi}(\phi_1-\phi_2)-i\,h\,\mathcal{O}_h(\phi_1)\,\delta'_{2\pi}(\phi_1-\phi_2)\,.
\end{equation}
This commutator follows from the OPE
\begin{equation}
     \overleftarrow{\mathcal{I}}\{T(w_1)\,\mathcal{O}_{h}(w_2)\} =-\frac{1}{2\pi}\biggl[\frac{h\,\mathcal{O}_h(w_1)}{4\sin^{2}(\frac{w_1-w_2}{2})}-\frac{(h-1)\,\partial_{w_1} \mathcal{O}_h(w_1)}{2\tan(\frac{w_1-w_2}{2})}\biggr] + (\text{finite})\,,
\end{equation}
via the formula \eqref{eq:commutatorlimit} and the identities \eqref{eq:limitinversetan} and \eqref{eq:Hidentity1}.

\paragraph{Primary-primary commutator.} The transformation property \eqref{primarytransformation} of the primary operator fixes 2-point functions of primaries in the vacuum state $\ket{0}$ which are invariant under the $\widetilde{SL}(2,\mathbb{R})$ subgroup of $\widetilde{\text{Diff}}_+S^1$ up to phase. More precisely, for $f_M \in \widetilde{SL}(2,\mathbb{R})$ of the form \eqref{eq:Sl2Rdiffeo}, we have $V_{f_M}\ket{0} = e^{-i(\theta-\chi)\,\frac{c}{24}}\ket{0}$ which follows from equation \eqref{eq:mobiusstatecoherentstate} with $k = 1$. Therefore we obtain the Ward identity
\begin{equation}
    \bra{0}\mathcal{O}_{h}(\phi_1)\,\mathcal{O}_{h}(\phi_2)\ket{0} = f'_M(\phi_1)^{h}\,f'_M(\phi_2)^{h}\bra{0}\mathcal{O}_{h}(f_M(\phi_1))\,\mathcal{O}_{h}(f_M(\phi_2))\ket{0}
\end{equation}
where we inserted $ 1 = V_{f_M}^{\dagger}V_{f_M}$ into the left-hand side and used \eqref{primarytransformation}. Note that the phase $e^{-i(\theta-\chi)\,\frac{c}{24}}$ is cancelled. One can check that the Ward identity is satisfied by the 2-point function
\begin{equation}
    \bra{0}\mathcal{O}_{h}(\phi_1)\,\mathcal{O}_{h}(\phi_2)\ket{0} =  \lim_{\varepsilon\rightarrow 0^+}\frac{1}{(2\pi)^2}\frac{1}{\bigl[2i\sin{\bigl(\frac{\phi_1-\phi_{2}+i\varepsilon}{2}\bigr)}\bigr]^{2h}}
    \label{eq:primary2pointfunctionapp}
\end{equation}
by inserting \eqref{eq:Sl2Rdiffeo}. This 2-point function follows from the OPE
\begin{equation}
    \overleftarrow{\mathcal{I}}\{\mathcal{O}_{h}(w_1)\,\mathcal{O}_{h}(w_2)\} = \frac{1}{(2\pi)^2}\frac{1}{\bigl[2i\sin{\bigl(\frac{w_1-w_2}{2}\bigr)}\bigr]^{2h}} + (\text{divergent}) + (\text{finite})\,,
\end{equation}
where ``$(\text{divergent}) + (\text{finite})$'' denotes a series of operators whose expectation values in the vacuum state are zero, ``divergent'' includes those operators whose coefficients diverge in the $w_2\rightarrow w_1$ limit (but are subleading to the leading diverge) while ``finite'' denotes terms that are finite as $w_2\rightarrow w_1$. 

When $h = 1$, the most general OPE compatible with the $SL(2,\mathbb{R})$ transformation properties, locality of the commutator and involving only the operator itself in the divergent terms takes the form
\begin{equation}
    \overleftarrow{\mathcal{I}}\{\mathcal{O}_{1}(w_1)\,\mathcal{O}_{1}(w_2)\} = \frac{1}{(2\pi)^2}\biggl[-\frac{1}{4\sin^2{\bigl(\frac{w_1-w_2}{2}\bigr)}} +\frac{2\pi\,b\,\mathcal{O}_1(w_1)}{2\tan{\bigl(\frac{w_1-w_2}{2}\bigr)}}  +(\text{finite})\biggr]\,,
\end{equation}
where $b$ is a constant. The result is
\begin{equation}
    [\mathcal{O}_{1}(\phi_1),\mathcal{O}_{1}(\phi_2)] = -i\,b\,\mathcal{O}(\phi_1)\,\delta_{2\pi}(\phi_1-\phi_2)-\frac{i}{2\pi}\,\delta'_{2\pi}(\phi_1-\phi_2)\,.
    \label{eq:scalarcommutatorapp}
\end{equation}
This is a Kac--Moody algebra and the parameter $1\slash b$ is the level of the algebra (it can be moved to the central term by a rescaling of the primary operator).

\paragraph{Primary mode operators.} Using the Fourier expansion of $\mathcal{O}_h(\phi)$, we obtain
\begin{equation}
\bra{0}\mathcal{O}_{h,n}\,\mathcal{O}_{h,m}\ket{0} = \int_{0}^{2\pi}d\phi_1 \int_{0}^{2\pi}d\phi_2\,e^{-in\phi_1-im\phi_2}\,\bra{0}\mathcal{O}_h(\phi_1)\,\mathcal{O}_h(\phi_2)\ket{0} \,.
\end{equation}
Substituting \eqref{eq:primary2pointfunctionapp} gives
\begin{equation}
    \bra{0}\mathcal{O}_{h,n}\,\mathcal{O}_{h,m}\ket{0} = \frac{1}{(2\pi)^2}\int_{0}^{2\pi}d\phi_1 \int_{0}^{2\pi}d\phi_2\,\frac{e^{-in\phi_1-im\phi_2}}{\bigl[2i\sin{\bigl(\frac{\phi_1-\phi_{2}+i\varepsilon}{2}\bigr)}\bigr]^{2h}}\,.
\end{equation}
and by using \eqref{eq:curlyFintegral} it can be written as
\begin{equation}
    \bra{0}\mathcal{O}_{h,n}\,\mathcal{O}_{h,m}\ket{0} =\frac{1}{2\pi}\,\delta_{n,-m}\int_{0}^{2\pi}d\phi\,\frac{e^{-in \phi}}{\bigl[2i\sin{\bigl(\frac{\phi+i\varepsilon}{2}\bigr)}\bigr]^{2h}}\,.
\end{equation}
Defining the complex variable $z = e^{i\phi}$, we get
\begin{equation}
    \bra{0}\mathcal{O}_{h,n}\,\mathcal{O}_{h,m}\ket{0} = \frac{1}{2\pi}\,\delta_{n,-m}\oint_{S^{1}}dz\,\frac{1}{iz^{n+1-h}\,(z-e^{\varepsilon})^{2h}}
\end{equation}
with the integration running counter-clockwise. Since $\varepsilon>0$, only the residue of the $z = 0$ pole is picked up and it is present only for $n\geq h$. The residues of this pole are given by
\begin{equation}
	\lim_{\varepsilon\to 0^+}2\pi i\,\text{Res}_{z = 0}\,\biggl(\frac{1}{iz^{n+1-h}\,(z-e^{\varepsilon})^{2h}}\biggr)
 =\frac{2\pi}{(n-h)!}\frac{(n+h-1)!}{(2h-1)!}\,.
\end{equation}
Hence, the result for the correlator is
\begin{align}	\label{eq:OnOmvevGeneralapp}\bra{0}\mathcal{O}_{h,n}\,\mathcal{O}_{h,m}\ket{0}  = \binom{n+h-1}{2h-1}\,\delta_{n,-m}\,\Theta(n-h)\,.
\end{align}

From this we can obtain the commutator
\begin{equation}
    \bra{0}[\mathcal{O}_{h,n},\mathcal{O}_{h,m}]\ket{0} = \binom{n+h-1}{2h-1}\,\delta_{n,-m}\,.
\end{equation}
As a consistency check, for $h = 1$ we see that
\begin{equation}
    \bra{0}[\mathcal{O}_{1,n},\mathcal{O}_{1,m}]\ket{0} = n\,\delta_{n,-m}\,,
\end{equation}
which is equivalent to the vacuum expectation value of \eqref{eq:scalarcommutatorapp} as expected.

\subsection{Adjoint action of the Virasoro group}

Let us derive the commutation relations responsible for the transformation law of the primary and the stress tensor operator under the adjoint action of the Virasoro group.

\paragraph{Primary operator.} We have
\begin{equation}
V_{f_t}^\dagger\,\mathcal{O}_h(\phi)\,V_{f_t} = V_{F_t}\,\mathcal{O}_h(\phi)\,V_{F_t}^{\dagger} = F'_t(\phi)^{h}\,\mathcal{O}_h(F_t(\phi)).
\label{primarytransformationapp}
\end{equation}
where the Virasoro unitary is explicitly
\begin{equation}
V_{F_t} = \overleftarrow{\mathcal{T}}\exp{\biggl[-i\int_{0}^{t} ds\int_{0}^{2\pi} d\phi\,(\dot{F}_s\circ F^{-1}_s)(\phi)\,T(\phi)\biggr]}\,.
\end{equation}
Taking the derivative with respect to $ t $ of this equation gives 
\begin{align}
&-i\int_{0}^{2\pi} d\phi_1\,(\dot{F}_t\circ F_t^{-1})(\phi_1)\,[T(\phi_1),\mathcal{O}_h(\phi)]\\
&= h\,\dot{F}_t'(\phi)\,F'_t(\phi)^{h-1}\,\mathcal{O}_h(F_t(\phi))+F'_t(\phi)^{h}\,\dot{F}_t(\phi)\,\mathcal{O}_h'(F_t(\phi))\,.
\end{align}
At $ t = 0 $ we have $F_0(\phi) = \phi$ and we define $\dot{F}_0(\phi) \equiv u(\phi)$. Setting $ t = 0 $ gives
\begin{equation}
-i\int_{0}^{2\pi} d\phi_1\,u(\phi_1)\,[T(\phi_1),\mathcal{O}_h(\phi)] = h\,u'(\phi)\,\mathcal{O}_h(\phi)+u(\phi)\,\mathcal{O}_h'(\phi)
\label{eq:midstepequationcommutator}
\end{equation}
The right-hand side can be written as
\begin{align}
&\int_{0}^{2\pi} d\phi_1\,\delta_{2\pi}(\phi_1-\phi)\,[h\,u'(\phi_1)\,\mathcal{O}_h(\phi_1)+u(\phi_1)\,\mathcal{O}_h'(\phi_1)]\\
&= \int_{0}^{2\pi} d\phi_1\,u(\phi_1)\,[-(h-1)\,\mathcal{O}_h'(\phi_1)\,\delta_{2\pi}(\phi_1-\phi)-h\,\mathcal{O}_h(\phi_1)\,\delta'_{2\pi}(\phi_1-\phi)]
\end{align}
where we have inserted a delta function and integrated by parts. Because \eqref{eq:midstepequationcommutator} is valid for all test functions $u(\phi)$, we obtain the local equation
\begin{equation}
[T(\phi_1),\mathcal{O}_h(\phi_2)] = -i\,(h-1)\,\mathcal{O}_h'(\phi_1)\,\delta_{2\pi}(\phi_1-\phi_2)-i\,h\,\mathcal{O}_h(\phi_1)\,\delta'_{2\pi}(\phi_1-\phi_2)\,.
\end{equation}

\paragraph{Stress tensor.} We have
\begin{equation}
V_{F_t}\,T(\phi)\,V_{F_t}^{\dagger} = F'_t(\phi)^{2}\,T(F_t(\phi)) - \frac{c}{24\pi}\{F_t(\phi),\phi\}\, .
\end{equation}
Taking the derivative with respect to $ t $ and setting $ t = 0 $ gives the equation
\begin{equation}
-i\int_{0}^{2\pi} d\phi_1\,u(\phi_1)\,[T(\phi_1),T(\phi)] = 2\,u'(\phi)\,T(\phi)+u(\phi)\,T'(\phi)-\frac{c}{24\pi}\,u'''(\phi)\,,
\label{eq:stresstensor_commutator_midstep}
\end{equation}
where we can write
\begin{align}
&2\,u'(\phi)\,T(\phi)+u(\phi)\,T'(\phi)-\frac{c}{24\pi}\,u'''(\phi)\\
&=\int_{0}^{2\pi} d\phi_1\,u(\phi_1)\,\biggl[-T'(\phi_1)\,\delta(\phi_1-\phi)-2\,T(\phi_1)\,\delta'(\phi_1-\phi)+\frac{c}{24\pi}\,\delta'''(\phi_1-\phi)\biggr]
\end{align}
as above. Setting $ \phi = \phi_2 $, equation \eqref{eq:stresstensor_commutator_midstep} thus implies
\begin{align}
[T(\phi_1),T(\phi_2)] &= -2i\,T(\phi_1)\,\delta'(\phi_1-\phi_2)-i\,T'(\phi_1)\,\delta(\phi_1-\phi_2)+\frac{ic}{24\pi}\,\delta'''(\phi_1-\phi_2)\,.
\label{eq:TTcommutator_app}
\end{align}
By the reversal of the arguments $ \phi_1 \leftrightarrow \phi_2 $, we obtain
\begin{equation}
[T(\phi_1),T(\phi_2)] = -2i\,T(\phi_2)\,\delta'(\phi_1-\phi_2)+i\,T'(\phi_1)\,\delta(\phi_1-\phi_2)+\frac{ic}{24\pi}\,\delta'''(\phi_1-\phi_2)\,.
\label{eq:TTcommutator_app_reversed}
\end{equation}
Adding \eqref{eq:TTcommutator_app} and \eqref{eq:TTcommutator_app_reversed} together and dividing by two gives
\begin{equation}
[T(\phi_1),T(\phi_2)] = -i\,\bigl(T(\phi_1)+T(\phi_2)\bigr)\,\delta'(\phi_1-\phi_2)+\frac{ic}{24\pi}\,\delta'''(\phi_1-\phi_2)\, .
\end{equation}

\subsection{Complex conjugate operators}\label{subapp:conjugate_rep}

We define the complex conjugate operators as $\overbar{T}(\phi) = T(-\phi)$ and $\overbar{\mathcal{O}}_h(\phi) = \mathcal{O}_h(-\phi)$. They have the mode expansions
\begin{equation}
    \overbar{T}(\phi) = \frac{1}{2\pi}\sum_{n=-\infty}^\infty L_{-n}\,e^{in\phi}\,,\quad \overbar{\mathcal{O}}_h(\phi) = \frac{1}{2\pi}\sum_{n=-\infty}^\infty \mathcal{O}_{h,-n}\,e^{in\phi}\,.
\end{equation}
Their commutation relations are the same as for $T(\phi)$ and $\mathcal{O}_h(\phi)$ (equations \eqref{eq:TTbarcommutator} and \eqref{eq:TOcommutator}), but with opposite sign for the imaginary unit $i\leftrightarrow -i$.

The complex conjugate representation $\overbar{V}_{f_t}$ of the Virasoro group is defined as in \eqref{eq:Vbar}. The generator appearing in the exponent is explicitly
\begin{equation}
    \overbar{G}(f_t) = \sum_{n=-\infty}^\infty u_{t,n}\,L_n\,,
\end{equation}
which differs from $G(f_t)$ by complex conjugation of $u_{t,n}$. The equality $\overbar{V}_{f_t} = V_{Z\circ f_t\circ Z}$ where $Z(\phi) = -\phi$ follows by an explicit computation. First, we can use $\partial_t(Z\circ f_t\circ Z) = -\dot{f}_t\circ Z $ to obtain
\begin{equation}
    \partial_t(Z\circ f_t\circ Z) \circ (Z\circ f_t\circ Z)^{-1} = -u_t\circ Z\,,%= -\dot{f}_t\circ Z\circ Z\circ f_t^{-1}\circ Z = -\dot{f}_t\circ f_t^{-1}\circ Z 
\end{equation}
where we have defined $u_t = \dot{f}_t\circ f_t^{-1}$. Therefore it follows that
\begin{equation}
    V_{Z\circ f_t\circ Z} = \overleftarrow{\mathcal{T}}\exp{\biggl(i\int_{0}^{t} ds\,\int_{0}^{2\pi}d\phi\,u_t(-\phi)\,T(\phi)\biggr)}\,.
\end{equation}
By a change of integration variable $\phi\rightarrow -\phi$, we get
\begin{equation}
    \int_{0}^{2\pi}d\phi\,u_t(-\phi)\,T(\phi) = \int_{-2\pi}^{0} d\phi\,u_t(\phi)\,\overbar{T}(\phi) =\int_{0}^{2\pi} d\phi\,u_t(\phi)\,\overbar{T}(\phi) \,,
\end{equation}
where the last equality follows by a further change of variable $\phi\rightarrow \phi + 2\pi$ and from the fact that the integrand is $2\pi$-periodic. Hence we have shown $V_{Z\circ f_t\circ Z} = \overbar{V}_{f_t}$. Using this formula, it follows that
\begin{equation}
	\overbar{V}_{f} \overbar{V}_{g} = e^{iB(Z\circ f\circ Z,Z\circ g\circ Z)}\,\overbar{V}_{f\circ g}\,.
\end{equation}
where the 2-cocycle is explicitly
\begin{equation}
    B(Z\circ f\circ Z,Z\circ g\circ Z) = \int_{0}^{2\pi}d \phi\,\frac{(Z\circ g\circ Z)''(\phi)}{(Z\circ g\circ Z)'( \phi)}\,\log{(Z\circ f\circ Z)'((Z\circ g\circ Z)(\phi))}\,.
    \label{eq:overbar_cocycle}
\end{equation}
We can evaluate this by using
\begin{equation}
    (Z\circ f\circ Z)'(Z(\phi)) = f'(\phi)\,,
    \label{eq:conjugate_diffeo_derivative}
\end{equation}
which further implies that $(Z\circ f\circ Z)''(\phi)  = -f''(Z(\phi)) $. By using in \eqref{eq:overbar_cocycle} gives
\begin{equation}
	B(Z\circ f\circ Z,Z\circ g\circ Z) 
	= -\int_{0}^{2\pi}d \phi\,\frac{g''(Z(\phi))}{g'(Z(\phi))}\,\log{ f'(g(Z(\phi)))}\\
	= -\int_{-2\pi}^{0}d \phi\,\frac{g''(\phi)}{g'(\phi)}\,\log{ f'(g(\phi))}\,.
\end{equation}
Change of variable $\phi\rightarrow \phi+2\pi$ and periodicity of the integrand gives $B(Z\circ f\circ Z,Z\circ g\circ Z) =-B(f,g) $ so that
\begin{equation}
	\overbar{V}_{f} \overbar{V}_{g} = e^{-iB(f,g)}\overbar{V}_{f\circ g}\,.
\end{equation}
The adjoint action of the complex conjugate representation on a conjugate primary operator is given by
\begin{equation}
\overbar{V}_f\,\overbar{\mathcal{O}}_h(\phi)\,\overbar{V}_f^{\dagger} = V_{Z\circ f\circ Z}\,\mathcal{O}_h(Z(\phi))\,V_{Z\circ f\circ Z}^\dagger = (Z\circ f\circ Z)'(Z(\phi))^h\,\mathcal{O}_h((Z\circ f)(\phi))\,.
\end{equation}
By using \eqref{eq:conjugate_diffeo_derivative} it follows that
\begin{equation} \overbar{V}_f\,\overbar{\mathcal{O}}_h(\phi)\,\overbar{V}_f^{\dagger} =f'(\phi)^h\,\overbar{\mathcal{O}}_h(f(\phi))\,.
\end{equation}
The stress tensor transforms similarly with $h = 2$ and with an additional Schwarzian of $f$ since $\{(Z\circ f\circ Z)(Z(\phi)),Z(\phi)\} = \{f(\phi),\phi\}$.

\section{Primary-deformed unitary operator}
\label{sec:appendix_splittingtheunitary}

In this appendix, we derive necessary mathematical results needed for the study of primary-deformed Virasoro circuits.

\subsection{Factorization}

Let us show that the unitary operator
\begin{align}
    U(t)=\overleftarrow{\mathcal{T}}\exp{\biggl(-i\int_{0}^{t} ds\,G(s)\biggr)}\,.
\end{align}
with the generator $G(t) = C(t) + \lambda P(t)$ factorizes as $U(t) = V(t)\,U_P(t)$ where the individual unitaries are given in \eqref{eq:unitaryfactors}. We can prove this by writing $V(t)^\dagger \,U(t) = U_P(t)$ and showing that $V(t)^\dagger \,U(t)$ satisfies the same first-order differential equation as $U_P(t)$ which is
\begin{equation}
    \partial_t U_P(t) = -i\lambda\,\widetilde{P}(t)\,U_P(t)\,,\quad U_P(0) = 1\,.
\end{equation}
Clearly $V(t)^\dagger \,U(t)$ satisfies the same boundary condition since $U(0) = V(0) = 1$. By an explicit computation
\begin{align}
& \partial_t[V(t)^\dagger\,U(t)]\nonumber\\
&= \overrightarrow{\mathcal{T}}\exp{\biggl(i\int_{0}^{t} ds\,C(s)\biggr)}\,i\,(C(t) -G(t) )\,\overleftarrow{\mathcal{T}}\exp{\biggl(-i\int_{0}^{t} ds\,G(s)\biggr)}\nonumber\\
&= \overrightarrow{\mathcal{T}}\exp{\biggl(i\int_{0}^{t} ds\,C(s)\biggr)}\,(-i\lambda P(t))\,\overleftarrow{\mathcal{T}}\exp{\biggl(-i\int_{0}^{t} ds\,G(s)\biggr)}\nonumber\\
&= V(t)^\dagger\,(-i\lambda P(t))\,V(t)\,[V(t)^\dagger\,U(t)]\,,
\end{align}
where $V(t)^\dagger\,(-i\lambda P(t))\,V(t) = -i\lambda \,\widetilde{P}(t)$ which proves the claim. As an aside, note that this factorization lies at heart of the definition of the interaction picture in quantum mechanics.

\subsection{Perturbative expansion}\label{subapp:perturbative_expansion}

We want to compute
\begin{equation}
\widetilde{Q}(t) = U_P(t)^{\dagger}\,Q(t)\,U_P(t)\,,
\end{equation}
where the unitary operator is given by
\begin{equation}
    U_P(t) = \overleftarrow{\mathcal{T}}\exp{\biggl(-i\int_0^t ds\, \lambda \widetilde{P}(s)\biggr)}\,,
\end{equation}
which satisfies the equation
\begin{equation}
\partial_tU_P(t) = -i\, \lambda\widetilde{P}(t)\,U_P(t)\,.
\label{eq:unitary_equation_app}
\end{equation}
Instead of writing the solution as a time-ordered exponential, the equation \eqref{eq:unitary_equation_app} can also be solved using the Magnus expansion
\begin{equation}
U_P(t) = \exp{\biggl(-i\,\lambda\int_{0}^{t} ds\,\widetilde{P}(s) -\frac{\lambda^2}{2}\int_{0}^{t} ds_1\int_0^{s_1} ds_2\,[\widetilde{P}(s_1),\widetilde{P}(s_2)] + \mathcal{O}(\lambda^3)\biggr)}\,.
\end{equation}
Since the commutator of $ \widetilde{P}(s) $ and $ [\widetilde{P}(s_1),\widetilde{P}(s_2)] $ is of order $ \mathcal{O}(\lambda^3) $, it follows from the BCH formula that the exponential factorizes at quadratic order
\begin{equation}
U_P(t)  = \exp{\biggl(-i\,\lambda\int_{0}^{t} ds\,\widetilde{P}(s)\biggr)}\,\exp{\biggl(-\frac{\lambda^2}{2}\int_{0}^{t} ds_1\int_0^{s_1} ds_2\,[\widetilde{P}(s_1),\widetilde{P}(s_2)]\biggr)} + \mathcal{O}(\lambda^{3})\,.
\end{equation}
Hence we can write
\begin{equation}
U_P(t)^{\dagger}\,Q(t)\,U_P(t) = e^{A_2(t)}e^{A_1(t)}\,Q(t)\,e^{-A_1(t)}e^{-A_2(t)} + \mathcal{O}(\lambda^{3})\,,
\end{equation}
where we have defined the operators
\begin{equation}
A_1(t) \equiv i\,\lambda\int_{0}^{t} ds\,\widetilde{P}(s),\quad A_2(t) \equiv \frac{\lambda^2}{2}\int_{0}^{t} ds_1\int_0^{s_1} ds_2\,[\widetilde{P}(s_1),\widetilde{P}(s_2)]\,.
\label{eq:A1A2operators}
\end{equation}
Now we have the expansion
\begin{equation}
e^{\lambda A}Be^{-\lambda A}  = A+\lambda\,[A,B]+\frac{\lambda^{2}}{2}\,[A,[A,B]] + \mathcal{O}(\lambda^{3})\,,%A+\sum_{n=1}^{\infty}\frac{\lambda^{n}}{n!}[A,[A,\cdots,[A,B]]]
\label{eq:adjoint_expansion_app}
\end{equation}
which gives
\begin{equation}
U_P(t)^{\dagger}\,Q(t)\,U_P(t) = e^{A_2(t)}\,\bigl(Q(t)+[A_1(t),Q(t)]+\frac{1}{2}\,[A_1(t),[A_1(t),Q(t)]] \bigr)\,e^{-A_2(t)} + \mathcal{O}(\lambda^{3})\,.
\end{equation}
Applying the expansion \eqref{eq:adjoint_expansion_app} a second time gives
\begin{equation}
U_P(t)^{\dagger}\,Q(t)\,U_P(t) = Q(t)+[A_1(t),Q(t)]+[A_2(t),Q(t)]+\frac{1}{2}\,[A_1(t),[A_1(t),Q(t)]]+ \mathcal{O}(\lambda^{3})\,.
\end{equation}
Substituting \eqref{eq:A1A2operators}, we get explicitly
\begin{align}
&U_P(t)^{\dagger}\,Q(t)\,U_P(t)\label{eq:Qadjoint_1}\\
&= Q(t)+ i\,\lambda\int_{0}^{t}ds\,[\widetilde{P}(s),Q(t)]-\frac{\lambda^2}{2}\int_{0}^{t} ds_1\int_{0}^{t} ds_2\,[\widetilde{P}(s_1),[\widetilde{P}(s_2),Q(t)]]\\
&+\frac{\lambda^2}{2}\int_{0}^{t} ds_1\int_{0}^{s_1} ds_2\,[[\widetilde{P}(s_1),\widetilde{P}(s_2)],Q(t)] + \mathcal{O}(
\lambda^{3}).
\end{align}
Here we can write
\begin{align}
    &\frac{1}{2}\int_{0}^{t} ds_1\int_{0}^{t} ds_2\,[\widetilde{P}(s_1),[\widetilde{P}(s_2),Q(t)]]\\
    &= \frac{1}{2}\int_{0}^{t} ds_1\int_{0}^{s_1} ds_2\,[\widetilde{P}(s_1),[\widetilde{P}(s_2),Q(t)]]+\frac{1}{2}\int_{0}^{t} ds_1\int_{0}^{s_1} ds_2\,[\widetilde{P}(s_2),[\widetilde{P}(s_1),Q(t)]]
\end{align}
where the commutator in the second term can be written using the Jacobi identity as
\begin{equation}
    [\widetilde{P}(s_1),[\widetilde{P}(s_2),Q(t)]]= [\widetilde{P}(s_2),[\widetilde{P}(s_1),Q(t)]]+[[\widetilde{P}(s_1),\widetilde{P}(s_2)],Q(t)]\,,
\end{equation}
which gives
\begin{align}
    &\frac{1}{2}\int_{0}^{t} ds_1\int_{0}^{t} ds_2\,[\widetilde{P}(s_1),[\widetilde{P}(s_2),Q(t)]]\\
    &= \int_{0}^{t} ds_1\int_{0}^{s_1} ds_2\,[\widetilde{P}(s_2),[\widetilde{P}(s_1),Q(t)]]+\frac{1}{2}\int_{0}^{t} ds_1\int_{0}^{s_1} ds_2\,[[\widetilde{P}(s_1),\widetilde{P}(s_2)],Q(t)]
\end{align}
Substituting to \eqref{eq:Qadjoint_1} gives finally
\begin{align}
&U_P(t)^{\dagger}\,Q(t)\,U_P(t)\\
&= Q(t)+ i\,\lambda\int_{0}^{t}ds\,[\widetilde{P}(s),Q(t)]-\lambda^2\int_{0}^{t} ds_1\int_{0}^{s_1} ds_2\,[\widetilde{P}(s_2),[\widetilde{P}(s_1),Q(t)]]+\mathcal{O}(\lambda^3)\nonumber\,.
\end{align}

\section{2-point functions of ring operators}

In this appendix, we compute 2-point functions of the chiral and coupled ring operators defined in Section \ref{sec:deformedcircuits}.

\subsection{Chiral case}
\label{app:2pt_chiral}

We will compute $\bra{0} \mathcal{R}_h(t_1)\,\mathcal{R}_h(t_2)\ket{0}$ where
\begin{equation}
    \mathcal{R}_h(t) = \int_0^{2\pi} d\phi\,S(\phi)\,\mathcal{O}_h(\phi-t)\,.
\end{equation}
Using the mode expansion of $\mathcal{O}_h(\phi)$ we get
\begin{equation}
    \mathcal{R}_h(t) =\sum_{n=-\infty}^\infty S_n^*\,\mathcal{O}_{h,n}\,e^{-in t}\,,
\end{equation}
where we have also expanded the spatial source as
\begin{equation}
    S(\phi) = \sum_{n=-\infty}^\infty S_n\,e^{in\phi}\,,\quad S_{n}^* = S_{-n}\,.
    \label{eq:jmodeexpansionapp}
\end{equation}

We obtain
\begin{equation}
    \bra{0} \mathcal{R}_h(t_1)\,\mathcal{R}_h(t_2)\ket{0} = \sum_{n,m=-\infty}^{\infty} S_n^*\,S_m^* \bra{0} \mathcal{O}_{h,n}\,\mathcal{O}_{h,m}\ket{0}\,e^{-int_1-imt_2}
\end{equation}
Substituting \eqref{eq:OnOmvevGeneralapp} gives
\begin{equation}
     \bra{0} \mathcal{R}_h(t_1)\,\mathcal{R}_h(t_2)\ket{0} = \sum_{n=h}^\infty\, \vert S_n\vert^2\;\binom{h+n-1}{n-h}\, e^{-in(t_1-t_2)}\,.
\end{equation}

\subsection{Coupled case}
\label{app:2pt_coupled}

We will compute $\bra{0} \mathcal{R}_{h\bar{h}}(t_1)\,\mathcal{R}_{h\bar{h}}(t_2)\ket{0}$ where
\begin{equation}
    \mathcal{R}_{h\bar{h}}(t) = \int_0^{2\pi} d\phi\,S(\phi)\,\mathcal{O}_h(\phi-t)\otimes \overbar{\mathcal{O}}_{\overbar{h}}(\phi+t)\,.
\end{equation}
We have the mode expansion
\begin{equation}
	\mathcal{O}_{h}(\phi-t)\otimes \mathcal{O}_{\overbar{h}}(-\phi-t)= \frac{1}{(2\pi)^{2}}\sum_{n,m} \mathcal{O}_{h,n}\otimes \mathcal{O}_{\overbar{h},m}\,e^{i(n-m)\phi}\,e^{-i(n+m)t}
\end{equation}
which gives
\begin{equation}
    \mathcal{R}_{h\bar{h}}(t) =\frac{1}{2\pi}\sum_{n,m} S_{n-m}^*\, \mathcal{O}_{h,n}\otimes \mathcal{O}_{\overbar{h},m}\,e^{-i(n+m)t}
\end{equation}
Thus we obtain
\begin{align}
    &\bra{0} \mathcal{R}_{h\bar{h}}(t_1)\,\mathcal{R}_{h\bar{h}}(t_2)\ket{0}\nonumber\\
    &= \frac{1}{(2\pi)^2}\sum_{p,q,k,l}S_{p-q}^*\,S_{k-l}^* \bra{0}\mathcal{O}_{h,p}\,\mathcal{O}_{h,k}\ket{0}\bra{0}\mathcal{O}_{\overbar{h},q}\,\mathcal{O}_{\overbar{h},l}\ket{0}\,e^{-i(p+q)t_1}\,e^{-i(k+l)t_2}
\end{align}
Substituting the previously derived mode representation of the chiral primary correlator \eqref{eq:OnOmvevGeneralapp} and assuming the spinless case $h=\bar{h}$ for the full primary leads to
\begin{align}
    &\bra{0} \mathcal{R}_{hh}(t_1)\,\mathcal{R}_{hh}(t_2)\ket{0}\nonumber\\
    &=\sum_{p,q}\frac{\vert S_{p-q}\vert^2}{(2\pi)^2}
    \binom{h+p-1}{p-h}
    \binom{h+q-1}{q-h}
    \,\Theta(p-h)\,\Theta(q-h)\,e^{-i(p+q)(t_1-t_2)}\\
    &=\sum_{n=-\infty}^\infty\frac{\vert S_{n}\vert^2}{(2\pi)^2}\,e^{in\Delta t}\sum_{p=-\infty}^\infty  \binom{h+p-1}{p-h}\binom{h+p-n-1}{p-n-h}\,\Theta(p-h)\,\Theta(p-n-h)\,e^{-2ip\,\Delta t}\,,\nonumber
\end{align}
where we have defined $n = p-q$ and used the shorthand notation $\Delta t\equiv t_1-t_2$. One can show that the sum that appears can be resummed as
\begin{align}
    &\sum_{p=-\infty}^\infty  \binom{h+p-1}{p-h}\binom{h+p-n-1}{p-n-h}\,\Theta(p-h)\,\Theta(p-n-h)\,z^p \nonumber\\&=
    \begin{dcases}
       \frac{ z^{h+n}\,(2 h+n-1)! \, _2F_1\left(2 h+n,2 h;n+1;z\right)}{(2 h-1)!\, n!}, &n> 0\\
        \frac{z^{h}\, (2 h-n-1)! \, _2F_1\left(2 h-n,2 h;1-n;z\right)}{(2 h-1)!\, (-n)!}, &n\leq 0
    \end{dcases}\,.
\end{align}
This lets us rewrite the correlator to 
\begin{align}
    &\bra{0} \mathcal{R}_{hh}(t_1)\,\mathcal{R}_{hh}(t_2)\ket{0}\nonumber\\
    &=\frac{1}{(2\pi)^2}\sum_{n=-\infty}^\infty\vert S_{n}\vert^2\,\frac{ (2 h+|n|-1)! }{(2 h-1)! \,|n|!}e^{-i(2h+|n|)\Delta t}\, _2F_1\left(2 h+|n|,2 h;1+|n|;e^{-2 i \Delta t}\right)\,.
    \label{eq:2pointmathematica_output}
\end{align}
This expression can be simplified using properties of the hypergeometric function.  Since only the absolute value $\vert n \vert$ appears, it is enough to consider $n\geq 0$. The first identity we use is given by
\begin{equation}
	_2F_1(a,b;a-b+1;z) = (1+z)^{-a}\, _2F_1\biggl(\frac{a}{2},\frac{a+1}{2};a-b+1;\frac{4z}{(1+z)^{2}}\biggr)\,,
\end{equation}
which holds for $|z| < 1$. This condition holds assuming $\Delta t$ has a small negative imaginary part and we get
\begin{align}
	&e^{-i(2h+n)\Delta t}\,_2F_1\bigl(2 h,2 h+n;n+1;e^{-2 i \Delta t}\bigr)\\
	&= \frac{[\sec{(\Delta t)}]^{2h+n}}{2^{2h+n}}\, _2F_1\biggl(h+\frac{n}{2},h+\frac{n+1}{2};n+1;\sec^{2}{(\Delta t)}\biggr)\,.
\end{align}
This has both a real and an imaginary part since $\sec^{2}{(\Delta t)} \in (1,\infty)$ for $\Delta t\in \mathbb{R}$. To make the real and imaginary parts explicit, we use the second identity
\begin{align}
	_2F_1(a,b;c;z) &=  \frac{\Gamma(b-a)\,\Gamma(c)}{\Gamma(b)\,\Gamma(c-a)}\,(-z)^{-a}\, _2F_1(a,a-c+1;a-b+1;z^{-1})\\
	&-\frac{\Gamma(a-b)\,\Gamma(c)}{\Gamma(a)\,\Gamma(c-b)}\,(-z)^{-b}\, _2F_1(b,b-c+1;b-a+1;z^{-1})\,,
\end{align}
which is valid when $z \notin (0,1)$ and $a-b \notin \mathbb{Z}$. We get
\begin{align}
	&\frac{[\sec{(\Delta t)}]^{2h+n}}{2^{2h+n}}\,_2F_1\biggl(h+\frac{n}{2},h+\frac{n+1}{2};n+1;\sec^{2}{(\Delta t)}\biggr)\label{eq:decompositionhyper}\\
	&=  \frac{(-1)^{h+n}\,i^{n}\,\sqrt{\pi}\,n!}{2^{2h+n}\,\Gamma(h + \frac{n+1}{2})\,\Gamma(\frac{n+2}{2}-h)}\, _2F_1\biggl(h+\frac{n}{2},h-\frac{n}{2};\frac{1}{2};\cos^{2}{(\Delta t)}\biggr)\nonumber\\
	&+\frac{(-1)^{h+n}\,i^{n+1}\,\sqrt{\pi}\,n!}{2^{2h+n-1}\,\Gamma(h + \frac{n}{2})\,\Gamma(\frac{n+1}{2}-h)}\,\cos{(\Delta t)}\, _2F_1\biggl(h+\frac{n+1}{2},h-\frac{n-1}{2};\frac{3}{2};\cos^{2}{(\Delta t)}\biggr)\,.\nonumber
\end{align}
In this expression, the hypergeometric functions are real since $\cos^{2}{(\Delta t)} \in (0,1)$ for $\Delta t\in \mathbb{R}$ so that the real and imaginary part are therefore controlled by the $i^n$, $i^{n+1}$ factors so that it is useful to consider even and odd $n$ separately. For even $n = 2m$ the first term gives the real part while for odd $n = 2m+1$ it is the second term. In both cases, the denominator contains $\Gamma(m+1-h)$ which has a pole when $m< h$. Therefore the real part vanishes when $m < h$:
\begin{equation}
    \text{Re}\,e^{-i(2h+n)\Delta t}\,_2F_1\bigl(2 h,2 h+n;n+1;e^{-2 i \Delta t}\bigr) = 0\,,\quad \begin{dcases}
        n< 2h\,,&\text{even }n\\
        n<2h+1\,,&\text{odd }n
    \end{dcases}\,.
\end{equation}
When $m \geq  h$ corresponding to $n\geq 2h$ ($n$ even) and $n\geq 2h+1$ ($n$ odd), we get
\begin{align}
	&\frac{(2 h+n-1)!}{(2 h-1)!\, n!}\,\text{Re}\,e^{-i(2h+n)\Delta t}\,_2F_1\bigl(2 h+n,2 h;n+1;e^{-2 i \Delta t}\bigr)\\
	&=\frac{(-1)^{h+m}(m+h)!}{(2h-1)!\,(m-h)!}\times
 \begin{dcases}
		\frac{1}{2\,(m+h)}\, _2F_1\biggl(h+m,h-m;\frac{1}{2};\cos^{2}{(\Delta t)}\biggr)\,, &n = 2m \\
		\cos{(\Delta t)}\, _2F_1\biggl(h+m+1,h-m;\frac{3}{2};\cos^{2}{(\Delta t)}\biggr)\,, &n = 2m+1
	\end{dcases}\nonumber
\end{align}
The real part of the 2-point function becomes
\begin{align}
	&\text{Re}\bra{0} \mathcal{R}_{hh}(t_1)\,\mathcal{R}_{hh}(t_2)\ket{0}\\
	&=\frac{1}{(2\pi)^2}\sum_{m=h}^\infty \frac{(-1)^{h+m}\,(m+h)!}{(2h-1)!\,(m-h)!}\,\biggl[\frac{\vert S_{2m}\vert^2}{2\,(m+h)}\, _2F_1\biggl(h-m,h+m;\frac{1}{2};\cos^{2}{(\Delta t)}\biggr)\biggr.\nonumber\\
	&\hspace{5cm}\biggl.+\vert S_{2m+1}\vert^2\,\cos{(\Delta t)}\, _2F_1\biggl(h-m,h+m+1;\frac{3}{2};\cos^{2}{(\Delta t)}\biggr)\biggr]\,.\nonumber
\end{align}
For $m\geq h$ the hypergeometric functions are finite polynomials of their arguments
\begin{equation}
    _2F_1(h-m,b;c;z) = \sum_{k=0}^{m-h}(-1)^k\,\binom{m-h}{k}\frac{(b)_k}{(c)_k}\,z^k\,.
\end{equation}
This implies that the real part is finite at $\Delta t = 0$, because the values of the hypergeometric functions are finite there $_2F_1(h-m,b;c;1) = \frac{(c-b)_{m-h}}{(c)_{m-h}}$. The coincident limit can be obtained by using the expansion
\begin{align}
    &\frac{(2 h+\vert n\vert -1)!}{(2 h-1)!\, \vert n\vert !}\,e^{-i(2h+\vert n\vert )\Delta t}\,_2F_1\bigl(2 h+\vert n\vert ,2 h;\vert n\vert +1;e^{-2 i \Delta t}\bigr)\label{eq:niceexpansion}\\
    &= -\frac{i}{\Delta t^{4h-1}}\frac{(4h-2)!}{2^{4h-1}(2h-1)!^2}\, [1+ \mathcal{O}(\Delta t^2)]+\frac{1}{2}\frac{\Gamma(2h+\vert n\vert )}{\Gamma(4h)\,\Gamma(\vert n\vert -2h+1)}\,[1+ \mathcal{O}(\Delta t^2)]\,,\nonumber
\end{align}
when $ \Delta t\rightarrow 0$. The coefficients of the subleading terms are real so that the divergent terms in the first series are purely imaginary, while the terms in the second series are real. Due to the poles of the Gamma function, the real part vanishes when $\vert n\vert < 2h$. Thus the coincident limit of the real part of the 2-point function \eqref{eq:2pointmathematica_output} is given by
\begin{equation}
    \label{eq:coupled_re_coincidencelimit_app}
    \lim_{t_2\rightarrow t_1}\Re\bra{0} \mathcal{R}_{hh}(t_1)\,\mathcal{R}_{hh}(t_2)\ket{0}=\frac{1}{(2\pi)^2}\sum_{n=2h}^\infty \binom{2h+n -1}{ n -2h}\,\vert S_{n}\vert^2\,.
\end{equation}
The expansion \eqref{eq:niceexpansion} also implies the imaginary part $\Im\bra{0} \mathcal{R}_{hh}(t_1)\,\mathcal{R}_{hh}(t_2)\ket{0}$ is divergent in the coincident limit.

\end{appendix}

\bibliographystyle{JHEP}
\bibliography{bib}

\end{document}